\documentclass[%
 reprint,
 amsmath,amssymb,
 aps,
longbibliography]{revtex4-1}

\usepackage{graphicx}
\usepackage{dcolumn}
\usepackage{bm}
\usepackage{braket}
\newcommand{\qop}{{\bf q}}

\newcommand{\Pop}{{\bf p}}

\newcommand{\PP}{{\bf P}}

\newcommand{\bsigma}{\boldsymbol{\sigma}}
\newcommand{\rhoo}{\boldsymbol{\rho}}

\newcommand{\D}{{\bf D}}

\newcommand{\HH}{{\bf H}}

\newcommand{\UU}{{\bf U}}
\newcommand{\OO}{{\bf O}}

\newcommand{\II}{{\bf 1}}

\newcommand{\LL}{{\bf L}}
\newcommand{\MM}{{\bf M}}

\newcommand{\rop}{{\bf r}}

\usepackage{soul}
\usepackage{siunitx}

\newcommand{\sigmax}{{\bf \bsigma}_x}
\newcommand{\sigmay}{{\bf \bsigma}_y}
\newcommand{\sigmaz}{{\bf \bsigma}_z}
\newcommand{\sigmam}{{\bf \bsigma_-}}
\newcommand{\sigmap}{{\bf \bsigma_+}}

\usepackage{hyperref}

\begin{document}

\preprint{APS/123-QED}

\title{Robust suppression of  noise propagation in GKP  error-correction}

\author{C. Siegele}
\affiliation{Laboratoire de Physique de l’Ecole normale supérieure, Mines-Paristech, Inria, ENS-PSL, Université PSL, CNRS, Sorbonne Université, Paris, France}
 \author{P. Campagne-Ibarcq}%
 \email{philippe.campagne-ibarcq@inria.fr}
\affiliation{Laboratoire de Physique de l’Ecole normale supérieure, Mines-Paristech, Inria, ENS-PSL, Université PSL, CNRS, Sorbonne Université, Paris, France}

\date{\today}

\begin{abstract}

Straightforward logical operations  contrasting with  complex state preparation  are the hallmarks of the bosonic encoding proposed by Gottesman, Kitaev and Preskill (GKP). The recently reported  generation and error-correction of GKP qubits in trapped ions and superconducting circuits thus holds great promise for the future of quantum computing architectures based on such encoded qubits. However,  these experiments rely on   error-syndrome detection  via  an auxiliary physical qubit, whose noise may propagate and corrupt the encoded GKP qubit. We propose a simple module composed of two oscillators and a physical qubit, operated with two experimentally accessible quantum gates and elementary feedback controls to implement an error-corrected GKP qubit protected from  such propagating errors. In the idealized setting of periodic GKP states, we develop efficient numerical methods to optimize our protocol parameters and show  that errors of the encoded  qubit stemming  from flips of the physical qubit and diffusion of the oscillators state in phase-space may be exponentially suppressed as the  noise strength over individual operations is decreased. Our approach circumvents the main roadblock towards fault-tolerant quantum computation with GKP qubits.
\end{abstract}

\maketitle


\section{Introduction}

In their seminal paper~\cite{gottesman2001encoding,grimsmo2021quantum}, Gottesman, Kitaev, and Preskill proposed to encode, within the vast Hilbert space of a harmonic oscillator, a qubit  robustly against position and momentum shifts of  the embedding oscillator. Clifford operations on encoded GKP qubits are straightforward to implement and do not amplify small shift errors. Therefore, concatenation of  the GKP code into  the surface code recently attracted interest~\cite{fukui2018high,vuillot2019quantum,terhal2020towards,noh2020fault,noh2022low} as, beyond the potentially enhanced coherence of  GKP qubits compared to faulty physical qubits,  analog information from the GKP error-correction layer may be decoded to improve the  surface code threshold. Crucially, these desirable features rely on the assumption that noise-induced shifts of the  embedding  oscillators are short and can be detected before they accumulate. This hypothesis is not valid in current experimental implementations with superconducting circuits~\cite{campagne2020quantum,sivak2023real}. In order to comprehend this serious limitation, one needs to delve  into the code structure and error-correction techniques employed in these experiments.\\

	\begin{figure}[htbp] 
		\centering
		\includegraphics[width=1.0\columnwidth]{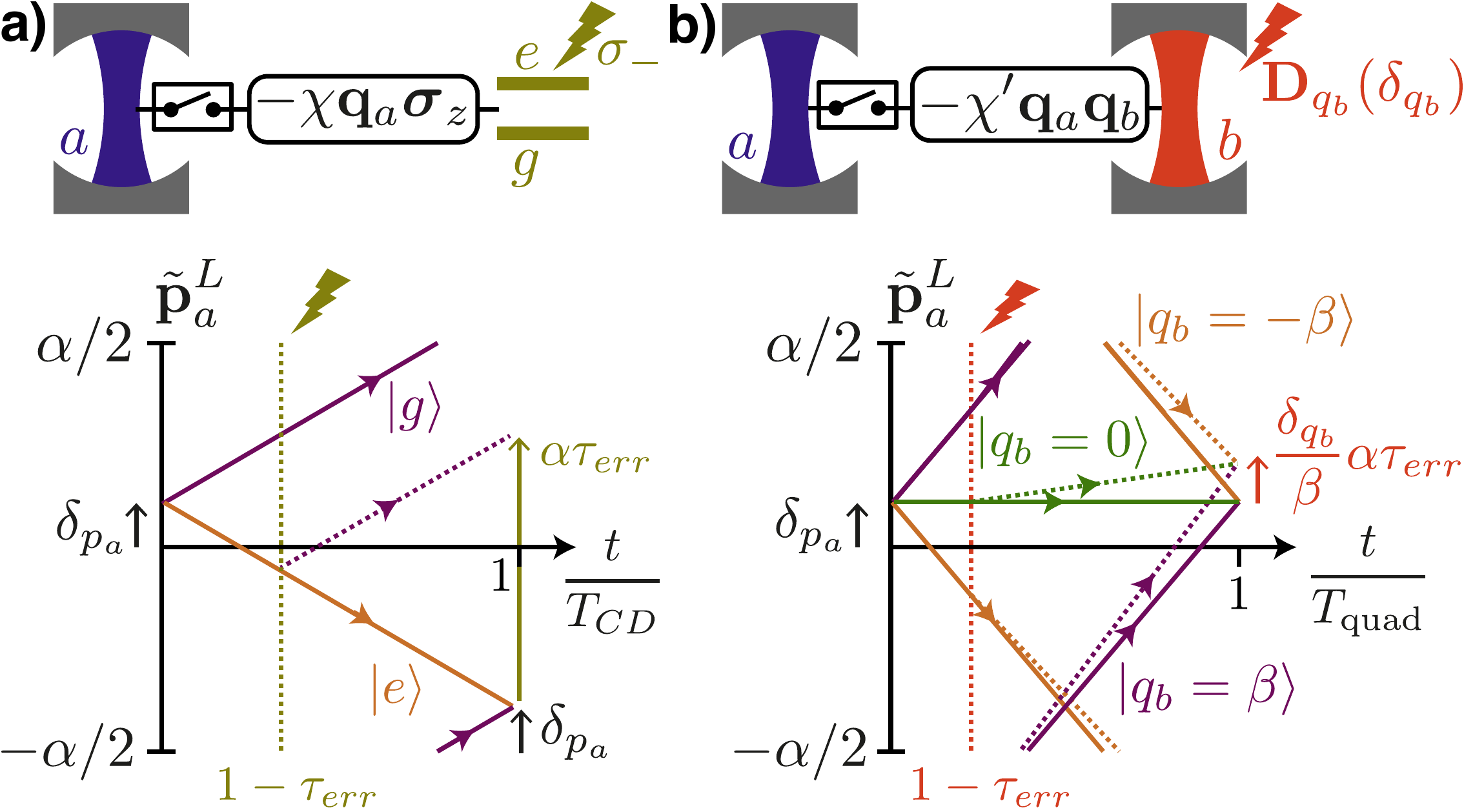}
		\caption{ {\bf Error propagation during  a)  qubit-based error-correction.} A Rabi-type interaction at rate $\chi$ is activated for a duration $T_{CD}=\alpha/(2\chi)$ to map the value of the target  modular stabilizer $\tilde{\qop}^S_a$ onto the phase of an auxiliary qubit prepared in $|+x\rangle=(|g\rangle+|e\rangle)/\sqrt{2}$ beforehand. As a backaction,  the modular logical operator $\tilde{\Pop}^L_a$ gets shifted conditioned on the qubit state (conditional trajectories shown in color for an oscillator initially in an eigenstate of $\tilde{\Pop}^L_a$ with eigenvalue $\delta_{p_a}$). At $T_{CD}$ and in absence of error,  $\tilde{\Pop}^L_a$  retrieves its initial value but for a deterministic shift by $\alpha/2$. Relaxation of the qubit to $|g\rangle$  at time $(1-\tau_{err}) T_{CD}$ (with $\tau_{err}\in [0,1]$, gold lightning symbol) propagates as a long shift of $\tilde{\Pop}^L_a$ by $\alpha \tau_{err}$ (dashed trajectory). {\bf b) Steane-type error-correction.} A quadrature interaction at rate $\chi'$ with an auxiliary oscillator $b$, initially  in  $|\o\rangle=\sum_n |q_b=n\beta\rangle$, maps the value of $\tilde{\qop}^S_a$ onto the  modular stabilizer $\tilde{\Pop}_b$. As a backaction,   $\tilde{\Pop}^L_a$ gets shifted conditioned on the  position of the auxiliary oscillator (conditional trajectories shown in color for three position states). At $T_{\mathrm{quad}}=\alpha/(\beta \chi')$ and in absence of error, $\tilde{\Pop}^L_a$  retrieves its initial value. A  shift of $\tilde{\qop}_b$ by $ \delta_{q_b}\ll \beta$ occurring at $(1-\tau_{err}) T_{\mathrm{quad}}$ (red lightning symbol) propagates as a short shift of $\tilde{\Pop}^L_a$ by  $ \frac{\delta_{q_b}}{\beta}\alpha \tau_{err}$ (dashed trajectories).} 
		\label{fig1}
	\end{figure}

In reduced phase-space coordinates $(q_a,p_a)$~\footnote{operators \( \hat{q}_a \) and \( \hat{p}_a \) with equal fluctuations and verifying \( [\hat{q}_a, \hat{p}_a] = i \)}, the basis states of the square GKP code are superpositions of periodically spaced position eigenstates 
  \begin{equation}
	 \label{infiniteStates}
        |+Z\rangle=\sum_{n\in \mathbb{Z}}|q_a= n \alpha \rangle  \qquad
         |-Z\rangle=\D_{q_a}(\frac{\alpha}{2}) |+Z\rangle.
 \end{equation}
where $\alpha=2\sqrt{\pi}$ and  the  operator   $\D_{r_a}(\delta)$  displaces the oscillator state by $\delta$ along $r_a$, for $r_a=q_a$ or $r_a=p_a$. The logical states $|\pm X \rangle$  are obtained by  a $\pi/2$ rotation in phase-space of $|\pm Z \rangle$. Note that   infinitely delocalized states are unrealistic, but the 
   essential properties  and control techniques considered in our work apply to states  normalized by a broad Gaussian envelope in phase-space~\cite{gottesman2001encoding,campagne2020quantum,royer2020stabilization,de2022error}. One may measure the GKP qubit in the $|\pm Z\rangle$ or $|\pm X\rangle$ basis by detecting the \emph{modular logical operators} $\tilde{\qop}^L_a= \qop_a~\mathrm{mod}~\alpha$ and $\tilde{\Pop}^L_a= \Pop_a~\mathrm{mod}~\alpha$. Crucially, a code state $|\Psi\rangle$  shifted in position and momentum can still be  correctly decoded as long as the shifts are shorter than $\alpha/4$. Moreover, these shifts can be detected without revealing the GKP qubit state by measuring the two commuting  \emph{ modular stabilizers} $\tilde{\qop}^S_a=\qop_a~\mathrm{mod}~\alpha/2$ and $\tilde{\Pop}^S_a=\Pop_a~\mathrm{mod}~\alpha/2$.\\
  
 Measuring the modular stabilizers  without extracting logical information is the main  challenge in   GKP error-correction~\cite{kitaev1995quantum,travaglione2002preparing,Pirandola2006,svore2013faster,terhal2016encoding,motes2017encoding,weigand2020realizing}. It was only recently achieved experimentally with trapped ions~\cite{fluhmann2018sequential,fluhmann2018encoding,de2022error} and  superconducting circuits~\cite{campagne2020quantum,sivak2023real}. In these experiments, the target oscillator is coupled to an auxiliary qubit  via a controllable Rabi-type interaction static in the interaction picture $-\chi \rop_a \sigmaz$ (where $\rop_a=\qop_a$ or $\rop_a=\Pop_a$, see Fig.~\ref{fig1}a), in order to implement a conditional displacement gate $\UU^{CD}_{r_a}=e^{i \frac{\alpha}{2} \rop_a \sigmaz}$ that rotates the qubit phase by   $-\alpha\tilde{\rop}^S_a$ around the $z$-axis of its Bloch sphere. The gate is named after its backaction on the oscillator, which is displaced by $\pm \frac{\alpha}{2}$ along the $\pi/2$-rotated quadrature $r^{\perp}_a$ conditioned on the qubit state (see Fig.~\ref{fig1}a). This evolution deterministically shifts the logical operator $\tilde{\rop}^{\perp L}_a$ by $\alpha/2$---accounted for in software---but otherwise leaves all modular operators unchanged. However, if a  bit-flip of the qubit occurs during the evolution, for instance due to energy relaxation as in Fig.~\ref{fig1}, the  displacement takes a  value  uniformly sampled in $[-\frac{\alpha}{2},\frac{\alpha}{2}]$ depending on the  unknown instant of the flip (see Appendix~\ref{sec:noise}). This randomizes the value of  $\tilde{\rop}^{{\perp} L}_a$ and the error propagates at the logical level with probability 1/2. These propagating errors, which become more frequent as the error-correction clock rate increases, are a serious bottleneck towards fault-tolerant quantum computation with GKP qubits. 
Various strategies were proposed~\cite{kapit2018error,puri2019stabilized,shi2019fault,ma2020path,vy2013error} and experimentally tested~\cite{rosenblum2018fault,reinhold2020error} to mitigate this advert effect, but either provide only a first-order protection against auxiliary oscillator errors~\cite{kapit2018error, shi2019fault,ma2020path,vy2013error} or rely on a biased-noise auxiliary qubit~\cite{puri2019stabilized} whose development is not yet mature enough~\cite{grimm2020stabilization,frattini2022squeezed} to 
unleash the full potential of GKP qubits.  \\

 This roadblock is not present in the so-called Steane-type error-correction scheme~\cite{gottesman2001encoding, glancy2006error}, where the target mode is probed via a quadrature interaction $-\chi' \rop_a \qop_b$  with an auxiliary oscillator $b$  to implement a quadrature gate $\UU^{\mathrm{quad}}_{r_a} = e^{i\frac{\alpha}{\beta}\rop_a \qop_b}$. The auxiliary oscillator is itself  prepared in a rectangular GKP state 
   \begin{equation}
	 \label{qunaught}
        |\o{}\rangle=\sum_{n\in \mathbb{Z}}|q_b=n \beta \rangle
 \end{equation}
 prior to the interaction. Since this state is employed as a displacement sensor~\cite{duivenvoorden2017single} and does not encode logical information, we  define only modular stabilizers $\tilde{\qop}_b=\qop_b~\mathrm{mod}~\beta$ and $\tilde{\Pop}_b=\Pop_b~\mathrm{mod}~2\pi/\beta$, of whom  $|\o{}\rangle$ is the single joint eigenstate with eigenvalue 0~\cite{gottesman2001encoding}. The quadrature gate displaces the auxiliary oscillator along $p_b$ conditioned on the value of $\rop_a$ while, reciprocally, the target oscillator  is shifted along $r^{\perp}_a$ conditioned on the value of $\qop_b$ (see Fig.~\ref{fig1}b). We  summarize the gate effect on modular operators as~\footnote{With the definition $\tilde{\Pop}^{\perp L}_a=-\tilde{\qop}^L_a$}
 \begin{equation}
 \label{stabevolve}
 \frac{\tilde{\Pop}_b}{2\pi/\beta} \rightarrow \frac{\tilde{\Pop}_b}{2\pi/\beta} + \frac{\tilde{\rop}^S_a}{\alpha/2} \qquad \quad \frac{\tilde{\rop}^{\perp L}_a}{\alpha} \rightarrow \frac{\tilde{\rop}^{\perp L}_a}{\alpha} + \frac{\tilde{\qop}_b}{\beta}
 \end{equation}

 The crucial difference with physical qubit-based error-correction lies in the  noise model, assumed to  only generate short shifts of its state. A  shift by $\delta_{q_b}$ along $\qop_b$, occurring before or during the gate, propagates to the target oscillator as a  shift shorter than $\frac{\delta_{q_b}}{ \beta}\alpha $   ($\tau_{err} \in [0,1]$ in Fig.~\ref{fig1}b), correctable if $\delta_{q_b}\ll\beta$. However, if the auxiliary qubit is prepared  through a series of qubit-based measurements of its stabilizers, bit-flips of the qubit may induce  shifts along $q_b$ covering the whole $[-\frac{\beta}{2},\frac{\beta}{2}]$ interval, propagating as  shifts of the target oscillator covering $[-\frac{\alpha}{2},\frac{\alpha}{2}]$ irrespective of the value of $\beta$. In GKP-surface code architectures, these \emph{structureless} shifts cancel the benefits of GKP qubits with respect to physical qubits. Therefore, a central question for the viability of Steane-type error-correction is: how can we ensure a supply of  auxiliary oscillator states $|\o{}\rangle$  whose errors do not propagate as long shifts of the target oscillator?

\section{Asymmetric  preparation of auxiliary oscillator}

\begin{figure}[htbp] 
		\centering
		\includegraphics[width=1.0\columnwidth]{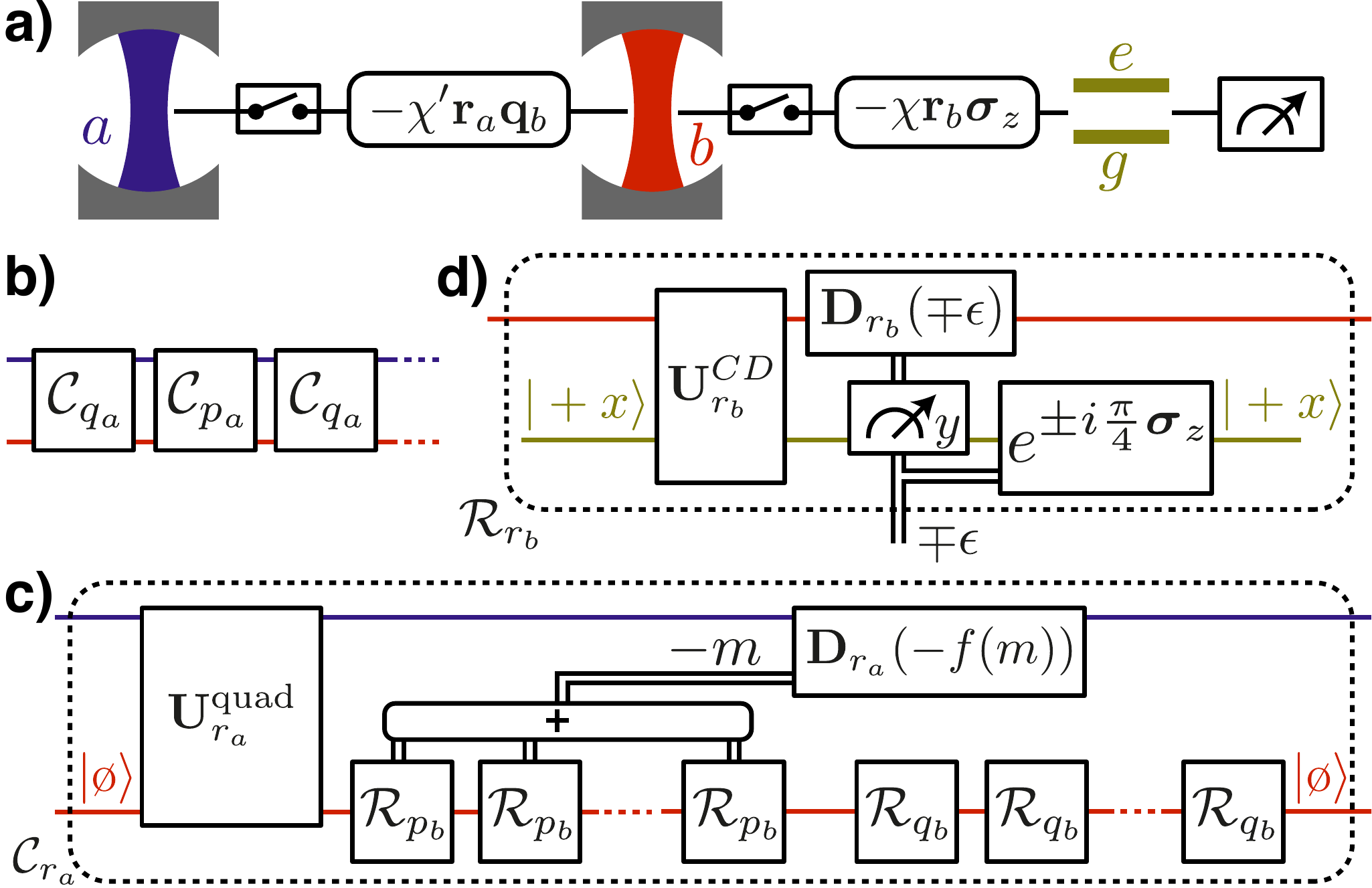}
		\caption{ {\bf a)} In our proposed architecture, the target oscillator $a$ couples to an auxiliary oscillator $b$ via a controlled quadrature interaction. The auxiliary state is prepared and measured via a physical qubit. {\bf b)} Alternating $\mathcal{C}_{q_a}$ and $\mathcal{C}_{p_a}$ correction cycles  protect the GKP qubit. {\bf c)} A $\mathcal{C}_{r_a}$ correction cycle ($r_a=q_a$ or $r_a=p_a$) starts with the auxiliary oscillator  prepared in $|\o\rangle$. The quadrature gate maps the value of the target stabilizer  $\tilde{\rop}^S_a$ to the  stabilizer $\tilde{\Pop}_b$. The auxiliary state is then measured and prepared for the next cycle by a series of $\mathcal{R}_{p_b}$ rounds \emph{followed} by a series of $\mathcal{R}_{q_b}$ rounds,   robustly suppressing  propagating errors. The measurement record from $\mathcal{R}_{p_b}$ rounds is  summed to estimate the value of $\tilde{\Pop}_b$ following the quadrature gate as detailed in Appendix~\ref{sec:phasestimation} (double black lines represent classical communication channels). The result $-m$ controls a  displacement by $-f(m)$ on the target oscillator (details on the feedback law $f$ in Appendix~\ref{sec:gradient}).  {\bf d)} A $\mathcal{R}_{r_b}$ round ($r_b=q_b$ or $r_b=p_b$) starts with the qubit prepared in $|+x\rangle$. A conditional displacement gate maps the value of $\tilde{\rop}_b$ to the qubit phase. The final qubit measurement along $\sigmay$ controls a proportional feedback displacement by $\mp \epsilon$, a conditional flip of the qubit to reset it in $|+x\rangle$, and the outcome is stored for further processing.  }
		\label{fig2}
	\end{figure}

We consider the  module depicted in Fig.~\ref{fig2}a where the target oscillator  interacts  with an  auxiliary oscillator,  itself coupled to a physical qubit. The target oscillator is corrected by repeated Steane-type correction \emph{cycles} denoted $\mathcal{C}_{r_a}$,  alternating $r_a=q_a$ and $r_a=p_a$ (Fig.~\ref{fig2}b). Each cycle starts with the  auxiliary oscillator  prepared in  $|\o\rangle$  (Fig.~\ref{fig2}c), possibly shifted due to preparation errors. A quadrature gate $\UU_{r_a}^{\mathrm{quad}}$ maps 
the value of $\tilde{\rop}^S_a$ onto the stabilizer $\tilde{\Pop}_b$. The auxiliary oscillator  is then measured and re-prepared through a sequence of  preparation \emph{rounds}  labeled $\mathcal{R}_{r_b}$ (for $r_b=q_b$ or $r_b=p_b$). Each round is built around a conditional displacement gate $\UU_{r_b}^{CD}$ mapping the value of $\tilde{\rop}_b$ onto the phase of the qubit, prepared beforehand in an eigenstate of $\sigmax$ and subsequently measured along $\sigmay$ (Fig.~\ref{fig2}b). Each qubit measurement  controls a proportional feedback displacement by $\pm \epsilon$ along $r_b$. As detailed below, repeated $\mathcal{R}_{r_b}$ rounds corral the auxiliary state  toward $\tilde{r}_b=0$~\cite{campagne2020quantum}. We  further store  the measurement record outputted  by the $\mathcal{R}_{p_b}$  rounds as it encodes the value of $\tilde{\Pop}_b$ following the quadrature gate, i.e. the target error-syndrome(see Appendix~\ref{sec:phasestimation}). After straightforward decoding, a corrective feedback displacement is applied to the target oscillator, concluding the correction cycle.\\

\label{sec:ancillaprepmain}
	\begin{figure}[htbp] 
		\centering
		\includegraphics[width=1.0\columnwidth]{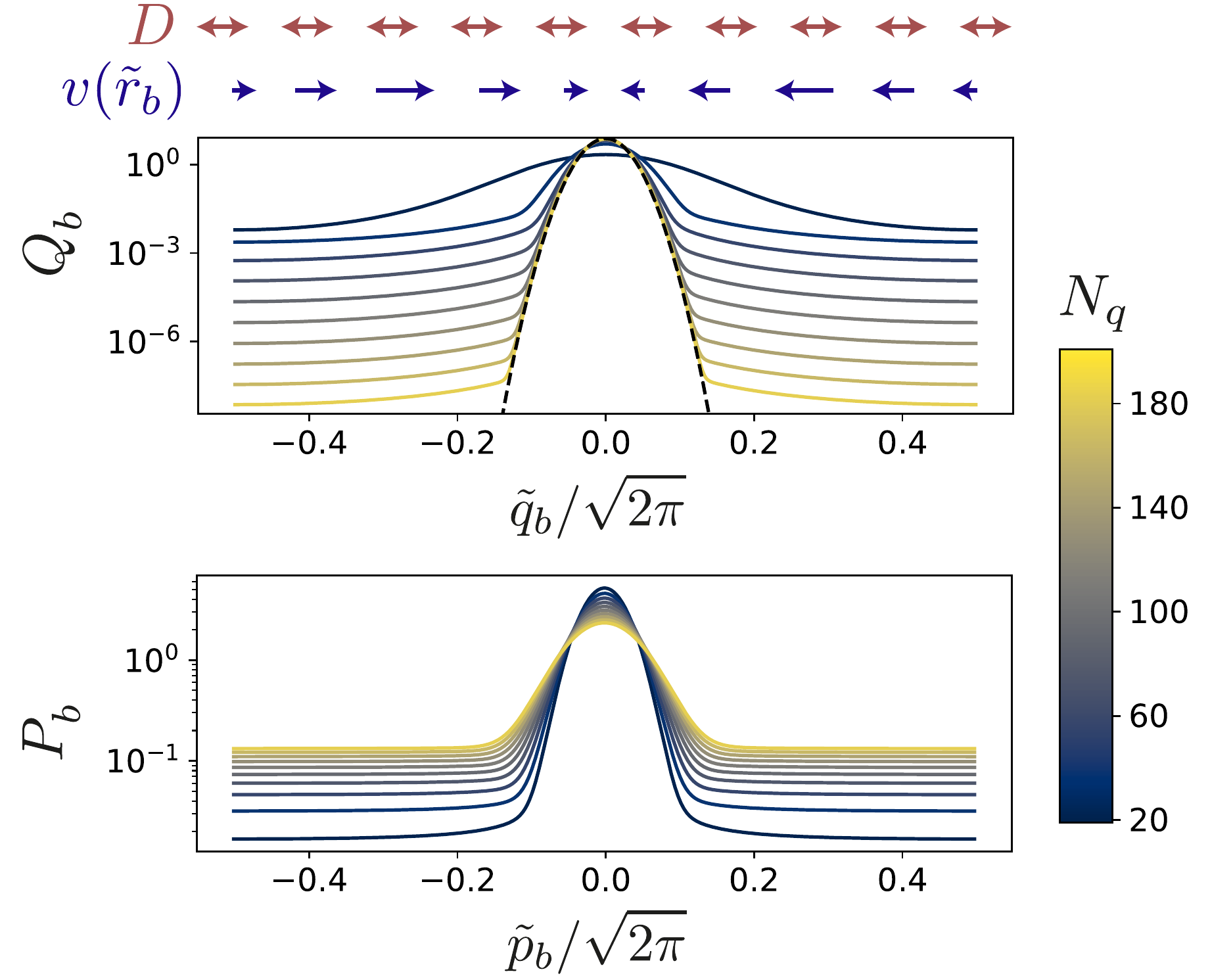}
		\caption{ Wrapped distributions $Q_b$ and $P_b$ of the  stabilizers $\tilde{q}_b$ and $\tilde{p}_b$ (for a square lattice corresponding to  $\beta=\sqrt{2\pi}$) prepared from a uniform distribution by  $N_p+N_q$  preparation rounds ($N_p=50$, varying $N_q$ encoded in color), in presence of quadrature noise at rate  $\kappa=(10^5~T_{\mathrm{round}})^{-1}$ and physical qubit flips with probabilities $p^{BF}=2 p^{PF}=0.002$ per round.  Arrows  above the top panel schematically represent the drift velocity (single-headed blue arrows) and  diffusion constant (double-headed brown arrows) of each distribution $R_b$ ($R_b=Q_b$ or $R_b=P_b$)   during corresponding $\mathcal{R}_{r_b}$ rounds. Vanishing drift velocity in the neighborhood of $\tilde{r}_b=0.5$ results in persisting tails of the distributions. The length  of feedback displacements $\epsilon_j$  is varied throughout the preparation to mitigate this effect while maintaining a minimal variance  for the central peak (see text, the black dashed line is a Gaussian with variance $V_{min}$), ensuring robust suppression of error propagation to the target mode. As $N_q$ increases, bit-flips of the qubit entail more frequent shifts along $p_b$ elevating the tails of $P_b$, and the  central peak of $P_b$ deflates due to quadrature noise. }
		\label{fig3}
	\end{figure}
 
Our proposal to suppress error-propagation is based on two observations. First, if the  oscillators only interact via the $\qop_b$ quadrature operator of the auxiliary one (see Fig.~\ref{fig2}a), only shifts along this quadrature  directly propagate to the target oscillator (second term in Eq.~\eqref{stabevolve}). As a consequence, the auxiliary state may be asymmetrically prepared, with a focus on preparing a probability distribution sharply peaked near 0 for  the stabilizer $\tilde{\qop}_b$ values - wrapped  distribution denoted $Q_b$, at the expense of a broader distribution for $\tilde{\Pop}_b$ -wrapped  distribution denoted $P_b$. Admittedly,  a broad $P_b$ distribution yields blurred error-syndromes (first term in Eq.~\eqref{stabevolve}), but these  errors are mitigated by  cycle repetition. Second, during  qubit-based preparation of the auxiliary oscillator,  flips of the qubit only trigger  long shifts  along $q_b$  if they  occur during $\mathcal{R}_{p_b}$ rounds. Based on these two observations, we propose to prepare the auxiliary state with a large number $N_p$ of $\mathcal{R}_{p_b}$  rounds  \emph{followed} by a large number $N_q$ of  $\mathcal{R}_{q_b}$  rounds (see  Fig.~\ref{fig2}c), allowing the latter to correct long shifts  induced by the former.\\ 

The detailed analysis of  this preparation sequence is  facilitated by the periodicity of the auxiliary state along both quadratures,  preserved by the applied controls and by our noise model. This model combines bit and phase flips of the physical qubit, with respective small probabilities  $p^{BF}$ and $p^{PF}$ during each round, and quadrature noise of the oscillators at rate $\kappa$ - equivalent to photon loss and gain at equal rate - inducing uniform state diffusion in phase-space. Under these assumptions, we show in Appendix~\ref{sec:zak} that the  density matrix of the auxiliary oscillator remains diagonal at all time in the Zak basis~\cite{zak1967finite}, whose base vectors are GKP states displaced in $[-\frac{\beta}{2},\frac{\beta}{2}]$ along $q_b$ and in $[\frac{\pi}{\beta},\frac{\pi}{\beta}]$ along $p_b$. Thus, its state is encoded by a 2D-wrapped probability distribution, and may be viewed as a classical particle living on a torus with coordinates $\tilde{q}_b$ and $\tilde{p}_b$. Furthermore, we show that this 2D-distribution is separable into two 1D-distributions, respectively defined along $\tilde{q}_b$ and denoted $Q_b$, and along $\tilde{p}_b$ and denoted  $P_b$. In this picture, repeated $\mathcal{R}_{r_b}$ rounds ($\tilde{r}_b=\tilde{q}_b$ or $\tilde{r}_b=\tilde{p}_b$) induce a classical random walk of the particle along $\tilde{r}_{b}$, whose  steps by  $ \pm\epsilon$ are biased toward $\tilde{r}_b=0$. In the limit of short steps, the corresponding $R_b$ distribution evolves with a position-dependent drift velocity $v(\tilde{r}_b)=- \frac{  \epsilon p^{NF} }{T_{\mathrm{round}}} \mathrm{sin}(2\pi \frac{\tilde{r}_b}{r_0})$ and a uniform diffusion constant $D=\frac{\epsilon^2}{ T_{\mathrm{round}}}+ \kappa$, where  $T_{\mathrm{round}}$ is the   round duration and $p^{NF}=1-p^{BF}-2p^{PF}$ is close to 1.  The steady-state of this dynamic approaches a wrapped normal distribution whose variance depends on $\epsilon$ and reaches a minimum  $V_{min}=( \kappa T_{\mathrm{round}})^{1/2}r_0/(2\pi p^{NF})$  for $\epsilon_{min}=(\kappa T_{\mathrm{round}})^{1/2}$, where $r_0=\beta$ when $\tilde{r}_b=\tilde{q}_b$ and $r_0=2\pi/\beta$ when $\tilde{r}_b=\tilde{p}_b$. However, the vanishing drift velocity in the vicinity of $\tilde{r}_b=r_0/2$ and small diffusion constant for $\epsilon=\epsilon_{min}$ ($\kappa T_{\mathrm{round}} < 10^{-4}$ considered in this work)  result in a long convergence time and persisting tails of the $R_b$ distribution at this optimal value.  We mitigate this advert effect by varying the feedback displacement length $\epsilon_j$ as a function of the round index $j$, starting with $\epsilon_j \sim r_0/2$ to suppress the tails of $R_b$  and ending with $\epsilon_j \sim \epsilon_{min}$ to limit its central peak width. We exactly compute the evolution of the distributions throughout this preparation, compactly encoded in the form of $(2n_F+1)$-Fourier coefficient vectors
($n_F~\sim30-60$ throughout this work). In Fig.~\ref{fig3} (top panel), we represent the  $Q_b$ distribution obtained after a given number  $N_q$ of $\mathcal{R}_{q_b}$ rounds.   Its tails are exponentially suppressed as $N_q$ increases while its  central peak has a constant variance $V_{min}$, ensuring robust suppression of error-propagation to the target mode. \\ 

As the $R_b$ distribution is being sculpted by repeated $\mathcal{R}_{r_b}$ rounds, long shifts triggered by  bit-flips of the qubit  uniformize the  distribution along the conjugate quadrature, and quadrature noise deflates its central peak. In our asymmetric preparation scheme, the $Q_b$ distribution is sculpted last and its final value is not impacted by these errors. On the other hand, they  have a dramatic effect on  $P_b$ which becomes near uniform for large values of $N_q$ (Fig.~\ref{fig3}, bottom panel) as  the probability $(1-p_{BF})^{N_q}$ that no bit-flip  occurred during the $\mathcal{R}_{q_b}$ rounds approaches 0. Thus, $N_q$ cannot be  arbitrarily large for the auxiliary state to be a resource for Steane-type error-correction, even in the limit of weak intrinsic noise of the oscillators (see Appendix~\ref{sec:optround}).

\section{Target mode error-correction}
\label{sec:targetcormain}
	\begin{figure}[htbp] 
		\centering
		\includegraphics[width=1.0\columnwidth]{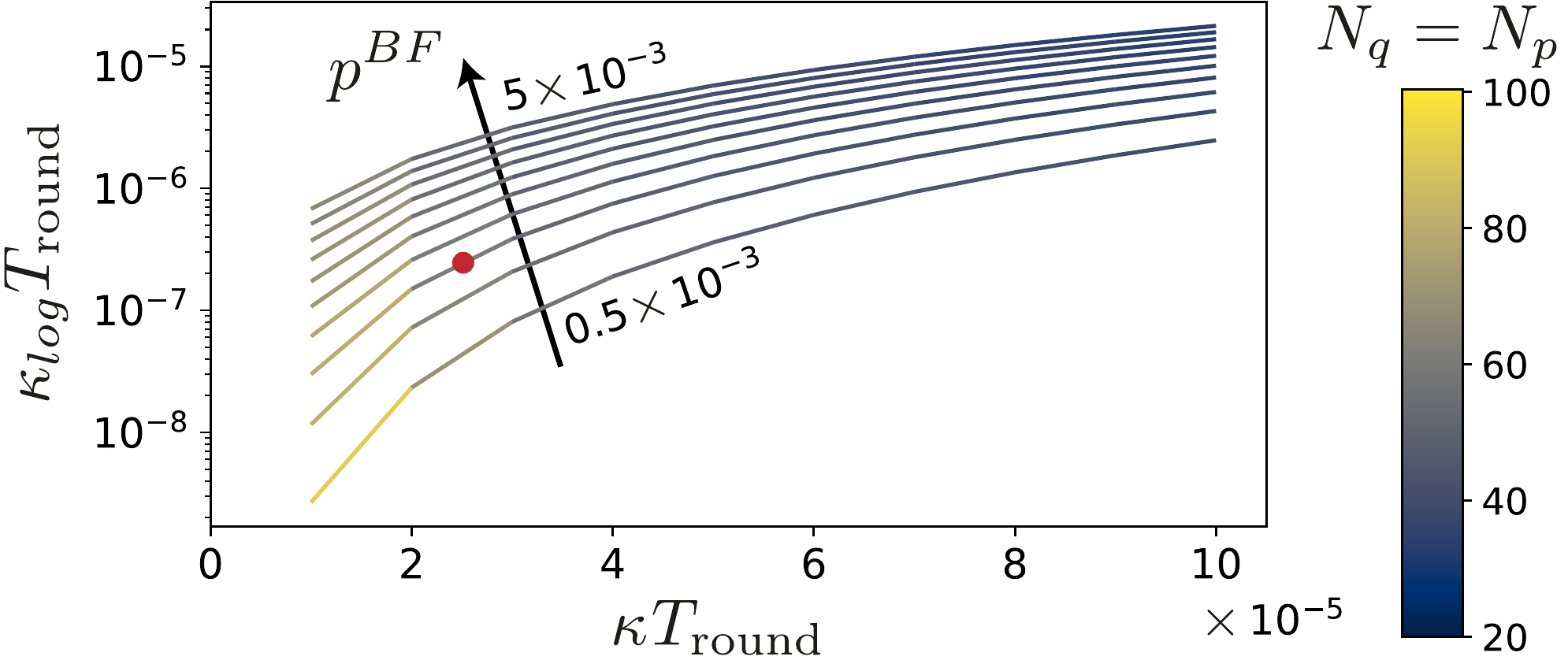}
		\caption{    Decay rate of the $z$-component of the  GKP qubit Bloch vector $\kappa_{log}$  as a function of the oscillators quadrature noise rate $\kappa$ and  of the probability  for flips of the physical qubit during each round $p^{BF}=2p^{PF}$,  linearly varied. For each noise value, the number of preparation rounds $N_q$ and $N_p$ are swept together (allowing different values did not significantly improve error-correction performances) and the cycle feedback parameters optimized by gradient ascent. We report the minimum  error rate with the corresponding round number, encoded in color. The red circle marks state-of-the-art hardware parameters for $T_{\mathrm{round}}=1.5~\mu$s (see text).}  
		\label{fig4}
	\end{figure}

We  now consider the evolution of the target oscillator over alternating $\mathcal{C}_{q_a}$ and $\mathcal{C}_{p_a}$ error-correction cycles. As for the auxiliary oscillator during preparation,  the target oscillator state remains periodic (see Appendix~\ref{sec:targeterror}). In order to estimate the decay rate  of the $z$-component of the GKP qubit Bloch vector $\kappa_{log}$---the $x$-component decays at the same rate and the $y$-component twice faster in the square code---we consider the evolution of the wrapped distribution of the logical operator $\tilde{q}^L_a$ only, denoted $Q_a$. We compactly represent it as an $(2n_F+1)$-Fourier coefficient vector   and encode the system evolution over a pair of $\mathcal{C}_{q_a}$ and $\mathcal{C}_{p_a}$ cycles in an $(2n_F+1)\times (2n_F+1)$ evolution  matrix, which accounts for  realistic auxiliary state preparation and $\tilde{\Pop}_b$ detection (see Fig.~\ref{fig2} and Appendix~\ref{sec:efficient} for details). The only approximation made in this formalism is to model  noise as effective quantum channels applied in-between perfect gates, with negligible impact on the estimate of error-correction performances~(see Appendix~\ref{sec:noise}). Choosing, as an initial guess, a simple \emph{sine} function for the feedback law $f$ controlling  displacements applied to the target oscillator (see Fig.~\ref{fig2}c), we observe that the $Q_a$ distribution  converges over a few cycles from an arbitrary initial state to a meta-stable state with two peaks centered at $\tilde{q}^L_a=0$ and $\tilde{q}^L_a=\alpha/2$ (shown in Fig.~\ref{fig:eigenvectors}), as expected from a state close to the GKP code manifold. A slow dynamic then comes into play, following which the respective amplitudes of the two peaks equilibrate as the GKP qubit relaxes to the fully mixed logical state.\\

For given numbers of preparation rounds $N_p$ and $N_q$ and noise values $p^{BF},~p^{PF}$, $\kappa$, we efficiently extract  $\kappa_{log}$ by spectral analysis of the evolution matrix (see Appendix~\ref{sec:spectral}). Moreover, we adjust the cycle feedback parameters (auxiliary state displacements $\epsilon_j$ and Fourier coefficients $f_k$ of a general feedback law $f$ on the target) by gradient ascent in order to minimize  $\kappa_{log}$. Finally, we select the preparation round number yielding the smallest error rate, assuming a quadrature gate time $T_{\mathrm{quad}}=5T_{\mathrm{round}}$---a longer gate time does not  impact significantly the performances as long as it does not dominate the overall cycle duration. In Fig.~\ref{fig4}, we report the  rate $\kappa_{log}$ obtained after this optimization.   Strikingly,  $\kappa_{log}$  decreases exponentially as the system noise strength---or equivalently the gates duration---decreases. This is in stark contrast with the linear scaling found for simple qubit-based error-correction---even considering multimode GKP codes~\cite{royer2022encoding}---and, following the argument given at the end of Sec.~I, for Steane-type error correction with a symmetrically prepared auxiliary state. Assuming that the protocol presented in this work may be adapted to protect finite-energy GKP states against photon loss with similar performances, we find that for a state-of-the-art system ($30~\mathrm{ms}$ photon lifetime in the oscillators~\cite{milul2023superconducting},  $T_1\sim T_2\sim 500~\mu$s for the physical qubit~\cite{place2021new,wang2022towards} and a preparation round duration of $1.5~\mu$s~\cite{campagne2020quantum,eickbusch2022fast}),  the coherence time of the GKP qubit could surpass that of the embedding hardware by two orders of magnitude (red circle in Fig.~\ref{fig4}).


\section{Conclusion and outlook}

In this letter, we  proposed a simple architecture controlled with two elementary gates to robustly protect an encoded GKP qubit. The conditional displacement gate is now routinely employed in superconducting circuit experiments. As for the quadrature gate, it may be decomposed into a sequence of single-mode squeezing gates and a beam-splitter gate~\cite{glancy2006error,terhal2016encoding,tzitrin2020progress} or activating simultaneously  a beam-splitter Hamiltonian and a two-mode squeezing Hamiltonian~\cite{zhang2019engineering}. Recent  progress toward the implementation of these operations in parametrically driven superconducting circuits~\cite{eriksson2023universal,lu2023high} gives reason to hope that a quadrature gate suited for continuous-variable quantum computing will be experimentally demonstrated in the near-future. This gate will anyhow be needed to perform operations on encoded GKP qubits. In that sense, the module we consider does not  unnecessarily increase the complexity of a GKP qubit-based computing platform. \\

Numerical simulations assuming a simplified noise model indicate that the lifetime of the GKP  qubit protected by our protocol is exponentially enhanced as the noise strength during each gate decreases. Extending  this result to normalized GKP code states and more realistic noise models will be the subject of future work.  In particular, we have not considered the impact of photon loss nor of imperfect quadrature gates in this work, but do not expect these errors to qualitatively impact our results if they can be mapped to short displacements of the target oscillator state during each correction cycle.
 While entering the regime of strong suppression of logical errors requires both oscillators and the qubit to be at the
state-of-the-art, a substantial margin for improvement exists by refining the feedback law beyond the short memory model considered in this work,  allowing longer conditional displacements whose lengths are multiples of the GKP lattice period~\cite{kitaev1995quantum,svore2013faster}, and considering more extensive hardware with multiple auxiliary oscillators and physical qubits in order to multiplex error-syndrome detection.  Given that Clifford operations in the GKP code rely on the same controls considered in this letter, our work opens a clear path toward fault-tolerant quantum computation with GKP qubits.

\begin{acknowledgments}
 The authors warmly thank M. Mirrahimi and P. Rouchon for carefully reviewing the manuscript,
 and M. H. Devoret, A. Eickbusch and S. Touzard for discussions that motivated this work. This work was supported by the Agence Nationale de la Recherche (ANR, project SYNCAMIL), the Paris Ile-de-France Region in the framework of DIM SIRTEQ, the European Research Council (ERC grant agreements No. 101042304 and  No. 884762),  the Plan France 2030 through the project ANR-22-PETQ-0006 and  by the Army Research Office (ARO) under grant No. W911NF-18-1-0212.
\end{acknowledgments}

\appendix
\section{Effective noise channels}
 \label{sec:noise}

 In this work,  we consider  a simplified noise model in which instantaneous and perfect conditional displacement and quadrature gates are  followed by the application  of an effective noise channel accounting for the errors having occurred during the gates. This effective model considerably reduces numerical simulation complexity, but leads to  approximations. In this section, we describe the effective noise channel applied after each gate, and argue that these approximations should not impact significantly the error correction performances estimated for our protocol.

	\subsection{Bit-flips of the physical qubit}

We consider  bit-flips induced by  qubit relaxation and excitation at respective rates $\Gamma_{+}$ and $\Gamma_{-}$. Their effect on the system density matrix is modeled by Lindblad dissipators $\sqrt{\Gamma_+}\mathcal{D}[\sigmap]$ and $\sqrt{\Gamma_-}\mathcal{D}[\sigmam]$, where $\sigmap$ and $\sigmam$ are respectively the raising and lowering operators of the qubit. Each dissipator $\mathcal{D}[\LL]$ yields, over an infinitesimal time-step $dt$, an evolution of the density matrix of the system 
	\begin{equation}
	        d\rhoo = dt \mathcal{D}[\LL](\rhoo)= dt \big(\LL \rhoo \LL^{\dagger}-\frac{1}{2}(\LL^{\dagger} \LL \rhoo + \rhoo \LL^{\dagger} \LL  ) \big)
	\end{equation}
 We  focus on the case   $\Gamma_+=\Gamma_-=\Gamma_1/2$ and briefly describe the most general  case $\Gamma_+ \neq \Gamma_-$ at the end of this section. Note that this particular case of equal rates of qubit excitation and relaxation  applies to current   experiments with superconducting circuits, as argued below.\\

We  consider the effect of bit-flips during the application of a conditional displacement gate   $\UU^{CD}_{q_b}=e^{i \frac{\pi}{\beta} \qop_b \sigmaz}$ along the $p_b$ quadrature of an oscillator---the calculation is directly adaptable to the case of a conditional displacement along $q_b$. We assume this gate to be performed by the application of a Rabi-type Hamiltonian---also known as longitudinal coupling Hamiltonian
\begin{equation}
\HH^{CD}_{q_b}=-\chi \qop_b \sigmaz
\end{equation}
with constant rate $\chi=\frac{\pi}{\beta T_{CD}}$ over the gate duration $T_{CD}$ (the coupling Hamiltonian is then turned off until the following gate).  We unravel the effect of bit-flips as stochastic collapses onto the ground state $|g\rangle$ or the excited state $|e\rangle$~\cite{wiseman2009quantum}. Intuitively, the trajectories so unraveled are obtained by detecting, with perfect efficiency, photon emission into the environment, and photon absorption from the environment. At each infinitesimal time step $dt$, the Kraus operators of this evolution are $\PP^-=\sqrt{\frac{\Gamma_1 dt}{2}} |g\rangle \langle e| $, $\PP^+=\sqrt{\frac{\Gamma_1 dt}{2}} |e\rangle \langle g|$---respectively modelling a jump to $|g\rangle$ or $|e\rangle$---and the  no-jump operator ${\bf 1}-i \HH^{NJ} dt $ where 
\begin{equation}
\begin{aligned}
\HH^{NJ}dt&=\HH_{q_b}^{CD} dt-\frac{i}{2} (\PP^{-\dagger}\PP^-+\PP^{+\dagger}\PP^+)\\
&=\HH_{q_b}^{CD} dt-\frac{i\Gamma_1 dt }{4}\II
\end{aligned}
\end{equation}
may be viewed as a non-Hermitian Hamiltonian.\\

We place ourselves in the weak noise limit ($\Gamma_1 T_{CD}  \ll 1$) and neglect the possibility of trajectories with two jumps. The evolution through the gate in absence of jumps is given by the operator 
\begin{equation}
    \OO^{NJ}=e^{-\frac{\Gamma_1 T_{CD}}{4}}\UU_{q_b}^{CD}
\end{equation}
and, from a state encoded by the density matrix $\rhoo$, the system evolves to the non-normalized density matrix
\begin{equation}
    \rhoo^{NJ}=\OO^{NJ}\rhoo~ \OO^{NJ\dagger}=e^{-\frac{\Gamma_1 T_{CD}}{2}}\UU_{q_b}^{CD} \rhoo \UU_{q_b}^{CD\dagger}
\end{equation}
We find the probability of no  bit-flip having occurred to be $1-p^{BF}=e^{-\frac{\Gamma_1 T_{CD}}{2}}\simeq 1+\frac{\Gamma_1 T_{CD}}{2} $, and the normalized density matrix conditioned on no jump having occurred is the same as for an evolution through a perfect gate. We detail the evolution of the auxiliary state through the remaining steps of the preparation round in Sec.~\ref{sec:ancillaprep} in this no-jump case. Note that the value of $p^{BF}$ is simply understood as the probability of a qubit excitation and relaxation over a small time step $dt$ being $n_e \Gamma_- dt + (1-n_e) \Gamma_+ dt= \Gamma_1 dt/2$, where $n_e$ is the expectation value of $|e\rangle \langle e|$. \\

We now focus on the trajectories during which the qubit has flipped. The evolution  through the gate when the qubit excites or relaxes in a time-interval of duration $dt$ around   $t$ (with $0< t< T_{CD}$) reads 
\begin{equation}
\begin{aligned}
\OO_{t}^{\pm}&= e^{-\frac{\Gamma_1 (T_{CD}-t)}{4}} e^{i\chi (T_{CD}-t)\qop_b \sigmaz}~ \PP^{\pm}~e^{-\frac{\Gamma_1 t}{4}} e^{i\chi t\qop_b \sigmaz}\\
&=  e^{-\frac{\Gamma_1 T_{CD}}{4}} e^{\mp i \frac{\pi}{\beta} (1-\frac{2t}{T_{CD}}) \qop_b}~\PP^{\pm}\\
&=  e^{-\frac{\Gamma_1 T_{CD}}{4}} \D_{t}^{\mp}~\PP^{\pm},
\end{aligned}
\label{eq:unravel}
\end{equation}
where we have defined $\D_{t}^{\mp}=\D_{p_b}(\mp  \frac{\pi}{\beta} (1-\frac{2t}{T_{CD}}))$. 
 $\OO_{t}^{\pm}$ thus collapses the qubit onto $|g\rangle$ or $|e\rangle$ and displaces the  oscillator state along $p_b$ by  $\pm \frac{\pi}{\beta} (1-\frac{2t}{T_{CD}}) \in I=[-\frac{\pi}{\beta}, \frac{\pi}{\beta}]$. In the protocol described in this work, conditional displacement gates employed for auxiliary state preparation are immediately followed by a measurement of the $\sigmay$ Pauli operator of the qubit whose  outcome controls a feedback displacement by $\pm \epsilon$ along $q_b$ and a qubit rotation resetting it   in $|+x\rangle$ (see Fig.~\ref{fig2}d). Recombining all trajectories during which the qubit has flipped, the system state then reads
\begin{widetext}
\begin{equation}
\begin{aligned}
\rhoo^{J,FB}&=e^{-\frac{\Gamma_1 T_{CD}}{2}} \sum_{\substack{s=+,-\\ u=+,- }} \int_{t=0}^{T_{CD}} \D_{q_b}(s \epsilon)  |+x\rangle \langle s y| \OO^u_{t} \rhoo \OO^{u \dagger}_{t}  |s y\rangle \langle + x| \D_{q_b}(s \epsilon)^{\dagger} \\
&=e^{-\frac{\Gamma_1 T_{CD}}{2}} \sum_{\substack{s=+,- }} \int_{t=0}^{T_{CD}} \frac{\Gamma_1 dt}{2} \D_{q_b}(s \epsilon)  |+x\rangle \langle s y|  \Big( \D_{t}^{-} |e\rangle \langle g| \rhoo |g\rangle \langle e| \D_{t}^{-\dagger} + \D_{t}^{+} |g\rangle \langle e| \rhoo |e\rangle \langle g| \D_{t}^{+\dagger} \Big) |s y\rangle \langle + x| \D_{q_b}(s \epsilon)^{\dagger} \\
&=e^{-\frac{\Gamma_1 T_{CD}}{2}} \sum_{\substack{s=+,- }} \int_{t=0}^{T_{CD}} \frac{\Gamma_1 dt}{2} \D_{q_b}(s \epsilon)  |+x\rangle \langle s y| \D_{t}^{-} \Big(  |e\rangle \langle g| \rhoo |g\rangle \langle e|  +  |g\rangle \langle e| \rhoo |e\rangle \langle g|  \Big) \D_{t}^{-\dagger} |s y\rangle \langle + x| \D_{q_b}(s \epsilon)^{\dagger} \\
&=e^{-\frac{\Gamma_1 T_{CD}}{2}} \sum_{\substack{s=+,- }} \int_{t=0}^{T_{CD}} \frac{\Gamma_1 dt}{4} \D_{q_b}(s \epsilon)    \D_{t}^{-} |+x\rangle \Big(   \langle g| \rhoo |g\rangle + \langle e| \rhoo |e\rangle   \Big) \langle + x| \D_{t}^{-\dagger}  \D_{q_b}(s \epsilon)^{\dagger} \\
&=e^{-\frac{\Gamma_1 T_{CD}}{2}} \sum_{\substack{s=+,- }} \int_{t=0}^{T_{CD}} \frac{\Gamma_1 dt}{4} \D_{q_b}(s \epsilon)    \D_{t}^{-} \rhoo_b  \D_{t}^{-\dagger}  \D_{q_b}(s \epsilon)^{\dagger} \otimes |+x\rangle  \langle + x|
\end{aligned}
\label{eq:long}
\end{equation}
\end{widetext}
where we have used that $\D_{T_{CD}-t}^{+}=\D_{t}^{-}$ from the second to the third line, and $\rhoo_b$ denotes the  density matrix  of the auxiliary state found after tracing out the  qubit mode on the last line. Thus, when a qubit flip occurs---which happens with probability $e^{-\frac{\Gamma_1 T_{CD}}{2}}\frac{\Gamma_1 T_{CD}}{2}\simeq p^{BF}$---the auxiliary state is randomly shifted in the whole interval $I$ along $p_b$, and randomly shifted by $\pm \epsilon$ along $q_b$.\\

We now argue that the \emph{echoed conditional displacement} gate employed in current superconducting circuit experiments~\cite{campagne2020quantum,sivak2023real}, performed on a qubit affected by relaxation only ($\Gamma_-=\Gamma_1$, $\Gamma_+=0$), is equivalent to a simple conditional displacement gate performed on a qubit affected by excitation and relaxation at equal rates  ($\Gamma_+ = \Gamma_- = \Gamma_1/2$). The echoed conditional displacement  includes a $\pi$-rotation  of the qubit around $\sigmay$ together with a change of sign for the oscillator-qubit interaction at $T_{CD}/2$ 
  \begin{equation}
   \HH_{q_b}^{ECD} =
    \begin{cases}
        \HH_{q_b}^{CD}  & \text{if $t< \frac{T_{CD}}{2}$ }\\
        -\HH_{q_b}^{CD}  & \text{if $t> \frac{T_{CD}}{2}$ }
    \end{cases}   
\end{equation}
Thus, for a noiseless qubit (perfect gate), the evolution reads
  \begin{equation}
  \begin{aligned}
  \UU_{q_b}^{ECD}&=e^{-i \frac{\pi}{2\beta} \qop_b \sigmaz} e^{-i\frac{\pi}{2}\sigmay}e^{i \frac{\pi}{2\beta} \qop_b \sigmaz}\\
  &= e^{-i\frac{\pi}{2}\sigmay}\UU_{q_b}^{CD}
  \end{aligned}
  \end{equation}
  
  Formally, the final $\pi$-rotation around $\sigmay$ does not impact the subsequent qubit measurement along $\sigmay$ so that, after applying a feedback displacement and tracing out the qubit, this evolution is equivalent to that obtained after a simple conditional displacement gate. \\

Here again,  we unravel the effect of qubit relaxation at rate $\Gamma_-=\Gamma_1$ as stochastic jumps, assuming that the jump probability is small ($\Gamma_1 T_{CD}  \ll 1$). In particular, this implies that double jump trajectories may be neglected. At each infinitesimal time-step $dt$, the Kraus operators are $\sqrt{\Gamma_1 dt} |g\rangle \langle e|=\sqrt{2}\PP^-$ and the corresponding no-jump operator ${\bf 1}-i \tilde{\HH}^{NJ} dt $ with
 \begin{equation}
 \begin{aligned}
    \tilde{\HH}^{NJ} dt&=\HH^{ECD}_{q_b}dt-i \PP^{-\dagger}\PP^- \\
    &=\HH^{ECD}_{q_b}dt-\frac{i\Gamma_1}{4}({\bf 1}-\sigmaz)dt.
\end{aligned}
    \label{eq:HJT}
\end{equation}
 
First focusing the no-jump evolution, it reads
  \begin{equation}
  \begin{aligned}
  &\OO^{ECD, NJ}\\
  &=e^{- \frac{\Gamma_1 T_{CD}}{8}(1-\sigmaz)}e^{-i \frac{\pi}{2\beta} \qop_b \sigmaz} e^{-i\frac{\pi}{2}\sigmay} e^{- \frac{\Gamma_1 T_{CD}}{8}(1-\sigmaz)} e^{i \frac{\pi}{2\beta} \qop_b \sigmaz}\\
  &= e^{-i\frac{\pi}{2}\sigmay} e^{-\frac{\Gamma_1 T_{CD}}{4}}\UU_{q_b}^{CD}
  \end{aligned}
  \end{equation}
  where the various terms have simply been ordered using  that the two terms in the non-Hermitian Hamiltonian~\eqref{eq:HJT} commute. Thus we recover the same no-jump evolution as for the simple conditional displacement gate.\\

Now focusing on the case where a jump occurred at time $t$ during the gate, the evolution operator  reads
 
 \begin{widetext}
  \begin{equation}
   \tilde{\OO}_{t} =
    \begin{cases}
        \tilde{\OO}^<_{t}&= \sqrt{2}~ e^{-i \frac{\pi}{2\beta}  \qop_b \sigmaz} e^{-\frac{\Gamma_1 T_{CD}}{8}  ({\bf 1 }-\sigmaz)}  e^{-i\frac{\pi}{2}\sigmay}e^{i \frac{\pi}{2\beta}(1-\frac{2t}{T_{CD}}) \qop_b \sigmaz} e^{-\frac{\Gamma_1 (T_{CD}-2t)}{8}  ({\bf 1 }-\sigmaz)} \PP^- e^{i \frac{\pi}{2\beta}\frac{2t}{T_{CD}} \qop_b \sigmaz} e^{-\frac{\Gamma_1 2t}{8}  ({\bf 1 }-\sigmaz)} \\  & \text{if $t< \frac{T_{CD}}{2}$ }\\
        \tilde{\OO}^>_{t}&=\sqrt{2} ~ e^{-i \frac{\pi}{2\beta}(2-\frac{2t}{T_{CD}}) \qop_b \sigmaz} e^{-\frac{\Gamma_1 (2T_{CD}-2t)}{8}  ({\bf 1 }-\sigmaz)}  \PP^- e^{-i \frac{\pi}{2\beta}(\frac{2t}{T_{CD}}-1) \qop_b \sigmaz} e^{-\frac{\Gamma_1 (2t-T_{CD})}{8}  ({\bf 1 }-\sigmaz)}
        e^{-i\frac{\pi}{2}\sigmay}e^{i \frac{\pi}{2\beta} \qop_b \sigmaz} e^{-\frac{\Gamma_1 T_{CD}}{8}  ({\bf 1 }-\sigmaz)}  \\ &  \text{if $t> \frac{T_{CD}}{2}$ }
    \end{cases}   
\end{equation}
 \end{widetext}
 After commuting the projectors and the qubit rotation operator through the conditional displacements, we find that 
  \begin{equation}
   \tilde{\OO}_{t} =
    \begin{cases}
        \tilde{\OO}^<_{t}= \sqrt{2} e^{-\frac{\Gamma_1t}{2}} e^{-i\frac{\pi}{2}\sigmay} \OO_{t}^- & \text{if $t< \frac{T_{CD}}{2}$ }\\
        \tilde{\OO}^>_{t}= -\sqrt{2} e^{-\frac{\Gamma_1(t-T)}{2}} e^{-i\frac{\pi}{2}\sigmay} \OO^+_{t} & \text{if $t> \frac{T_{CD}}{2}$ }
    \end{cases}   
\end{equation}
where $\OO^{\pm}$ is defined as in Eq.~\eqref{eq:unravel}. $\tilde{\OO}_{t}$ thus collapses the qubit onto  $|e\rangle$ and  displaces the  oscillator state along $p_b$ by  $ \frac{\pi}{\beta} (1-2\frac{t}{T_{CD}}) \in I_>=[0, \frac{\pi}{\beta}]$ if $t < \frac{T_{CD}}{2}$, and collapses the qubit onto  $|g\rangle$ and  displaces the  oscillator state along $p_b$ by  $ -\frac{\pi}{\beta} (1-2\frac{t}{T_{CD}}) \in I_>$ if $t > \frac{T_{CD}}{2}$. Notice that for a periodic state with period $\frac{2\pi}{\beta}$, the former displacement is equivalent to a displacement by $ -\frac{\pi}{\beta} (1+2\frac{t}{T_{CD}}) \in I_<$.
In both cases, measuring the $\sigmay$ Pauli operator of the qubit following the evolution yields a random outcome, and thus a random feedback displacement by $\pm \epsilon$ along $q_b$. The integrated probability for the trajectories during which a jump occurred is---at first order in $\Gamma_1 T_{CD}$---$\Gamma_1 T_{CD}/2=p^{BF}$, and the non normalized density matrix conditioned on a jump having occurred is the same as the one found in Eq.~\eqref{eq:long} -  at first order in $\Gamma_1 T_{CD}$.\\

The most notable difference with the case of a simple conditional displacement in presence of transmon excitation and relaxation at equal rates lies in the  distribution of displacements entailed by a jump. After the qubit measurement and feedback, the non-normalized  system state conditioned on a jump having occurred differs from Eq.~\eqref{eq:long} at second order
\begin{equation}
\begin{aligned}
&\rhoo^{ECD,J,FB}=e^{-\frac{\Gamma_1 T_{CD}}{2}} |+x\rangle  \langle + x|\\
&\sum_{\substack{s=+,- }} \int_{t=0}^{\frac{T_{CD}}{2}} \frac{\Gamma_1 dt}{2} \mathrm{cosh}(\Gamma_1 t)\D_{q_b}(s \epsilon)    \D_{t}^{+} \rhoo_b  \D_{t}^{+\dagger}  \D_{q_b}(s \epsilon)^{\dagger} 
\end{aligned}
\label{eq:long2}
\end{equation}
where the $\mathrm{cosh}(\Gamma_1 t)$ slightly favors shortly displaced states over states displaced by a long distance (the oscillator modular position is no longer uniformly sampled in $I$).   We insist that this correction is of second order in $\Gamma_1 T_{CD}$, and is given here to show that such second order correction should not impact significantly the performance of our protocol.\\

To conclude this section, we mention that in the  case where a simple conditional displacement is applied to a qubit in presence of  excitation and relaxation at different rates, one finds a modified no-jump evolution by which the qubit partially collapses onto $|g\rangle$. This partial collapse slightly unbalances the relative amplitudes of probability for the two conditionally displaced copies of $\rhoo$, thereby reducing  the contrast of the subsequent qubit measurement. The impact on the performances of our protocol is expected to be similar to that of the qubit phase-flips, which is described in the next section.

	\subsection{Phase-flips of the physical qubit}

By comparison with bit-flips, phase-flips of the qubit are simpler to model. Indeed, in the quantum trajectory approach described above, they correspond to $\sigmaz$  gates randomly applied to the qubit over any time-interval of duration $dt$ with probability $\frac{\Gamma_{\phi}}{2}dt$, where $\Gamma_{\phi}$ is the qubit pure dephasing time. Since $\sigmaz$ commutes with the interaction Hamiltonian,  phase-flips are equivalently modeled as a $\sigmaz$ gate applied after the gate with probability $p^{PF}=\frac{\Gamma_{\phi}}{2}T_{CD}$ (in the weak noise limit). By flipping the sign of the subsequently measured $\sigmay$ Pauli operator, this error results in an erroneously applied feedback displacement. We set $p^{PF}=p^{BF}/2$,  typical of superconducting circuit experiments, in all simulations performed in this work.  \\

Note that qubit readout errors have an impact similar to phase-flips, but may cause more damage when the qubit is actively reset based on the measurement outcome, yielding a qubit  erroneously prepared in $|-x\rangle$ for the subsequent auxiliary state preparation round (see Fig.~\ref{fig2}d). Experimentally, such reset errors  may be mitigated by  repeating the reset procedure in $|g\rangle$, assuming the measurement to be Quantum Non Demolition for the $|g\rangle$ state~\cite{riste2012feedback,campagne2013persistent}. Readout errors are not modeled in this work.\\

	\subsection{Quadrature noise}

	Quadrature noise at rate $\kappa$ is modeled by two Lindblad dissipators $\sqrt{\kappa}\mathcal{D}[\qop]$ and $\sqrt{\kappa}\mathcal{D}[\Pop]$, inducing uniform diffusion of an oscillator state in phase-space. Its effect  can equivalently be modeled by the application of  stochastic evolution operators
	 \begin{equation}
	 \begin{split}
	     \UU^q_{dt}=e^{i \sqrt{\kappa} dW_q \qop }\\
	     \UU^p_{dt}=e^{i \sqrt{\kappa} dW_p \Pop }
	 \end{split}
	 \end{equation}
	where $dW_q$ and $dW_p$ are independent Wiener processes characterized by $\overline{dW_q}=\overline{dW_p}=0$ and $dW_q^2=dW_p^2=dt$~\cite{wiseman2009quantum}

	 \subsubsection{Effective noise channel after a conditional displacement gate}
	 
	 We here consider a conditional displacement gate applied on the $q$ quadrature of the auxiliary oscillator reading
	 \begin{equation}
	     \UU^{CD}_{q}=e^{i \frac{\pi}{\beta} \qop \sigmaz}
	 \end{equation}
  where  we dropped the subscript $b$ to designate the auxiliary oscillator quadrature. It is straightforward to adapt the following calculation to the case of a conditional displacement along the $p$ quadrature. \\
	
    When the gate is applied in finite time $T_{CD}$ and in presence of quadrature noise, we use Trotter decomposition over $N=\frac{T_{CD}}{dt}\gg 1$ steps to write the  stochastic evolution over a single trajectory
    \begin{equation}
	     \tilde{\UU}^{CD}_{q_a}=\prod_{j=1}^N \big( e^{i \frac{\pi}{ N\beta} \qop \sigmaz} e^{i \sqrt{\kappa} dW^j_q \qop } e^{i \sqrt{\kappa} dW^j_p \Pop } \big)
	 \end{equation}
    where all Wiener processes $dW^j_q$, $dW^j_p$ are independent. Using  Baker-Cambpbell-Hausdorff formula, we  reorder this product to put the noise terms in front
       \begin{equation}
       \begin{aligned}
	     \tilde{\UU}^{CD}_{q_a}=& \prod_{j=1}^N \big(  e^{-i \sqrt{\kappa} dW^j_p \frac{j \pi}{N \beta}\sigmaz }  \big) \prod_{j=1}^N \big(  e^{i \sqrt{\kappa} dW^j_q \qop } ~ e^{i \sqrt{\kappa} dW^j_p \Pop } \big) \\ &\qquad \qquad \qquad   \times~\prod_{j=1}^N \big(  e^{i \frac{\theta}{N} \qop \sigmaz }  \big)
      \end{aligned}
      \label{eq:BKHCD}
	 \end{equation}
	The center and rightmost products correspond to a quadrature noise channel applied for a duration $T_{CD}$ after an error-free conditional displacement gate $\UU^{CD}_{q}$,  corresponding to our  effective noise model. We thus neglect the leftmost term, which rotates  the qubit Bloch vector around its $\sigmaz$ axis conditioned on the value of $dW^j_p$, i.e. the particular value of the  $q$-shifts induced by noise during the interaction.  Its physical interpretation is clear:  $q$-shifts of the oscillator that occur at the beginning of the gate ($j\rightarrow N$) leave an imprint on the qubit phase similarly to shifts having occurred before the gate, while shifts  that occur  toward the end of the gate ($j\rightarrow 1$) impact negligibly the qubit phase. Thus, if we were to exactly model the system evolution with all the terms in Eq.~\eqref{eq:BKHCD}, the recentering feedback displacement   following the gate and qubit readout would partially correct for $q$-quadrature noise during the gate. By neglecting the leftmost term in  Eq.~\eqref{eq:BKHCD}, we carry over this noise to the next round, resulting in a slightly broadened  $q$-probability distribution for the auxiliary state. Similar arguments can be made for the  $p$ distribution. For the noise figures considered in this work ($\kappa T_{CD}<\kappa T_{\mathrm{round}}<10^{-4}$), we expect this approximation to have a marginal impact on the estimated performances of our error-correction protocol, which are only slightly underestimated.

	 \subsubsection{Effective noise channel after a quadrature gate}
  \label{sec:quadraturenoise}
	 
	 We follow a similar reasoning for the quadrature gate
	 \begin{equation}
	     \UU^{\mathrm{quad}}_{q_a}=e^{i \theta \qop_a \qop_b}
	 \end{equation}
    where $\theta=\alpha/\beta$. Here, for simplicity, we consider $q_a$ and $q_b$ quadrature noise only, corresponding to stochastic terms of the form $e^{i \sqrt{\kappa} dW_{p_a} \Pop_a}$ and $e^{i \sqrt{\kappa} dW_{p_b} \Pop_b}$---the terms inducing shifts along $p_a$ and $p_b$  commute trivially through the gate. We decompose the noisy gate  over $N=\frac{T_{\mathrm{quad}}}{dt}\gg 1$ steps as
    \begin{equation}
	     \tilde{\UU}^{\mathrm{quad}}_{q}=\prod_{j=1}^N \big( e^{i \frac{\theta}{N} \qop_a \qop_b} e^{i \sqrt{\kappa} dW^j_{p_a} \Pop_a } e^{i \sqrt{\kappa} dW^j_{p_b} \Pop_b } \big)
	 \end{equation}
    where all Wiener processes $dW^j_{p_a}$, $dW^j_{p_b}$ are independent. Using  Baker-Cambpbell-Hausdorff formula, we  reorder this product to place the noise terms in front
    \begin{widetext}
       \begin{equation}
       \begin{split}
	     \tilde{\UU}^{\mathrm{quad}}_{q}&= e^{i\phi} \prod_{j=1}^N \big(  e^{i \sqrt{\kappa} dW^j_{p_a} \Pop_a } ~ e^{-i \sqrt{\kappa} dW^j_{p_b} \frac{j \theta}{N}\qop_a } \big) \prod_{j=1}^N \big(  e^{i \sqrt{\kappa} dW^j_{p_b} \Pop_b } ~ e^{-i \sqrt{\kappa} dW^j_{p_a} \frac{j\theta}{N}\qop_b } \big)  \prod_{j=1}^N \big(  e^{i \frac{\theta}{N} \qop_a \qop_b }  \big)\\
      &=e^{i\phi'} \prod_{j=1}^N e^{-i \sqrt{\kappa} dW^j_{p_b} \frac{j \theta}{N}\qop_a }  \prod_{j=1}^N e^{-i \sqrt{\kappa} dW^j_{p_a} \frac{j\theta}{N}\qop_b }  \prod_{j=1}^N e^{i \sqrt{\kappa} dW^j_{p_a} \Pop_a }   \prod_{j=1}^N   e^{i \sqrt{\kappa} dW^j_{p_b} \Pop_b }     \prod_{j=1}^N   e^{i \frac{\theta}{N} \qop_a \qop_b }  
      \end{split}
	 \end{equation}
  \end{widetext}
  where $\phi$ and $\phi'$ are  irrelevant global phases that can be omitted. Our simplified model  includes the  fifth product (equal to the noiseless evolution $\UU^{\mathrm{quad}}_{q_a}$) and the third and fourth products (corresponding to quadrature noise channels applied on idling oscillators).  The first product corresponds to
  an extra $p_a$ quadrature noise term, correlated to the auxiliary mode $q_b$ noise during the gate. Since, in our protocol, we discard the result of measurements of  the modular stabilizer $\tilde{\qop}_b$ and reset the auxiliary state  at the end of each cycle,  this correlated noise boils down to random displacements of the target state along $p_a$, with zero mean value and variance $\overline{(\frac{\sqrt{\kappa }\theta}{N}\sum_{j}j dW^j_{p_b})^2} \rightarrow \kappa \frac{\theta^2}{3} T_{\mathrm{quad}} $, where the last limit is taken for $N\rightarrow \infty$. We account for this term by renormalizing the $p_a$ quadrature noise rate during the gate following $\kappa \rightarrow \kappa (1+ \frac{\theta^2}{3})$. Note that its effect could be partially mitigated by decoding the information yielded by  the $\tilde{\qop}_b$ measurements  at the end of the cycle. Similarly, the second product describes shifts along $p_b$ correlated to the target mode $q_a$ noise during the gate. In analogy to the case of the conditional displacement gate detailed in the previous section, it is interpreted as partial mapping of the target oscillator shift errors occurring during the gate onto the auxiliary oscillator (shifts occurring at the beginning of the evolution leave a stronger imprint than those occurring toward the end). By neglecting this term in our simplified model, we carry over to the following cycle errors that would have been partly corrected in a more accurate model,  and thereby expect to slightly underestimate the performances of our protocol.

    \section{Oscillators dynamics in the Zak basis}
   \label{sec:zak}
	\subsection{The Zak basis}
		
  The dynamics of our system is conveniently described in the Zak basis~\cite{zak1967finite} of the oscillators, which is the basis formed by displaced GKP states within one unit cell of the GKP lattice. Equivalently, the Zak basis we will consider for the target oscillator can be seen as the joint eigenbasis of the modular logical operator   $\tilde{\qop}^L_a$ and of the modular  stabilizer $\tilde{\Pop}^S_a$, and the Zak basis for the auxiliary oscillator as the joint eigenbasis of the modular stabilizers $\tilde{\qop}_b$ and  $\tilde{\Pop}_b$. Formally, the Zak states are defined as 
  
\begin{equation}
\begin{split}
|u,v\rangle_a &=e^{-i u \Pop_a + i v \qop_a}|+Z\rangle=e^{\frac{i}{2}uv}\sum_{n\in \mathbb{Z}} e^{i n v \alpha} |n\alpha + u\rangle_{q_a} \\
|u',v'\rangle_b &=e^{-i u' \Pop_b + i v' \qop_b}|\o{} \rangle=e^{\frac{i}{2}u'v'}\sum_{m\in \mathbb{Z}} e^{i m v' \beta} |m\beta + u'\rangle_{q_b} 
\end{split}
\end{equation}
where we use the convention  $u\in [-\frac{\alpha}{2},\frac{\alpha}{2}]$, $v\in [-\frac{2\pi}{\alpha},\frac{2\pi}{\alpha}]$ , $u'\in [-\frac{\beta}{2},\frac{\beta}{2}]$ and $v'\in [-\frac{2\pi}{\beta},\frac{2\pi}{\beta}]$ and denote by $|r_0\rangle_{r}$ an eigenstate of the operator $\rop$ with eigenvalue $r_0$. \\

We will later use the following properties :\\

\emph{Momentum basis representation}
\begin{equation}
\begin{split}
|u,v\rangle_a &=e^{-\frac{i}{2}uv}\sum_{n\in \mathbb{Z}} e^{-i n u \frac{2\pi}{\alpha}} |n \frac{2\pi}{\alpha} + v \rangle_{p_a} \\
|u',v'\rangle_b &=e^{-\frac{i}{2}u'v'}\sum_{m\in \mathbb{Z}} e^{-i m u' \frac{2\pi}{\beta}} |m \frac{2\pi}{\beta} + v' \rangle_{p_b}
\end{split}
\end{equation}

\emph{Displacements} (for Zak states of either mode)
\begin{equation}
\begin{split}
e^{-i w \Pop} |u,v\rangle &=e^{-\frac{i}{2}wv} |u+w,v\rangle\\
e^{+i w \qop} |u,v\rangle &=e^{\frac{i}{2}wu} |u,v+w\rangle
\end{split}
\end{equation}
where $u+w$  and $v+w$ are to be considered as modular coordinates (respectively modulo $\alpha$ or $\beta$ and modulo $\frac{2\pi}{\alpha}$ or $\frac{2\pi}{\beta}$). \\

We will now show that, if the target mode is initialized in the $|\pm Z\rangle$ logical basis, the states of both the auxiliary and target modes are described by diagonal density matrices in their respective Zak bases throughout auxiliary state preparation and Steane-type error correction. Therefore, they can be represented by  wrapped probability distributions $Q_a$ and $P_a$ for the target oscillator,  $Q_b$ and $P_b$ for the auxiliary oscillator. Moreover, these distributions are separable between the two parameters of each Zak basis:
\begin{equation}
\begin{split}
\rhoo_a=&\int_u \int_v Q_a(u) P_a(v) |u,v\rangle \langle u,v|_a \\
\rhoo_b=&\int_{u'} \int_{v'} Q_b(u') P_b(v') |u',v'\rangle \langle u',v'|_b 
\end{split}
\label{rhodistrib}
\end{equation}
We also give evolution rules for these distributions throughout correction rounds and cycles, on which numerical simulations used in this work are based.

\subsection{auxiliary state preparation}
\label{sec:ancillaprep}
We here describe the evolution of the auxiliary state through a  $\mathcal{R}_{q_b}$ preparation round. The results can be directly adapted to the case of  $\mathcal{R}_{p_b}$ rounds. Moreover, we drop the subscript $b$ to simplify notations.\\

A $\mathcal{R}_{q}$  round labeled $j$ ($N_p+1 \leq j \leq N_p+N_q$) starts with a qubit initialization in the +1 eigenstate of its Pauli operator $\sigmax$, followed by a conditional displacement gate $\UU_{q}^{CD}=e^{i \theta \qop \sigmaz}$   where $\theta=\frac{\pi}{\beta}$. The qubit is then measured along $\sigmay$ , and a feedback displacement by $\pm \epsilon_j$ is applied along $q$ depending on the outcome. The Kraus operators corresponding to the two possible outcomes are~\cite{campagne2020quantum}
	 \begin{equation}
  \label{eq:krausancilla}
  \begin{aligned}
	 \mathbf{M_+}&=e^{-i \epsilon_j \Pop}\text{cos}(\theta  \qop + \frac{\pi}{4}) \\
  \mathbf{M_-}&=e^{+i \epsilon_j \Pop}\text{cos}(\theta \qop - \frac{\pi}{4})
  \end{aligned}
	 \end{equation} 
If, before the round, the auxiliary state is  of the form \eqref{rhodistrib} with probability distributions 
\begin{align}
 Q_{j-1}(u) \qquad\qquad P_{j-1}(v) 
\end{align}
the conditional states  after the qubit readout and feedback displacement are of the same form. In detail, if no qubit flip occurred during the gate, which happens with probability $1-p^{PF}-p^{BF}$ (in the limit of small flip probability), the non normalized conditional probability distributions read 
\begin{equation}
\begin{aligned}
&Q_{j-1}^{\pm, NF}(u)=\\
&~~(1-p^{PF}-p^{BF}) \Big( \frac{1}{2} \pm \frac{1}{2} \text{sin}\big(\frac{2\pi}{\beta} (u \pm \epsilon_j )\big)\Big)  Q_{j-1}(u \pm \epsilon_j) \\
&P_{j-1}^{\pm, NF}(v) =(1-p^{PF}-p^{BF}) P_{j-1}(v)
\end{aligned} 
\end{equation}
As detailed in  Sec.~\ref{sec:noise}, phase flips of the qubit during the gate, occurring with probability $p^{PF}$,  lead to an erroneously applied feedback displacement, yielding non normalized conditional probability distributions
\begin{equation}
\begin{aligned}
Q_{j-1}^{\pm, PF}(u)&=p^{PF} \Big( \frac{1}{2} \pm \frac{1}{2} \text{sin}\big(\frac{2\pi}{\beta} (u \mp \epsilon_j )\big)\Big)  Q_{j}(u \mp \epsilon_j)\\  P_{j-1}^{\pm, PF}(v) &=p^{PF} P_{j-1}(v)
\end{aligned} 
\end{equation}
while bit flips of the qubit during the gate, occurring with probability  $p^{BF}$,  result in a randomly applied feedback displacement by $\pm \epsilon_j$ along $q_b$ and a long  displacement along $p_b$ uniformly sampled in $[-\frac{\pi}{\beta},\frac{\pi}{\beta}]$. The corresponding non normalized conditional probability distributions read
\begin{equation}
\begin{aligned}
Q_{j-1}^{ BF}(u)&=\frac{p^{BF}}{2} \Big ( Q_{j-1}(u + \epsilon_j) +  Q_{j-1}(u - \epsilon_j) \Big)\\
 P_{j-1}^{ PF}(v) &=p^{BF} \frac{\beta}{2\pi}.
\end{aligned} 
\end{equation}

After recombining all  conditional probability distributions to model the proportional (memoryless) feedback strategy, the summed distributions read 
\begin{equation}
\label{eq:shiftancilla}
\begin{split}
Q_{j-1}^{FB}(u)&= \Big( \frac{1}{2} + \frac{p^{NF}
}{2} \text{sin}\big(\frac{2\pi}{\beta} (u + \epsilon_j )\big)\Big)  Q_{j-1}(u + \epsilon_j)\\
&\quad + \Big( \frac{1}{2} - \frac{p^{NF}
}{2} \text{sin}\big(\frac{2\pi}{\beta} (u - \epsilon_j )\big)\Big)  Q_{j-1}(u - \epsilon_j) \\
P_{j-1}^{FB}(v) &=(1- p^{BF}) P_{j-1}(v) + p^{BF} \frac{\beta}{2\pi}
\end{split}
\end{equation} 
where we defined $p^{NF}=1-2p^{PF}-p^{BF}$. At the end of the round, we apply an effective quadrature noise channel, which convolves the probability distributions with  wrapped normal distributions $G_q$ and $G_p$, respectively defined on $[-\frac{\beta}{2},\frac{\beta}{2}]$ and $[-\frac{\pi}{\beta},\frac{\pi}{\beta}]$,  both with variance $\sigma^2=\kappa T_{\mathrm{round}}$. We thus get at the beginning of the following round a state  of  the form \eqref{rhodistrib} with probability distributions 
\begin{equation}
\begin{aligned}
 Q_{j}(u)&= Q_{j-1}^{FB} \ast G (u) \\
 P_{j}(v)&= P_{j-1}^{FB} \ast G (v) 
\end{aligned}
 \label{eq:ancillanoise}
\end{equation}

The evolution of the auxiliary state through a $\mathcal{R}_{p}$ round is simply obtained by the exchange $q\rightarrow p$  in the above formulas. In our simulations, we initialize  the  auxiliary state of the form \eqref{rhodistrib} with uniform  $Q_0$ and $P_0$ distributions  before preparation (see Sec.~\ref{sec:phasestimation}  for a justification of this hypothesis). It follows from the above analysis  that the auxiliary state remains of this form throughout  preparation. Note that the Kraus map defined by the operators \eqref{eq:krausancilla}, as well as  quadrature noise,  suppress both off-diagonal terms of the Zak basis density matrix and classical correlations between the values of the modular operators $\tilde{\qop}$ and  $\tilde{\Pop}$. Therefore, a state which is not initially of the form \eqref{rhodistrib} becomes so after a long sequence of preparation rounds.

\subsection{Target mode error-correction }
\label{sec:targeterror}

We here describe the evolution of the target oscillator state through a    $\mathcal{C}_{q_a}$ correction cycle (see Fig.~\ref{fig2}c). A cycle starts with the auxiliary state prepared as described in the previous section. A quadrature  gate $\UU^{\mathrm{quad}}_{q_a}=e^{i \theta \qop_a \qop_b}$   with $\theta=\frac{\alpha}{\beta}$ is applied to the oscillators, followed by detection of the  $\tilde{\Pop}_b$ stabilizer with  outcome $m \in [-\frac{\pi}{\beta},\frac{\pi}{\beta}]$. In this section, we assume this detection to be perfect, and detail how to model its finite accuracy in  Sec.~\ref{sec:phasestimation}.  Finally,  the target oscillator is displaced by $-f(m)$ along $q_a$, and an effective noise channel is applied to the target oscillator state to account for quadrature noise throughout the cycle. \\

\begin{widetext}
We suppose the target state to be of the form \eqref{rhodistrib} when the $j-th$ cycle begins. After the auxiliary state preparation, which  yields a  state of the form \eqref{rhodistrib} with probability distributions $Q_{b,N_p+N_q}$ and $P_{b,N_p+N_q}$, abbreviated to $Q_b$ and $P_b$ for simplicity, the joint state of the system reads
\begin{equation}
\rhoo^0_{j-1}=\int_u \int_v \int_{u'} \int_{v'} Q_{a_{j-1}}(u) P_{a_{j-1}}(v) Q_b(u') P_b(v')~ |u,v\rangle \langle u,v|_a~ |u',v'\rangle \langle u',v'|_b  ~du dv du' dv'
\end{equation}
After the quadrature gate, the state reads
\begin{equation}
\rhoo^1_{j-1}=\int_u \int_v \int_{u'} \int_{v'} Q_{a_{j-1}}(u) P_{a_{j-1}}(v) Q_b(u') P_b(v')~ |u,v+\theta u'\rangle \langle u,v +\theta u'|_a~ |u',v'+\theta u\rangle \langle u',v'+\theta u|_b  ~du dv du' dv'
\end{equation}
Detection of the  $\tilde{\Pop}_b$ stabilizer yielding an outcome $m$  is modeled by the application of the Kraus operator $\MM_{m} = \int_{u'} |u',m\rangle \langle u',m|_b $. After tracing out the auxiliary state, the non normalized target oscillator density matrix conditioned on the outcome $m$  reads
\begin{equation}
\rhoo^{m}_{a_{j-1}}=\int_u \int_v \int_{u'}  Q_{a_{j-1}}(u) P_{a_{j-1}}(v) Q_b(u') P_b(m - \theta u)~ |u,v+\theta u'\rangle \langle u,v +\theta u'|_a  ~du dv  dv'
\end{equation}
After a feedback displacement by $-f(m)$ and summing over $m$ to model our  memoryless feedback strategy (in the sense that the measurement records are not carried to the following cycle), we get
\begin{equation}
\rhoo^{FB}_{a_{j-1}}=\int_u \int_v \int_{u'} \int_{m}  Q_{a_{j-1}}(u) P_{a_{j-1}}(v) Q_b(u') P_b(m - \theta u)~ |u-f(m),v+\theta u'\rangle \langle u-f(m),v +\theta u'|_a  ~du dv  dv' dm
\end{equation}
\end{widetext}
Given that the probability distributions are wrapped functions and that the integrals are defined over their whole domains, we find that this state is of the form \eqref{rhodistrib} with probability distributions
\begin{equation}
\begin{aligned}
\label{eq:targetFB}
 Q^{FB}_{a_{j-1}}(u)&= \int_{m} Q_{a_{j-1}}(u+f(m))P_b\big(m - \theta (u+f(m))\big)dm \\
 P^{FB}_{a_{j-1}}(v)&= \int_{u'} P_{a_{j-1}}(v-\theta u')Q_b(u') ~du'
\end{aligned}
\end{equation}

Finally, we apply the effective  noise channel accounting for quadrature  noise during the gate and the following auxiliary state preparation rounds. It convolves the probability distribution $Q_a$ 
with a wrapped normal distribution $G_a$ of  variance $\sigma^2=\kappa \big(T_{\mathrm{quad}} + (N_q+N_p) T_{\mathrm{round}} \big) $, and the probability distribution $P_a$ 
with a wrapped normal distribution $\tilde{G}_a$ with slightly larger variance to account for the renormalized quadrature noise $\kappa \rightarrow \tilde{\kappa}$ during the quadrature gate (Sec.~\ref{sec:quadraturenoise}). We thus get the target state at the beginning of the following cycle, also of   the form \eqref{rhodistrib}, with probability distributions 
\begin{equation}
\begin{aligned}
 Q_{a_{j}}(u)&= Q^{FB}_{a_{j-1}}(u) \ast G_a (u) \\
 P_{a_{j}}(v)&= P^{FB}_{a_{j-1}}(v) \ast \tilde{G_a} (v)
\end{aligned}
 \label{eq:targetgaussian}
 \end{equation}

The evolution of the target oscillator state during a    $\mathcal{C}_{p_a}$ correction cycle is derived with  similar calculations, inverting the role of $Q_{a_j}$ and $P_{a_j}$. It also transforms a state of the form \eqref{rhodistrib} into a state of the same form. Therefore, if the target is initialized in a state of this form, e.g. when prepared in $|+Z\rangle$, it remains so indefinitely. In order to extract the decay rate of the $z$-component of the GKP qubit Bloch vector under a particular set of error-correction parameters, one only needs to compute the evolution of $Q_a$ through successive $\mathcal{C}_{q_a}$ and $\mathcal{C}_{p_a}$ cycles. After some number of cycles $N_c$, the GKP qubit is decoded and its  $z$ Bloch sphere coordinates reads
\begin{equation}
z(N_c)=\int_{u}Q_{a_{N_c}}(u) \Theta(u) du 
\end{equation}
where $\Theta$ is a step function with value 1 on $[-\frac{\alpha}{4},\frac{\alpha}{4}]$ and $-1$ elsewhere. By fitting the decay of $z(N_c)$ with an exponential function, one extracts the  decay rate of the $z$-component of the GKP qubit Bloch vector $\kappa_{log}$. In Sec.~\ref{sec:efficient} we present a more efficient method to extract this same rate. \\

Note that with the Zak basis we chose,  constructed from the logical $|+ Z\rangle$  state, we cannot directly simulate the decay of other logical Pauli operators. One could do so by considering alternative Zak basis definitions. However, the square GKP code symmetry properties ensure that the three components of the logical Bloch vector decay with respective rates $\kappa_z=\kappa_x=\kappa_y/2=\kappa_{log}$.

\subsection{Detection of the modular stabilizer}
\label{sec:phasestimation}

In the previous section, we considered the detection of the $\tilde{\Pop}_b$ stabilizer as perfect and instantaneous. Since this measurement can be destructive for the auxiliary oscillator state, homodyne detection of $p_b$ is typically considered in the literature. However, letting the  field leak out of the auxiliary resonator to be detected requires to wait at least a few $1/\kappa$. This is not a viable option for  error-correction, which requires $\kappa T_{\mathrm{cycle}}<<1$. One could partly circumvent the issue by mapping the value of $p_b$ to a supplementary, low-Q resonator via a quadrature gate, but we found that, for a quadrature-quadrature interaction strength of the same order as that  activated between the target and auxiliary oscillators, the operation would here again dominate the error-correction cycle duration. Moreover,  combined photon collection and homodyne detection efficiencies are in practice limited to $\eta \lesssim 1/2$ in all experimental platforms, which would result in a too low  detection accuracy. Alternatively, we consider detecting the modular operator $\tilde{\Pop}_{b}$ through repeated physical qubit-based measurement rounds. Indeed, the outcome of the $\mathcal{R}_{p_b}$  rounds  preparing the auxiliary state for the following cycle can be straightforwardly decoded to estimate the value of $\tilde{\Pop}_{b}$ prior to re-preparation, with sufficient accuracy for error-correction. This method belongs to the class of \emph{phase-estimation} protocols~\cite{kitaev1995quantum,svore2013faster, travaglione2002preparing,Pirandola2006,terhal2016encoding,motes2017encoding,weigand2020realizing}. Indeed, measuring the value of the modular operator $\tilde{\Pop}_{b}$ is equivalent to estimating the phase of $e^{i\beta \Pop_b}$. Note however that here, the measurement is not Quantum Non Demolition (QND) in the sense that the phase of $e^{i\beta \Pop_b}$ is modified during each $\mathcal{R}_{p_b}$ round by the applied feedback displacements. As detailed below, the memory of the initial phase is fully erased after a few tens of rounds.  \\

To justify this approach and estimate the $\tilde{p}_b$ detection accuracy, we  suppose that the auxiliary oscillator is in a Zak-diagonal state of the form \eqref{rhodistrib} with a $P_b$ probability distribution  Dirac-peaked in $p_0$, whose value we want to estimate. Over a number $N_p$ of $\mathcal{R}_{p_b}$ preparation rounds, this distribution is on average shifted and broadened by the  feedback displacements $\{\pm \epsilon_j\}_{1\leq j \leq N_p}$ applied at the end of each round. Denoting  $\mathcal{S}=\{s_j\}_{1\leq j \leq N_p}$ a particular measurement record (with $s_j=\pm 1$ for each round $j$) and $-m(\mathcal{S})=\sum_{j=1}^{N_p} s_j \epsilon_j$  the total applied displacement, the auxiliary state $p$ distribution after re-preparation reads
\begin{equation}
P^{\mathcal{S}}_{p_0}(p)=\delta_{\frac{2\pi}{\beta}}(p-p_0+m(\mathcal{S})).
\end{equation}
 where $\delta_{\frac{2\pi}{\beta}}$ is a Dirac comb of period ${\frac{2\pi}{\beta}}$. Averaging over all possible measurement outcomes, the auxiliary state distribution  after re-preparation reads
\begin{equation}
P_{p_0}(p)=\sum_{\mathcal{S}}  \mathcal{P}_{p_0}(\mathcal{S} ) \delta_{\frac{2\pi}{\beta}}(p-p_0+m(\mathcal{S})) 
\end{equation}
where $\mathcal{P}_{p_0}(\mathcal{S})$ is the probability of the measurement record $\mathcal{S}$. $P_{p_0}$ becomes smooth for $N_p$ sufficiently large. \\

We simply propose to  estimate $p_0$ with $m(\mathcal{S})$ for a given measurement record $\mathcal{S}$. The accuracy of the $\tilde{\Pop}_{b}$ detection so performed is  characterized by the distribution $\mathcal{E}_{p_0}$ of the  error $e(\mathcal{S})=m(\mathcal{S})-p_0$. It reads
\begin{equation}
\mathcal{E}_{p_0}(e)= \sum_{\mathcal{S}}  \mathcal{P}_{p_0}(\mathcal{S} ) \delta_{\frac{2\pi}{\beta}}(e-(m(\mathcal{S})-p_0)) = P_{p_0}(-e)
\end{equation}
The  last equality simply means that the detection accuracy is as good as the auxiliary state re-preparation.\\

\begin{figure}[htbp] 
    \centering
    \includegraphics[width=1\columnwidth]{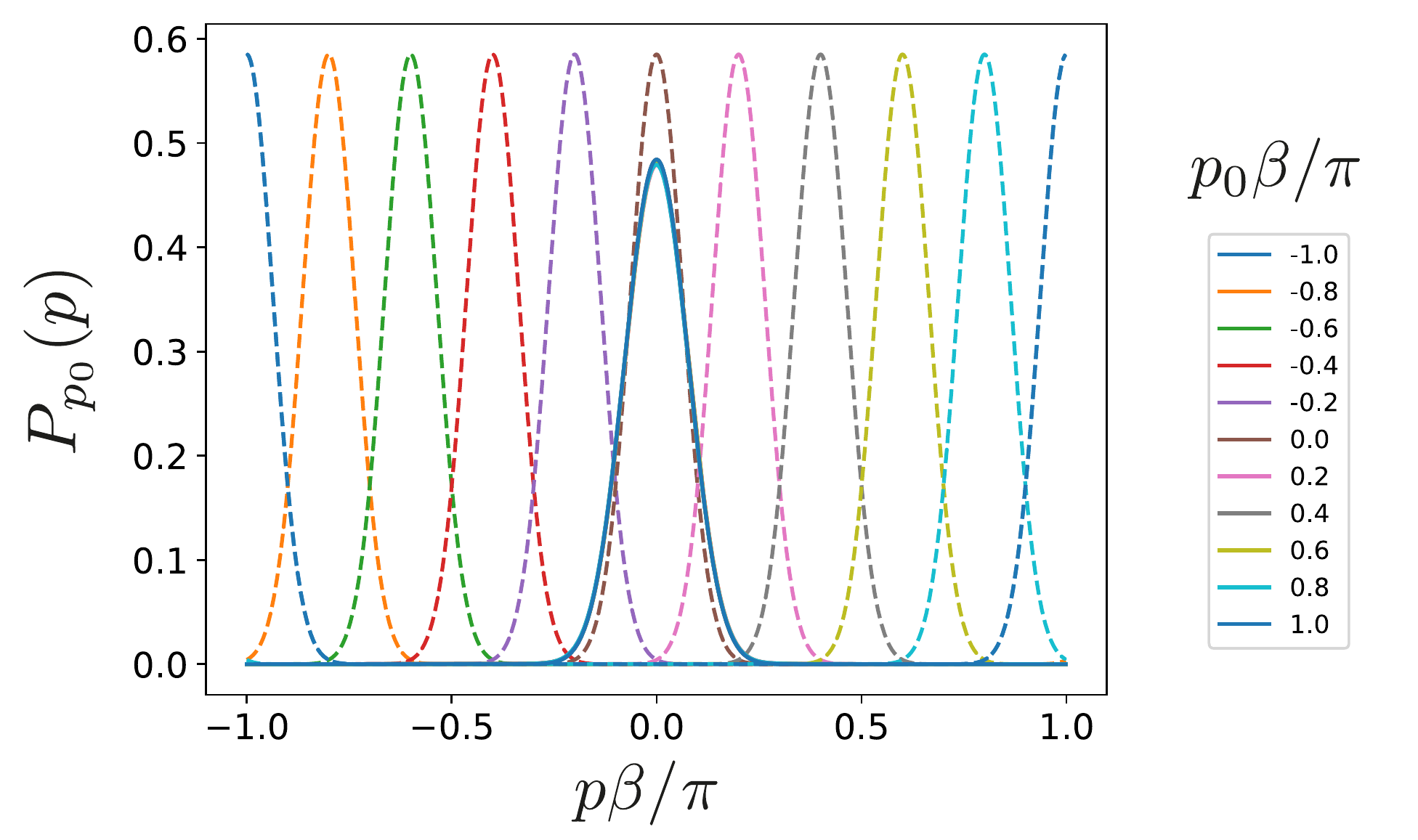}
    \caption{ {\bf Measurement and re-preparation of the auxiliary state}. We compute the $P_{p_0}$ distribution of the auxiliary oscillator (plain lines) prepared by a number $N_p=20$ of $\mathcal{R}_{p_b}$  rounds from a narrow Gaussian distribution (dashed lines, standard deviation $\sigma\sim 0.1)$ centered at $p_0$ (encoded in color). The feedback displacements applied after each round and the rectangularity parameter of the auxiliary oscillator are the ones returned by gradient ascent to minimize the logical error rate for $p^{BF}=2p^{PF}=0.005$ and $\kappa T_{\mathrm{round}}=2.10^{-5}$. We pick these example parameters - in particular the small number of preparation rounds - as the \emph{a priori} less favourable situation for the prepared auxiliary state not to depend on the initial condition $p_0$. We observe that the final distributions corresponding to different initial states do not differ significantly, justifying our approach to modular operator detection as detailed in Sec.~\ref{sec:phasestimation}.    }
    \label{fig:reprep}
\end{figure}

Crucially, we observe in Fig.~\ref{fig:reprep} that for all the re-preparation sequences used in this work, the auxiliary state distribution $P_{p_0}$ after the $\mathcal{R}_{p_b}$ rounds---and thus the error distribution $\mathcal{E}_{p_0}$---depends negligibly on $p_0$. This is simply understood as the long feedback kicks $\epsilon_j$ applied during the first few $\mathcal{R}_{p_b}$ rounds  quickly erase the memory of its prior state. This justifies \emph{a posteriori} the hypothesis made in Sec.~\ref{sec:ancillaprep} that the auxiliary state is  of the form \eqref{rhodistrib} with uniform distributions  prior to re-preparation: any initial state would yield the same prepared state.   As for the finite-accuracy of the $\tilde{\Pop}_{b}$ detection this method yields, it can be modeled by an ideal detection preceded by a convolution of the  $P_b$ probability distribution with the error function $\mathcal{E}=P_{N_p}$, where $P_{N_p}$ is the   distribution describing the auxiliary state prepared---from an arbitrary state---by a number  $N_p$ of   $\mathcal{R}_{p_b}$ rounds as detailed in Sec.~\ref{sec:ancillaprep} (we used that $P_{N_p}$ is an even distribution). Note that in Fig.~\ref{fig:reprep},  the auxiliary state distribution prior to re-preparation is a Gaussian centered in $p_0$ and with width $\sigma=0.1$ in rescaled coordinates. Taking $\sigma \rightarrow 0$ as in the above calculation led to numerical aberrations attributed to the  encoding of the wrapped distributions in the form of Fourier coefficient vectors of length $2n_F+1$ with $n_F=60$ (see Sec.~\ref{sec:ancillaprepFourier}). \\

In our reasoning, we have omitted shifts of the auxiliary state distribution entailed by flips of the qubit and intrinsic quadrature noise of the auxiliary oscillator during the $\mathcal{R}_{p_b}$ rounds. The former only entails shifts of the  $Q_b$ probability distribution during $\mathcal{R}_{p_b}$ rounds, and has no impact on $P_b$. The effect of the latter is to broaden the $P_b$ distribution as it is being measured and re-prepared. We model it by including quadrature noise in the numerical computation of $P_{N_p}$, by which we expect to slightly underestimate  the $\tilde{\Pop}_{b}$ detection accuracy. Indeed, by supposing that  $P_{N_p}$ is solely broadened by the stochastic nature of the applied feedback displacements  we overestimate the spread of $\mathcal{E}$.

\section{Efficient numerical estimate of the GKP qubit decoherence rate}
\label{sec:efficient}

Computing the evolution of the auxiliary and target state under the form of classical probability distributions $Q_a$, $P_a$, $Q_b$ and $P_b$ as detailed in the previous section greatly reduces the cost of numerical simulation compared to a full description in terms of density matrices. Typically, one keeps track of the distributions as two vectors of length $1000$. In this section, we further reduce simulation costs by representing the probability distributions in Fourier domain, as vectors of  $2n_F+1$ Fourier coefficients with $30 \leq n_F \leq 60$. After translating the calculations of sections \ref{sec:ancillaprep} and \ref{sec:targeterror} to the Fourier domain in sections \ref{sec:ancillaprepFourier} and \ref{sec:targetdynamicsFourier}, we encode the evolution of the target mode over a pair of $\mathcal{C}_{q_a}/\mathcal{C}_{p_a}$ cycle in an evolution matrix and show how to extract the decay rate of the GKP qubit by spectral analysis of this matrix in section \ref{sec:spectral}. We also show how to compute the gradient of this rate  with respect to the continuous parameters of the protocol (length of the feedback displacements $\epsilon_j$  applied after each  round, Fourier coefficients  of the target oscillator feedback law $f$, rectangularity $R=\sqrt{2\pi}/\beta$ of the auxiliary GKP state lattice), which greatly facilitates their optimization.\\

In order to simplify calculations,  we consider in the following re-scaled wrapped distributions $\Pi_{q_a}$, $\Pi_{q_b}$ and $\Pi_{p_b}$, and re-scaled feedback shifts $e_j$ at the end of each round,  $\tilde{\Pop}_{b}$ detection outcome $\psi$ at the end of each cycle and feedback law $F$ governing the feedback displacement applied to the target oscillator, all defined over $[-\pi,\pi]$:
	\begin{equation}
 \begin{split}
	\label{Anb}
	&\Pi_{q_a}(\phi)=\frac{1}{\sqrt{\pi}}Q_a(  \frac{\phi}{\sqrt{\pi}})\\
	&\Pi_{q_b}(\phi)=\frac{\beta}{2\pi}Q_b( \frac{\beta \phi}{2\pi })\\
	&\Pi_{p_b}(\phi)=\frac{1}{\beta}P_b( \frac{\phi}{ \beta})\\
 	&e_j=\epsilon_j \beta ~~\quad \text{for}~1\leq j \leq N_p\\
    &e_j=\epsilon_j \frac{2\pi}{\beta} \quad \text{for}~N_p+1\leq j \leq N_p+N_q\\
	&\psi=m \beta\\
	&F(\psi)=\sqrt{\pi}f(\frac{\psi}{\beta})\\
	\end{split}
 \end{equation}
We also define the Fourier coefficients of a $2\pi$-periodic  function $g$ as $g^{(k)}=\frac{1}{2\pi}\int_{-\pi}^{\pi}g(\phi)e^{-ik\phi}d\phi$.  Note that the evolution of the $P_a$ probability distribution is not considered here as the decay rate $\kappa_{log}$ of the GKP qubit. the z-Bloch vector component is computed from the evolution of $Q_a$ only.

\subsection{Auxiliary state preparation in Fourier domain}
\label{sec:ancillaprepFourier}
We revisit the auxiliary state preparation  described in Sec.~\ref{sec:ancillaprep} to translate it in the Fourier domain. The  distributions are uniform before preparation, with Fourier coefficients  $\Pi^{(k)}_{p_b,0}=\Pi^{(k)}_{q_b,0}=\delta_k/(2\pi)$, where $\delta$ is the Kronecker symbol.\\

During the $j$-th $\mathcal{R}_{p_b}$ round, the $\Pi_{p_b}$ distribution evolves after the physical qubit readout and application of a feedback displacement  following Eq.~\eqref{eq:shiftancilla}, which  reads in re-scaled coordinates 
\begin{equation}
\begin{aligned}
\Pi^{FB}_{p_b,{j-1}}(\phi)=& \Big( \frac{1}{2} + \frac{p^{NF}}{2} \text{sin}\big( \Phi + e_j \big)\Big)  \Pi_{p_b,{j-1}}(\phi + e_j)\\
&+ \Big( \frac{1}{2} - \frac{p^{NF}}{2} \text{sin}\big( \phi - e_j \big)\Big)  \Pi_{p_b,{j-1}}(\phi -e_j) 
\end{aligned}
\end{equation}
where we used the shorthand notation $p^{NF}=1-p^{BF}-2p^{PF}$.\\

Expanding this expression in powers of $e_j$, we get
\begin{equation}
\begin{aligned}
\Pi^{FB}_{p_b,{j-1}}(\phi)\simeq\frac{1}{2} &\sum_{n=0}^{n_T} \frac{e^n_j}{n!} \Big( \frac{\partial^n \Pi_{p_b,{j-1}}(\phi)}{\partial \phi^n}(1+(-1)^n)\\
&+ p^{NF}\frac{\partial^n (\Pi_{p_b,{j-1}}(\phi) \text{sin}(\phi))}{\partial \phi^n}(1-(-1)^n) \Big)
\end{aligned}
\label{eq:taylorancilla}
\end{equation}

Note that the term $n=1$ corresponds to a drift velocity $e_jp^{NF}\text{sin}(\phi)/T_{\mathrm{round}}$ and the term $n=2$ to a diffusion constant $e_j^2/(2T_{\mathrm{round}})$, quoted in Sec.~II in non-rescaled coordinates. Neglecting following terms, one obtains a Fokker-Planck equation, which is only valid for $e_j \rightarrow 0$.  For the numerical simulations performed in this work, we truncate the expansion at $n_T=30$.\\

In Fourier domain, this expression translates to 
\begin{equation}
\begin{aligned}
{\Pi_{p_b,{j-1}}^{FB~(k)}}=&\sum_{n~\text{even}} \frac{(i k e_j)^n}{n!} \Pi_{p_b,{j-1}}^{(k)}\\
&+ \sum_{n~\text{odd}} p^{NF}\frac{(i k e_j)^n}{n!}  \frac{1}{2i}(\Pi_{p_b,{j-1}}^{(k-1)}-\Pi_{p_b,{j-1}}^{(k+1)}).
\end{aligned}
\label{eq:ancillaround1}
\end{equation}
The distribution is then convolved with a Gaussian kernel modeling the effect of quadrature noise (see Eq.~\eqref{eq:ancillanoise}). In Fourier domain, it reads 
\begin{equation}
\label{eq:gaussianconv}
\Pi_{p_b,j}^{(k)}={\Pi_{p_b,{j-1}}^{FB~(k)}} e^{-\frac{1}{2} \kappa_p T_{\mathrm{round}} k^2}
\end{equation}
where $\kappa_p=2\pi \kappa/R^2$ is the re-scaled quadrature noise rate. After $N_p$ rounds, the   error function $\mathcal{E}$ for the $\tilde{\Pop}_{b}$ detection is inferred from  the distribution $\Pi_{p_b,N_p}$  (see Sec.~\ref{sec:ancillaprep}). The  $\Pi_{q_b,N_p}$ distribution is still uniform at this stage as we assume the distributions $\Pi_{q_b,0}$ and $\Pi_{p_b,0}$ prior to preparation to be uniform (see Sec.~\ref{sec:phasestimation}).\\

Through the sequence of $\mathcal{R}_{q_b}$ rounds, the $\Pi_{p_b}$ distribution evolves due to quadrature noise and random displacements induced by bit-flips of the qubit as
\begin{equation}
\Pi_{p_b,N_p+N_q}^{(k)}=(1-p_{tot}^{BF})\Pi_{p_b,N_p}^{(k)} e^{-\pi N_q \kappa_p k^2} + p_{tot}^{BF} \frac{\delta_k}{2\pi} 
\end{equation}
where $p_{tot}^{BF}=1-(1-p^{BF})^{N_q}$ is the probability for at least one bit-flip to have occurred. As for the $\Pi_{q_b}$ distribution, it evolves through  $\mathcal{R}_{q_b}$ rounds following the same rules as $\Pi_{p_b}$ through $\mathcal{R}_{p_b}$ rounds (Eqs.~(\ref{eq:ancillaround1},\ref{eq:gaussianconv})), albeit with a re-scaled quadrature noise rate $\kappa_q= 2\pi \kappa R^2$ for the Gaussian kernel convolution.\\

Overall, we thus compute the prepared auxiliary state under the form of two $(2n_F+1)$-vectors of Fourier coefficients ($-n_F\leq k \leq n_F$), and obtain the  error-function $\mathcal{E}$ for the $\tilde{\Pop}_{b}$ detection under the same form.  Moreover, it is straightforward to compute the gradient of each vector with respect to each feedback displacement length $e_j$, as well as with respect to the grid rectangularity parameter $R$, by taking the derivative of the formulas given above and applying chain rules.

\subsection{Target oscillator dynamics in Fourier domain}
\label{sec:targetdynamicsFourier}
We revisit the target oscillator evolution over a pair of $\mathcal{C}_{q_a}/\mathcal{C}_{p_a}$ cycles, labeled $j$ and $j+1$,  described in Sec.~\ref{sec:targeterror} to translate it in Fourier domain. The auxiliary state  distributions before the quadrature gate are $\Pi_{q_b,N_p+N_q}$ (abbreviated to $\Pi_{q_b}$) and $\Pi_{p_b,N_p+N_q}$, as computed in the previous section. As detailed in Sec.~\ref{sec:phasestimation}, we model the inaccuracy of the $\tilde{\Pop}_{b}$ detection by convolving $\Pi_{p_b,N_p+N_q}$  with an error distribution $\mathcal{E}=\Pi_{p_b,N_p}$ - which is a simple vector multiplication in Fourier domain - and denote the resulting distribution by $\Pi_{p_b}$.  \\

During the $\mathcal{C}_{q_a}$ cycle, the initial target oscillator distribution $\Pi_{q_a,j-1}$  is first evolved with the left expression in \eqref{eq:targetFB} modelling the quadrature gate followed by a measurement of $\tilde{\Pop}_{b}$ whose outcome controls a feedback displacement applied to the target oscillator. In re-scaled coordinates, this evolution reads
\begin{equation}
\Pi^{FB}_{q_a,{j-1}}(\phi)= \int_{-\pi}^{\pi} \Pi_{q_a,{j-1}} \big(\phi+F(\psi)\big)\Pi_{p_b}\big(\psi -  2(\phi+F(\psi))\big)d\psi 
\end{equation}
We now expand this expression in powers of the re-scaled feedback displacement $F(\psi)$ applied to the target oscillator, and truncate the series at $n_T$ ($n_T=30$ for all simulations performed in this work). We then get
\begin{equation}
\label{eq:taylortarget}
\begin{split}
\Pi^{FB}_{q_a,{j-1}}(\phi)&\simeq \int_{-\pi}^{\pi}  \sum_{n=0}^{n_T} \frac{F^n(\psi)}{n!}  \frac{\partial^n }{\partial \phi^n}\Big(\Pi_{q_a,{j-1}}(\phi) \Pi_{p_b}\big(\psi -  2\phi \big) \Big)d\psi\\
&=\sum_{n=0}^{n_T} \frac{1}{n!} \frac{\partial^n }{\partial \phi^n}\Big(D_n(\phi)\Pi_{q_a,{j-1}}(\phi)  \Big)
\end{split}
\end{equation}
where we  defined the generalized Fokker-Planck coefficient functions $D_n$
\begin{equation}
\begin{split}
D_n(\phi)&=\int_{-\pi}^{\pi} F^n(\psi) \Pi_{p_b}(\psi-2\phi)\\
&=(F^n\ast \Pi_{p_b})(2\phi)
\end{split}
\end{equation}
(we use that $\Pi_{p_b}$ is even in the last equality). In Fourier domain, this translates to
\begin{equation}
{\Pi^{FB~(k)}_{q_a,{j-1}}}=\sum_{n=0}^{n_T} \frac{(ik)^n}{n!}  \Big( \sum_{l=-N}^N D_n^{(k-l) } {\Pi_{q_a,{j-1}}}^{(l)} \Big)
\label{eq:targetfouerierevol}
\end{equation}
and the Fourier coefficients of $D_n$ are computed with
\begin{equation}
   D_n^{(k) } =
    \begin{cases}
        (F^n \ast \Pi_{p_b})^{(\frac{k}{2}) }= 2\pi {(F^{\tilde{\ast}^n})}^{(\frac{k}{2}) } \Pi_{p_b}^{(\frac{k}{2}) }  & \text{if $k$ even}\\
       0&\text{if $k$ odd}
    \end{cases}   
\label{eq:convol}
\end{equation}
where $\tilde{\ast}^n$ denotes the $n$-fold discrete convolution product defined as $(u\tilde{\ast} v)^{(k)}=\sum_{l=-N}^N u^{(k-l) } v^{(l) }$. In simulations, we truncate this  sum in order to maintain a $2n_F+1$ structure for the Fourier coefficient vectors.\\

The distribution is then convolved with a Gaussian kernel $G_a$ modeling the effect of quadrature noise  during the $\mathcal{C}_{q_a}$ cycle (left equation in \eqref{eq:targetgaussian}), then convolved with the $\Pi_{q_b}$ distribution to model the backaction of the quadrature gate in the following $\mathcal{C}_{p_a}$ cycle (right equation in \eqref{eq:targetFB}, replacing $P_a\rightarrow Q_a$), and again convolved with a Gaussian kernel $\tilde{G}_a$ modeling the effect of quadrature noise  during the $\mathcal{C}_{p_a}$ cycle (right equation in \eqref{eq:targetgaussian} replacing $P_a\rightarrow Q_a$). In Fourier domain, it reads 
\begin{equation}
\Pi_{q_a,j+1}^{(k)}=2\pi \Pi_{q_b}^{(k)}  e^{-\frac{k^2 \sigma^2_{tot}}{2} } {\Pi^{FB~(k)}_{q_a,{j-1}}}
\label{eq:fourierconv}
\end{equation}
with $\sigma^2_{tot}=2\kappa T_{\mathrm{cycle}}+\frac{\theta^2}{3}\kappa T_{\mathrm{quad}} $.\\

Combining Eq.~\eqref{eq:targetfouerierevol} and Eq.~\eqref{eq:fourierconv}, the evolution through the two cycles can be expressed under a  matrix form 
\begin{equation}
\Pi_{q_a,j+1}^{(k)}=\sum_{l=-N}^N M_{kl}\Pi_{q_a,{j-1}}^{(l)}.
\end{equation}
Note that $M$ is real when $F$ is odd, which is the case in the following.

\subsection{GKP qubit decoherence rate and convergence rate to the code manifold  by spectral analysis of the evolution matrix}
\label{sec:spectral}

The evolution matrix $M$ is the Fourier transform of a stochastic matrix. As such, it shares the same eigenspectrum $\{\lambda_i\}$ where we arrange the eigenvalues in decreasing magnitude order. In particular $\lambda_0=1$, and $|\lambda_i|\leq 1$ for  $i\geq 1$.\\

In the regime where the logical flip probability per cycle is small, we find that the spectrum is gaped with $|\lambda_j| \ll |\lambda_1|$ for $j>1$. Qualitatively, this gap indicates a fast convergence of the system to a 2D-manifold of probability vectors (distributions), at a rate \begin{equation}
\Gamma_{conv}=-\mathrm{log}(|\lambda_2|)/(2T_{\mathrm{cycle}})
\end{equation}
(remember that the evolution matrix corresponds to two error-correction cycles). We interpret this fast dynamics as a convergence of the target oscillator state to a metastable state in the vicinity the GKP code manifold. It is followed by a slow relaxation, within this manifold, to the system steady-state---the probability distribution $\Pi_0$ obtained by inverse Fourier transform of the eigenvector attached to $\lambda_0$---at a rate
\begin{equation}
\kappa_{\mathrm{log}}=-\mathrm{log}(\lambda_1)/(2T_{\mathrm{cycle}}).
\end{equation}
In this expression, we have used that, since $M$ is real and $\lambda_1$ does not have a conjugate eigenvalue, $\lambda_1$ is real. We interpret this slow dynamics as the relaxation of the GKP qubit towards the mixed logical state.\\

\begin{figure}[htbp] 
    \centering
    \includegraphics[width=1\columnwidth]{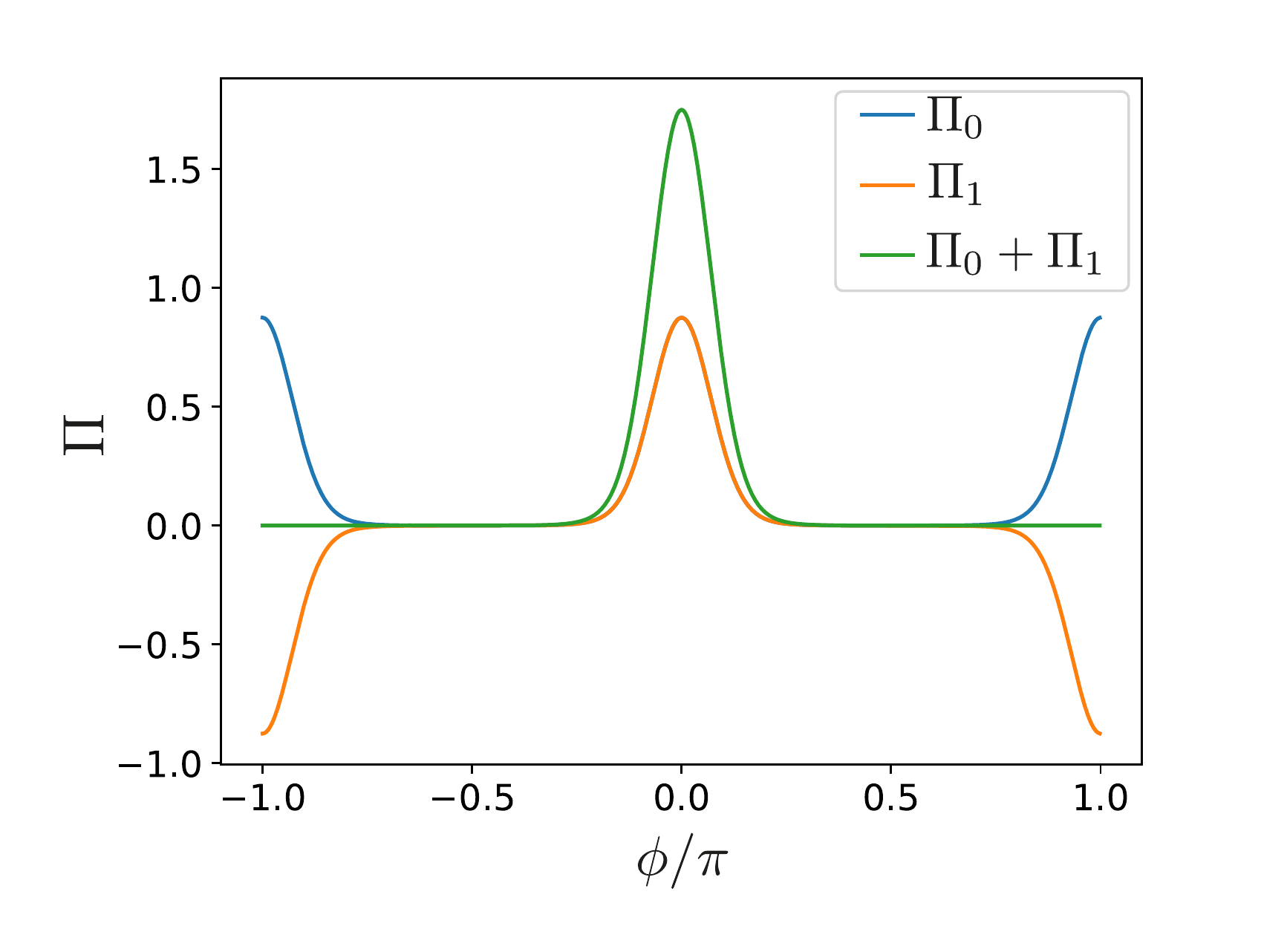}
    \caption{ {\bf Eigenvectors of  the  stochastic evolution matrix $M$}. For $\kappa T_{\mathrm{round}}=2e-5$, $p^{BF}=1e-3$, $N_p=N_q=60$ and all other parameters optimized by gradient ascent, we represent the inverse Fourier transform  of the eigenvectors of $M$  with largest eigenvalues  $\lambda_0=1$ and $\lambda_1=1-1.3\times10^{-5}$, respectively labeled $\Pi_0$ and $\Pi_1$. Rescaled to a unit $L_1$ norm, $\Pi_0$ is the probability distribution   of the target mode steady-state under error-correction ($\Pi_{q_a,j}$ with $j\rightarrow \infty$ in Sec.~\ref{sec:targetdynamicsFourier}). This state is close to the code manifold, with narrow peaks centered at $\phi=0~\text{mod}~\pi$ and is decoded as the fully mixed  state of the GKP qubit.  $\Pi_1$  has a null $L_1$ norm, and is here re-scaled to the same $L_{\infty}$ norm as $\Pi_0$.   Given that $\lambda_1$ is close to 1  and that a  gap exists with the next largest eigenvalue ($\lambda_2=0.55$), a general state  converges in a few correction cycles to a probability distribution $ \Pi_0 + \zeta \Pi_1$, where $\zeta$ is an excellent approximation of the $z$-component of the GKP qubit Bloch vector when the peaks of $\Pi_0$ and $\Pi_1$ are sufficiently narrow. }
    \label{fig:eigenvectors}
\end{figure}

We confirm this intuition by representing the probability distributions $\Pi_0$ and $\Pi_1$ corresponding to $\lambda_0$ and $\lambda_1$ in Fig.~\ref{fig:eigenvectors}, for cycle parameters allowing a robust protection of the GKP qubit. $\Pi_0$ displays two peaks of equal height centered in $\Phi=0$ and $\phi=\pi$, as expected from a state close to the code manifold and decoded as the fully mixed logical state. With the proper normalization, $\Pi_0+\Pi_1$ displays a single peak centered in  $\phi=0$, as expected from a state close to the code manifold and decoded as $|+Z\rangle$.\\

This spectral analysis in Fourier domain is a powerful tool to estimate the decay rate of the $z$-component of the GKP qubit Bloch vector. We compared its results to brute-force computation of the evolution of the target oscillator state, encoded as a probability vector, over a large number of error-correction cycles (see Sec.~\ref{sec:targeterror}) before fitting the decay of the decoded $z$-component of the GKP qubit Bloch vector. Both methods agree quantitatively when the oscillators state are encoded in a sufficiently long  Fourier vector of length $2n_F+1$, and when the Taylor expansion in Eq.~\eqref{eq:taylorancilla} and Eq.~\eqref{eq:taylortarget} is truncated at  a sufficiently high order $n_{T}$  (not shown). In practice, we found that  $n_F=n_{T}=30$   was sufficient for all numerical simulations  presented in this paper, except to estimate the smallest decay rates of Fig.~\ref{fig:noiseless} and to obtain the real-domain distributions with no visible ripples presented in Fig.~\ref{fig3}, for which $n_F=60$ was used. Given the small matrix size involved,  spectral analysis  in Fourier domain is significantly faster than brute-force simulation in real domain. It also allows us to estimate the convergence rate to the code manifold $\Gamma_{conv}$, as represented in Fig.~\ref{fig:noiseless}. \\

Furthermore, for a given feedback parameter set, the method  allows us to compute the gradient of $\lambda_1$  with respect to the cycle continuous parameters (length $e_j$ of the  feedback displacements on the auxiliary oscillator,  Fourier coefficients $F^{(k)}$ of the feedback function and rectangularity $R$ of the auxiliary state).  To this end, we first take the derivative of the evolution rules for the target and the auxiliary state probability distributions  (see Sec.~\ref{sec:ancillaprepFourier} and Sec.~\ref{sec:targetdynamicsFourier}), respectively through a cycle and through a round, and apply chain  rules to obtain the derivative of the evolution matrix $M$  with respect to a given parameter $x$. Each  component of the gradient is then given by
\begin{equation}
\frac{\partial \lambda_1}{\partial x}= \dfrac{P^L_1 \cdot M \cdot P^R_1}{P^L_1\cdot P^R_1}
\end{equation}
where $(\cdot)$ denotes the matrix product and $P^L_1$ and $P^R_1$ are respectively left and right eigenvectors of $M$ for the eigenvalue $\lambda_1$.

\section{Optimization of error-correction parameters}

\subsection{Optimizing continuous parameters by gradient ascent}
\label{sec:gradient}

\begin{figure*}[t] 
    \centering
    \includegraphics[width=1.4\columnwidth]{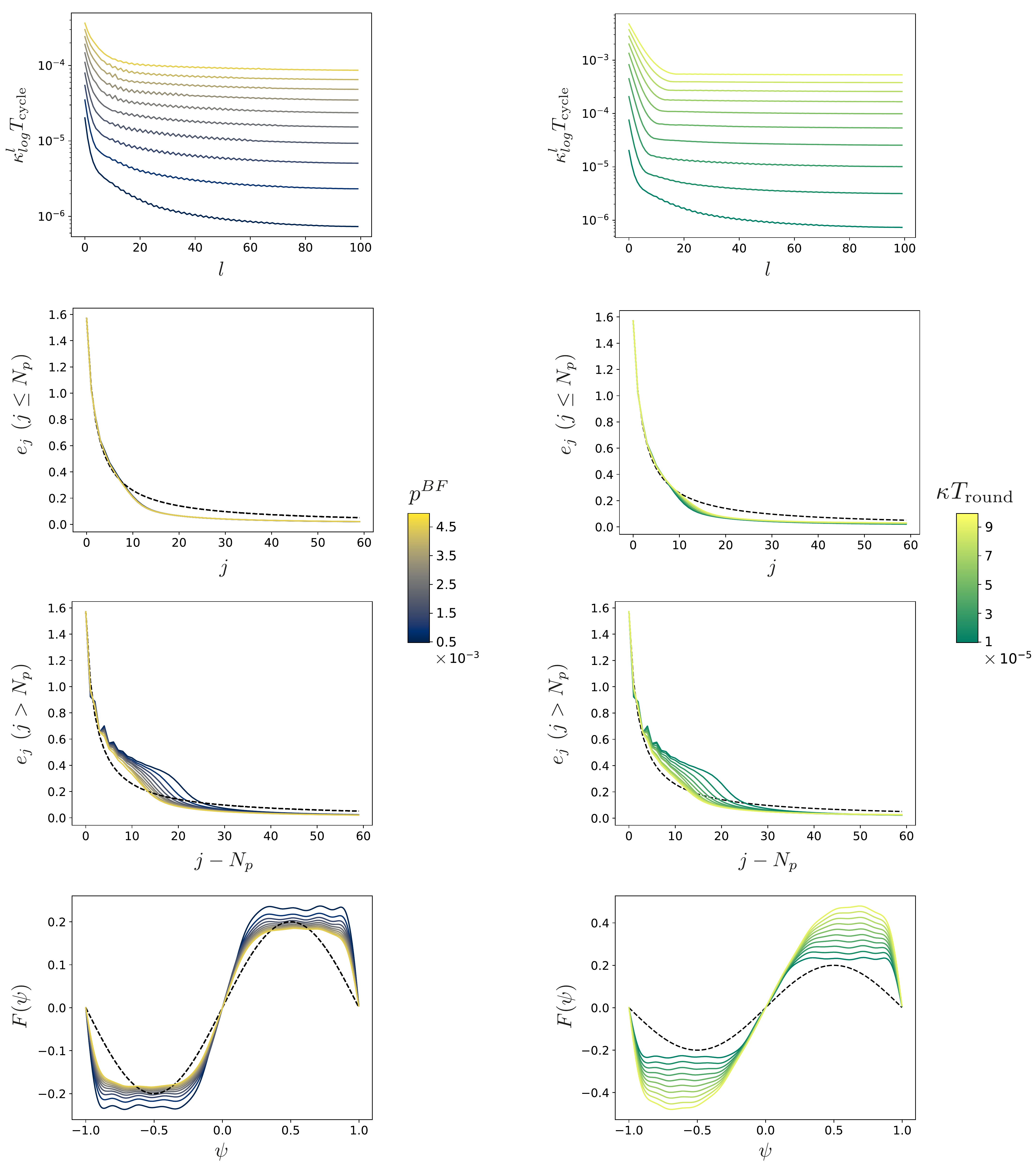}
    \caption{ {\bf Optimization of feedback parameters by gradient ascent}. The continuous control parameters of our protocol are optimized  by gradient ascent to minimize the  decay rate of the $z$-component  of the GKP qubit Bloch vector. The top two panels  represent this decay rate $\kappa^l_{log}$  as a function of the ascent step number $l$. On the left, we vary the physical qubit flip probability $p^{BF}=2p^{PF}$ (encoded in color) for a fixed value of the oscillators noise strength $\kappa T_{\mathrm{round}}=10^{-5}$ (left). On the right, we vary  $\kappa T_{\mathrm{round}}$ (encoded in color) for a fixed value of $p^{BF}=2p^{PF}=5.10^{-4}$. In the three next lines of panels, we represent, for the same noise figures and at the end of the gradient ascent  ($l=100$), the feedback displacements  applied after each $\mathcal{R}_{p_b}$  round as a function of the round index $j\leq N_p$ (second line), the feedback displacements  applied after each $\mathcal{R}_{q_b}$  round as a function of $N_p<j\leq N_p+N_q$ (third  line), and the feedback law $F$ controlling the displacements applied to the target oscillator as a function of the normalized outcome $\psi$ of the $\tilde{\Pop}_{b}$ estimation (fourth line).  For each parameter, the initial guess (before gradient ascent, $l=0$) is represented by a black dashed line.  The total number of  rounds is fixed to   $N_q=N_p=60$. } 
    \label{fig:paramgradient}
\end{figure*}

For a given set of noise values $\kappa T_{\mathrm{round}}$, $p^{BF}$ and $p^{PF}$ ($p^{PF}=p^{BF}/2$ throughout this work) and  given preparation round numbers $N_q$ and $N_p$, we optimize the remaining parameters of the error-correction cycle  by gradient ascent to maximize the value of $\lambda_1$. In detail, we consider the gradients 
\begin{equation}
\begin{split}
A&=\{\frac{\partial \lambda_1}{\partial e_j} \}_{\{ 1\leq j \leq N_p\}}\\
B&=\{\frac{\partial \lambda_1}{\partial e_j} \}_{\{ N_p+1\leq j \leq N_p + N_q\}}\\
C&=\{\frac{\partial \lambda_1}{\partial F_s^{(k)}} \}_{\{ 1\leq k \leq k_{max}} \}\\
D&=\{\frac{\partial \lambda_1}{\partial R} \}
\end{split}
\end{equation}
where we  defined $F_s^{(k)}=(F^{(k)}-F^{(-k)})/(2i)$. This choice constrains the feedback function $F$ to the odd sector, ensuring that the target probability distribution remains symmetric at all time (real evolution matrix $M$). We choose to limit the number of free Fourier coefficients of $F$ to $n'_F=10 < n_F $ to limit aberrations entailed by Fourier series truncation during the convolution step \eqref{eq:convol}. Pushing $n'_F$ to larger values---and increasing $n_F$ accordingly to avoid aberrations---did not lead to a significant  improvement in error-correction performances. \\

At each step $l$ of the gradient ascent - for a total number of steps $L=100$ - we update the parameter values following
\begin{equation}
\begin{split}
&\{e_j \}^{l+1}_{\{ 1\leq j \leq N_p\}}=\{e_j \}^{l}_{\{ 1\leq j \leq N_p\}} + a\frac{A}{|A|_{\infty}}\Delta_l \\
&\{e_j \}^{l+1}_{\{ N_p+1\leq j \leq N_p+N_q\}}=\{e_j \}^{l}_{\{ N_p+1\leq j \leq N_p+N_q\}} + b\frac{B}{|B|_{\infty}}\Delta_l \\
&\{  F_s^{(k)} \}^{l+1}_{\{ 1\leq k \leq k_{max} \}}=\{  F_s^{(k)} \}^{l}_{\{ 1\leq k \leq k_{max}\}} + c\frac{C}{|C|_{\infty}}\Delta_l \\
&\{ R \}^{l+1}=\{ R \}^{l} + d\frac{D}{|D|_{\infty}}\Delta_l
\end{split}
\end{equation}
where the step $\Delta_l$ decreases linearly from $1$ to $0.05$ when l varies from $1$ to $L$, the  parameters $a=b=0.005$, $c=0.02$ and $d=0.04$ were adjusted empirically such that the vectors $A$, $B$, $C$ and $D$ would converge at comparable speeds toward their final values.\\

As initial guess parameters, we set
\begin{equation}
\begin{split}
&e_j^{0}=  \dfrac{e_i e_f}{e_f +(e_i-e_f)\frac{j}{N_p}}  \quad \text{for } j\leq N_p\\
&e_j^{0}=  \dfrac{e_i e_f}{e_f +(e_i-e_f)\frac{j-N_p}{N_q}}  \quad \text{for } j> N_p\\
&  {F_s^{(k)}} ^{0}=f_1 \delta_{k-1} \\
& R ^{0}=1 
\end{split}
\end{equation}
In these expressions, the  initial guess for the  feedback displacements on the auxiliary oscillator (first two expressions) is a truncated $1/j$ function with large initial value $e_i=\pi/2$ in order to suppress the tails of the $\Pi_{p_b}$ and $\Pi_{q_b}$ distributions,  and small final value $e_f=0.05 \ll 2\pi$ to limit the distributions central peak width (see Sec.~II). The $1/j$ power law was chosen to maximize the reduction rate of the distributions central peak width, while ensuring that this width reaches 0 when $N_q,~N_p\rightarrow \infty$, in absence of intrinsic noise of the oscillator. The initial guess for the re-scaled feedback law $F$ is a simple sine function of amplitude $f_1=0.2$.\\

We observed that the final value of $\lambda_1$ (and hence the decay rate $\kappa_{log}$) and the  correction parameters returned by the gradient ascent algorithm depends slightly on the initial guess, indicating the existence of multiple local minima of $\lambda_1$ (not shown). The rugged aspect of $\{e_j \}^{l+1}_{\{ N_p+1\leq j \leq N_p+N_q\}}$ observed after gradient ascent for some noise values (see Fig.~\ref{fig:paramgradient}, third line) tends to confirm this complex structure. More refined gradient ascent techniques may avoid these issues, but were not attempted in this work.

\subsection{Optimizing the number of preparation rounds}
\label{sec:optround}

\begin{figure}[h] 
\includegraphics[width=1\columnwidth]{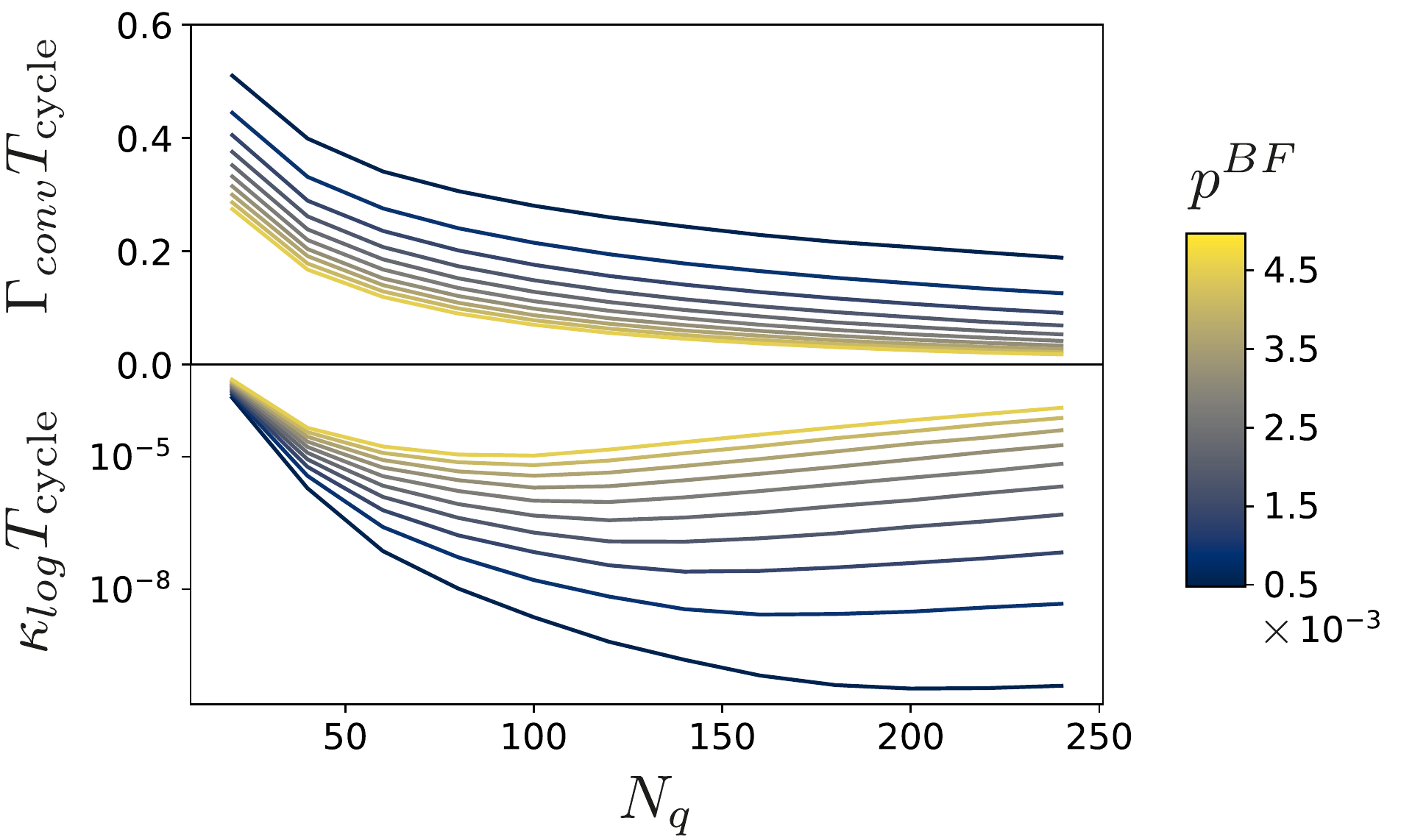}
    \caption{Convergence rate towards the GKP code manifold (top panel) and decay rate of the $z$-component of the GKP qubit Bloch vector  (bottom panel) as a function of the number $N_q$ of $\mathcal{R}_{q_b}$ preparation rounds in each cycle, in absence of  intrinsic oscillator noise ($\kappa=0$) and for various physical qubit flip probabilities per round (encoded in color, with $p^{PF}=p^{BF}/2$).  For each value of $N_q$, remaining control parameters are optimized by gradient ascent. The convergence rate towards the code manifold decreases with $N_q$ as the probability of a single qubit flip to have occurred during  the  $\mathcal{R}_{q_b}$ preparation rounds approaches 1, blurring error-syndromes extracted during the following error-correction cycle. As a consequence, there exists a  finite value of $N_q$ that minimizes the GKP qubit decay rate.
    }
    \label{fig:noiseless}
\end{figure}	

In this section, we analyze the impact of the number of auxiliary state preparation rounds, $N_q$ and $N_p$, on the  performances of our error-correction protocol, and comment on the existence of an optimal, finite value for $N_q$ and $N_p$, irrespective of the system noise strength.\\

We  first consider  the case of noiseless  oscillators ($\kappa=0$). Since the cycle duration is irrelevant in that limit, we allow  the auxiliary state to be prepared with a large number $N_p\rightarrow \infty$ of $\mathcal{R}_{p_b}$ rounds. Letting $e_j=e\rightarrow 0$,  the evolution of the $\Pi_{p_b}$ distribution over a $\mathcal{R}_{p_b}$ round is  given by Eq.~\eqref{eq:taylorancilla}, which can be truncated at $n_{T}=2$. We can approximate this discrete time evolution with a continuous time evolution   governed by a Fokker-Planck equation 

\begin{equation}
\frac{\partial \Pi_{p_b}}{\partial t}=-\frac{\partial (v(\phi) \Pi_{p_b})}{\partial \phi}+\frac{1}{2}\frac{\partial^2 (D \Pi_{p_b})}{\partial \phi^2}.
\end{equation}
where $v(\phi)=-e(1-p^{BF}-2p^{PF})/T_{\mathrm{round}}$ and $D=e^2/T_{\mathrm{round}}$. After an infinite number of $\mathcal{R}_{p_b}$ rounds, $\Pi_{p_b}$ reaches the  steady-state of this equation, which approaches a wrapped normal distribution with variance $e/(2(1-p^{BF}-2p^{PF}))$ for $e\rightarrow 0$. In other words, $\Pi_b$ is a Dirac distribution, and it follows from Sec.~\ref{sec:phasestimation} that the detection  of  $\tilde{\Pop}_{b}$ following the quadrature gate is perfect.  \\

We now consider the decay of the GKP qubit in this configuration. In Fig.~\ref{fig:noiseless}, we represent the decay rate $\kappa_{log}$ of the z-component of the GKP qubit Bloch vector and the convergence rate towards the code manifold $\Gamma_{conv}$, in units of $T_{\mathrm{cycle}}$,  as a function of $N_q$. For each value of $N_q$, the remaining feedback parameters (feedback displacements following $\mathcal{R}_{q_b}$ rounds, feedback law $F$ and  rectangularity $R$) are optimized by gradient ascent as detailed in the previous section. Surprisingly, even in this limit case of noiseless oscillators, we find that an optimal number of $\mathcal{R}_{q_b}$ rounds exists, which can be understood with the following arguments. When  $N_q\rightarrow 0$, the $Q_b$ distribution becomes widely spread (see Fig.~\ref{fig3}, bottom panel) and long shifts propagate through the quadrature gate, increasing $\kappa_{log}$. In the opposite limit  $N_q\rightarrow \infty$, $P_b$ is a near-uniform  distribution as the probability of at least one bit-flip during $\mathcal{R}_{q_b}$ rounds approaches 1, blurring the error-syndromes extracted from the target oscillator during the following correction cycle (see Fig.~\ref{fig3}, bottom panel). As a result, the convergence rate to the code manifold drops to 0 (see Fig.~\ref{fig:noiseless}, top panel), and small shifts propagating through the quadrature gate are sufficient to trigger logical errors. Admittedly, we still expect $\kappa_{log}$ to vanish for $N_q \rightarrow \infty$ (Dirac-peaked $Q_b$ distribution) and $F=0$ (no feedback displacement applied to the target oscillator) as the target mode  dynamics cancels, but this  regime is reached for round numbers far beyond the range considered here. \\


\begin{figure*}[t] 
    \centering
    \includegraphics[width=2\columnwidth]{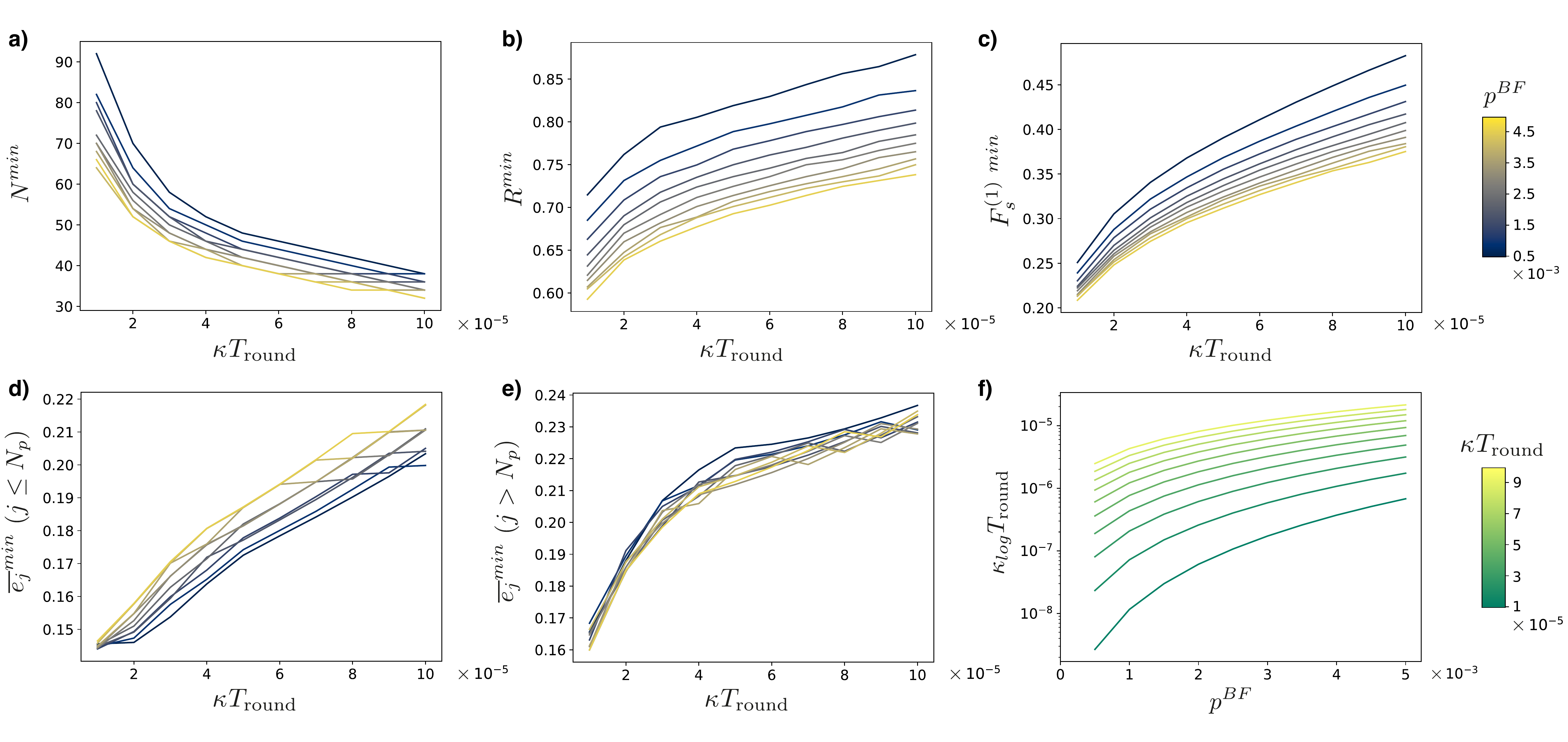}
    \caption{  Error correction parameters used for the simulations whose results are presented in Fig.~\ref{fig4}. For each considered noise figure  (oscillator noise strength  $\kappa T_{\mathrm{round}}$ varied along the x-axis of each panel, physical qubit flip probability per round $p^{BF}=2 p^{PF}$ encoded color) we sweep the number of preparation round $N_q=N_p$ and perform gradient ascent on the continuous correction parameters  to minimize the logical qubit decay rate  for each value of $N_q=N_p$. {\bf a)} Round number $N^{min}$ yielding the minimum logical error rate after gradient ascent. For this round number, {\bf b)} represents the auxiliary  lattice rectangularity $R^{min}$ found by gradient ascent, {\bf c)} the dominant Fourier coefficient $F^{(1)~min}$  of the feedback law  controlling the feedback displacement applied to the target oscillator {\bf d)} the average value of the  feedback displacements $\overline{e_j}^{min}$  ($j\leq N_p$) applied to the auxiliary oscillator during $\mathcal{R}_{p_b}$ rounds and  {\bf e)} the average value of the  feedback displacements $\overline{e_j}^{min}$  ($j> N_p$) applied to the auxiliary oscillator during $\mathcal{R}_{q_b}$ rounds. {\bf f)} Decay rate $\kappa_{log}$ of the $z$-component of the  GKP qubit Bloch vector  as a function of $p^{BF}$, with $\kappa T_{\mathrm{round}}$ encoded in color (different colorscale from other panels). This is the same data as presented in Fig.~\ref{fig4}, albeit plotted against different axes.}
    \label{fig:fig3sup}
\end{figure*}

Now considering the case of noisy oscillators ($\kappa>0$), the optimal value of $N_q$ is lower than in the noiseless case (see Fig.~\ref{fig4} and Fig.~\ref{fig:fig3sup}a). Indeed, quadrature noise on the auxiliary oscillator during $\mathcal{R}_{q_b}$ rounds causes  the $\Pi_{p_b}$ distribution to diffuse and homogenize, so that the convergence rate to the code manifold decreases faster with $N_q$ than in the noiseless case. It also impacts the accuracy of the $\tilde{\Pop}_b$ detection, with a similar effect (see Sec.~\ref{sec:phasestimation}). In Fig.~\ref{fig:fig3sup}a, we represent the optimal number of $\mathcal{R}_{q_b}$ rounds $N^{min}$ found when sweeping together $N_q=N_p$, for the same range of system noise strength considered in Fig.~\ref{fig4}. Note that here, $N^{min}$ is the value found to minimize $\kappa_{log}T_{\mathrm{round}}$ (and not $\kappa_{log}T_{\mathrm{cycle}}$ as in Fig.~\ref{fig:noiseless}), which tends to favor smaller numbers of preparation round. For completeness, we represent in  Fig.~\ref{fig:fig3sup}b-e the value of the other cycle parameters, found by gradient ascent at $N_q=N_p=N^{min}$, which  yield the GKP qubit decay rates presented in Fig.~\ref{fig4}. Finally, in order to confirm the exponential suppression of logical errors as a function of both the qubit error probability per round $p^{BF}$ and the oscillators noise strength $\kappa T_{\mathrm{round}}$, we plot in Fig.~\ref{fig:fig3sup}f the same data as in Fig.~\ref{fig4} (decay rate of the $z$-component of the  GKP qubit Bloch vector $\kappa_{log}$) but here as a function of $p^{BF}$, with $\kappa T_{\mathrm{round}}$  encoded in color.





\bibliography{biblio.bib}

\begin{thebibliography}{46}%
\makeatletter
\providecommand \@ifxundefined [1]{%
 \@ifx{#1\undefined}
}%
\providecommand \@ifnum [1]{%
 \ifnum #1\expandafter \@firstoftwo
 \else \expandafter \@secondoftwo
 \fi
}%
\providecommand \@ifx [1]{%
 \ifx #1\expandafter \@firstoftwo
 \else \expandafter \@secondoftwo
 \fi
}%
\providecommand \natexlab [1]{#1}%
\providecommand \enquote  [1]{``#1''}%
\providecommand \bibnamefont  [1]{#1}%
\providecommand \bibfnamefont [1]{#1}%
\providecommand \citenamefont [1]{#1}%
\providecommand \href@noop [0]{\@secondoftwo}%
\providecommand \href [0]{\begingroup \@sanitize@url \@href}%
\providecommand \@href[1]{\@@startlink{#1}\@@href}%
\providecommand \@@href[1]{\endgroup#1\@@endlink}%
\providecommand \@sanitize@url [0]{\catcode `\\12\catcode `\$12\catcode `\&12\catcode `\#12\catcode `\^12\catcode `\_12\catcode `\%12\relax}%
\providecommand \@@startlink[1]{}%
\providecommand \@@endlink[0]{}%
\providecommand \url  [0]{\begingroup\@sanitize@url \@url }%
\providecommand \@url [1]{\endgroup\@href {#1}{\urlprefix }}%
\providecommand \urlprefix  [0]{URL }%
\providecommand \Eprint [0]{\href }%
\providecommand \doibase [0]{http://dx.doi.org/}%
\providecommand \selectlanguage [0]{\@gobble}%
\providecommand \bibinfo  [0]{\@secondoftwo}%
\providecommand \bibfield  [0]{\@secondoftwo}%
\providecommand \translation [1]{[#1]}%
\providecommand \BibitemOpen [0]{}%
\providecommand \bibitemStop [0]{}%
\providecommand \bibitemNoStop [0]{.\EOS\space}%
\providecommand \EOS [0]{\spacefactor3000\relax}%
\providecommand \BibitemShut  [1]{\csname bibitem#1\endcsname}%
\let\auto@bib@innerbib\@empty
\bibitem [{\citenamefont {Gottesman}\ \emph {et~al.}(2001)\citenamefont {Gottesman}, \citenamefont {Kitaev},\ and\ \citenamefont {Preskill}}]{gottesman2001encoding}%
  \BibitemOpen
  \bibfield  {author} {\bibinfo {author} {\bibfnamefont {Daniel}\ \bibnamefont {Gottesman}}, \bibinfo {author} {\bibfnamefont {Alexei}\ \bibnamefont {Kitaev}}, \ and\ \bibinfo {author} {\bibfnamefont {John}\ \bibnamefont {Preskill}},\ }\bibfield  {title} {\enquote {\bibinfo {title} {Encoding a qubit in an oscillator},}\ }\href@noop {} {\bibfield  {journal} {\bibinfo  {journal} {Phys. Rev. A}\ }\textbf {\bibinfo {volume} {64}},\ \bibinfo {pages} {012310} (\bibinfo {year} {2001})}\BibitemShut {NoStop}%
\bibitem [{\citenamefont {Grimsmo}\ and\ \citenamefont {Puri}(2021)}]{grimsmo2021quantum}%
  \BibitemOpen
  \bibfield  {author} {\bibinfo {author} {\bibfnamefont {Arne~L}\ \bibnamefont {Grimsmo}}\ and\ \bibinfo {author} {\bibfnamefont {Shruti}\ \bibnamefont {Puri}},\ }\bibfield  {title} {\enquote {\bibinfo {title} {Quantum error correction with the {Gottesman-Kitaev-Preskill} code},}\ }\href@noop {} {\bibfield  {journal} {\bibinfo  {journal} {PRX Quantum}\ }\textbf {\bibinfo {volume} {2}},\ \bibinfo {pages} {020101} (\bibinfo {year} {2021})}\BibitemShut {NoStop}%
\bibitem [{\citenamefont {Fukui}\ \emph {et~al.}(2018)\citenamefont {Fukui}, \citenamefont {Tomita}, \citenamefont {Okamoto},\ and\ \citenamefont {Fujii}}]{fukui2018high}%
  \BibitemOpen
  \bibfield  {author} {\bibinfo {author} {\bibfnamefont {Kosuke}\ \bibnamefont {Fukui}}, \bibinfo {author} {\bibfnamefont {Akihisa}\ \bibnamefont {Tomita}}, \bibinfo {author} {\bibfnamefont {Atsushi}\ \bibnamefont {Okamoto}}, \ and\ \bibinfo {author} {\bibfnamefont {Keisuke}\ \bibnamefont {Fujii}},\ }\bibfield  {title} {\enquote {\bibinfo {title} {High-threshold fault-tolerant quantum computation with analog quantum error correction},}\ }\href@noop {} {\bibfield  {journal} {\bibinfo  {journal} {Physical review X}\ }\textbf {\bibinfo {volume} {8}},\ \bibinfo {pages} {021054} (\bibinfo {year} {2018})}\BibitemShut {NoStop}%
\bibitem [{\citenamefont {Vuillot}\ \emph {et~al.}(2019)\citenamefont {Vuillot}, \citenamefont {Asasi}, \citenamefont {Wang}, \citenamefont {Pryadko},\ and\ \citenamefont {Terhal}}]{vuillot2019quantum}%
  \BibitemOpen
  \bibfield  {author} {\bibinfo {author} {\bibfnamefont {Christophe}\ \bibnamefont {Vuillot}}, \bibinfo {author} {\bibfnamefont {Hamed}\ \bibnamefont {Asasi}}, \bibinfo {author} {\bibfnamefont {Yang}\ \bibnamefont {Wang}}, \bibinfo {author} {\bibfnamefont {Leonid~P}\ \bibnamefont {Pryadko}}, \ and\ \bibinfo {author} {\bibfnamefont {Barbara~M}\ \bibnamefont {Terhal}},\ }\bibfield  {title} {\enquote {\bibinfo {title} {Quantum error correction with the toric {Gottesman-Kitaev-Preskill} code},}\ }\href@noop {} {\bibfield  {journal} {\bibinfo  {journal} {Physical Review A}\ }\textbf {\bibinfo {volume} {99}},\ \bibinfo {pages} {032344} (\bibinfo {year} {2019})}\BibitemShut {NoStop}%
\bibitem [{\citenamefont {Terhal}\ \emph {et~al.}(2020)\citenamefont {Terhal}, \citenamefont {Conrad},\ and\ \citenamefont {Vuillot}}]{terhal2020towards}%
  \BibitemOpen
  \bibfield  {author} {\bibinfo {author} {\bibfnamefont {Barbara~M}\ \bibnamefont {Terhal}}, \bibinfo {author} {\bibfnamefont {Jonathan}\ \bibnamefont {Conrad}}, \ and\ \bibinfo {author} {\bibfnamefont {Christophe}\ \bibnamefont {Vuillot}},\ }\bibfield  {title} {\enquote {\bibinfo {title} {Towards scalable bosonic quantum error correction},}\ }\href@noop {} {\bibfield  {journal} {\bibinfo  {journal} {Quantum Science and Technology}\ }\textbf {\bibinfo {volume} {5}},\ \bibinfo {pages} {043001} (\bibinfo {year} {2020})}\BibitemShut {NoStop}%
\bibitem [{\citenamefont {Noh}\ and\ \citenamefont {Chamberland}(2020)}]{noh2020fault}%
  \BibitemOpen
  \bibfield  {author} {\bibinfo {author} {\bibfnamefont {Kyungjoo}\ \bibnamefont {Noh}}\ and\ \bibinfo {author} {\bibfnamefont {Christopher}\ \bibnamefont {Chamberland}},\ }\bibfield  {title} {\enquote {\bibinfo {title} {Fault-tolerant bosonic quantum error correction with the surface--{Gottesman-Kitaev-Preskill} code},}\ }\href@noop {} {\bibfield  {journal} {\bibinfo  {journal} {Physical Review A}\ }\textbf {\bibinfo {volume} {101}},\ \bibinfo {pages} {012316} (\bibinfo {year} {2020})}\BibitemShut {NoStop}%
\bibitem [{\citenamefont {Noh}\ \emph {et~al.}(2022)\citenamefont {Noh}, \citenamefont {Chamberland},\ and\ \citenamefont {Brand{\~a}o}}]{noh2022low}%
  \BibitemOpen
  \bibfield  {author} {\bibinfo {author} {\bibfnamefont {Kyungjoo}\ \bibnamefont {Noh}}, \bibinfo {author} {\bibfnamefont {Christopher}\ \bibnamefont {Chamberland}}, \ and\ \bibinfo {author} {\bibfnamefont {Fernando~GSL}\ \bibnamefont {Brand{\~a}o}},\ }\bibfield  {title} {\enquote {\bibinfo {title} {Low-overhead fault-tolerant quantum error correction with the surface-gkp code},}\ }\href@noop {} {\bibfield  {journal} {\bibinfo  {journal} {PRX Quantum}\ }\textbf {\bibinfo {volume} {3}},\ \bibinfo {pages} {010315} (\bibinfo {year} {2022})}\BibitemShut {NoStop}%
\bibitem [{\citenamefont {Campagne-Ibarcq}\ \emph {et~al.}(2020)\citenamefont {Campagne-Ibarcq}, \citenamefont {Eickbusch}, \citenamefont {Touzard}, \citenamefont {Zalys-Geller}, \citenamefont {Frattini}, \citenamefont {Sivak}, \citenamefont {Reinhold}, \citenamefont {Puri}, \citenamefont {Shankar}, \citenamefont {Schoelkopf} \emph {et~al.}}]{campagne2020quantum}%
  \BibitemOpen
  \bibfield  {author} {\bibinfo {author} {\bibfnamefont {Philippe}\ \bibnamefont {Campagne-Ibarcq}}, \bibinfo {author} {\bibfnamefont {Alec}\ \bibnamefont {Eickbusch}}, \bibinfo {author} {\bibfnamefont {Steven}\ \bibnamefont {Touzard}}, \bibinfo {author} {\bibfnamefont {Evan}\ \bibnamefont {Zalys-Geller}}, \bibinfo {author} {\bibfnamefont {Nicholas~E}\ \bibnamefont {Frattini}}, \bibinfo {author} {\bibfnamefont {Volodymyr~V}\ \bibnamefont {Sivak}}, \bibinfo {author} {\bibfnamefont {Philip}\ \bibnamefont {Reinhold}}, \bibinfo {author} {\bibfnamefont {Shruti}\ \bibnamefont {Puri}}, \bibinfo {author} {\bibfnamefont {Shyam}\ \bibnamefont {Shankar}}, \bibinfo {author} {\bibfnamefont {Robert~J}\ \bibnamefont {Schoelkopf}},  \emph {et~al.},\ }\bibfield  {title} {\enquote {\bibinfo {title} {Quantum error correction of a qubit encoded in grid states of an oscillator},}\ }\href@noop {} {\bibfield  {journal} {\bibinfo  {journal} {Nature}\ }\textbf {\bibinfo {volume} {584}},\ \bibinfo {pages} {368--372} (\bibinfo {year}
  {2020})}\BibitemShut {NoStop}%
\bibitem [{\citenamefont {Sivak}\ \emph {et~al.}(2023)\citenamefont {Sivak}, \citenamefont {Eickbusch}, \citenamefont {Royer}, \citenamefont {Singh}, \citenamefont {Tsioutsios}, \citenamefont {Ganjam}, \citenamefont {Miano}, \citenamefont {Brock}, \citenamefont {Ding}, \citenamefont {Frunzio} \emph {et~al.}}]{sivak2023real}%
  \BibitemOpen
  \bibfield  {author} {\bibinfo {author} {\bibfnamefont {VV}~\bibnamefont {Sivak}}, \bibinfo {author} {\bibfnamefont {Alec}\ \bibnamefont {Eickbusch}}, \bibinfo {author} {\bibfnamefont {Baptiste}\ \bibnamefont {Royer}}, \bibinfo {author} {\bibfnamefont {Shraddha}\ \bibnamefont {Singh}}, \bibinfo {author} {\bibfnamefont {Ioannis}\ \bibnamefont {Tsioutsios}}, \bibinfo {author} {\bibfnamefont {Suhas}\ \bibnamefont {Ganjam}}, \bibinfo {author} {\bibfnamefont {Alessandro}\ \bibnamefont {Miano}}, \bibinfo {author} {\bibfnamefont {BL}~\bibnamefont {Brock}}, \bibinfo {author} {\bibfnamefont {AZ}~\bibnamefont {Ding}}, \bibinfo {author} {\bibfnamefont {Luigi}\ \bibnamefont {Frunzio}},  \emph {et~al.},\ }\bibfield  {title} {\enquote {\bibinfo {title} {Real-time quantum error correction beyond break-even},}\ }\href@noop {} {\bibfield  {journal} {\bibinfo  {journal} {Nature}\ }\textbf {\bibinfo {volume} {616}},\ \bibinfo {pages} {50--55} (\bibinfo {year} {2023})}\BibitemShut {NoStop}%
\bibitem [{Note1()}]{Note1}%
  \BibitemOpen
  \bibinfo {note} {Operators \( \protect \hat {q}_a \) and \( \protect \hat {p}_a \) with equal fluctuations and verifying \( [\protect \hat {q}_a, \protect \hat {p}_a] = i \)}\BibitemShut {NoStop}%
\bibitem [{\citenamefont {Royer}\ \emph {et~al.}(2020)\citenamefont {Royer}, \citenamefont {Singh},\ and\ \citenamefont {Girvin}}]{royer2020stabilization}%
  \BibitemOpen
  \bibfield  {author} {\bibinfo {author} {\bibfnamefont {Baptiste}\ \bibnamefont {Royer}}, \bibinfo {author} {\bibfnamefont {Shraddha}\ \bibnamefont {Singh}}, \ and\ \bibinfo {author} {\bibfnamefont {SM}~\bibnamefont {Girvin}},\ }\bibfield  {title} {\enquote {\bibinfo {title} {Stabilization of finite-energy {Gottesman-Kitaev-Preskill} states},}\ }\href@noop {} {\bibfield  {journal} {\bibinfo  {journal} {Physical Review Letters}\ }\textbf {\bibinfo {volume} {125}},\ \bibinfo {pages} {260509} (\bibinfo {year} {2020})}\BibitemShut {NoStop}%
\bibitem [{\citenamefont {de~Neeve}\ \emph {et~al.}(2022)\citenamefont {de~Neeve}, \citenamefont {Nguyen}, \citenamefont {Behrle},\ and\ \citenamefont {Home}}]{de2022error}%
  \BibitemOpen
  \bibfield  {author} {\bibinfo {author} {\bibfnamefont {Brennan}\ \bibnamefont {de~Neeve}}, \bibinfo {author} {\bibfnamefont {Thanh-Long}\ \bibnamefont {Nguyen}}, \bibinfo {author} {\bibfnamefont {Tanja}\ \bibnamefont {Behrle}}, \ and\ \bibinfo {author} {\bibfnamefont {Jonathan~P}\ \bibnamefont {Home}},\ }\bibfield  {title} {\enquote {\bibinfo {title} {Error correction of a logical grid state qubit by dissipative pumping},}\ }\href@noop {} {\bibfield  {journal} {\bibinfo  {journal} {Nature Physics}\ }\textbf {\bibinfo {volume} {18}},\ \bibinfo {pages} {296--300} (\bibinfo {year} {2022})}\BibitemShut {NoStop}%
\bibitem [{\citenamefont {Kitaev}(1995)}]{kitaev1995quantum}%
  \BibitemOpen
  \bibfield  {author} {\bibinfo {author} {\bibfnamefont {A~Yu}\ \bibnamefont {Kitaev}},\ }\bibfield  {title} {\enquote {\bibinfo {title} {Quantum measurements and the abelian stabilizer problem},}\ }\href@noop {} {\bibfield  {journal} {\bibinfo  {journal} {arXiv preprint quant-ph/9511026}\ } (\bibinfo {year} {1995})}\BibitemShut {NoStop}%
\bibitem [{\citenamefont {Travaglione}\ and\ \citenamefont {Milburn}(2002)}]{travaglione2002preparing}%
  \BibitemOpen
  \bibfield  {author} {\bibinfo {author} {\bibfnamefont {BC}~\bibnamefont {Travaglione}}\ and\ \bibinfo {author} {\bibfnamefont {Gerard~J}\ \bibnamefont {Milburn}},\ }\bibfield  {title} {\enquote {\bibinfo {title} {Preparing encoded states in an oscillator},}\ }\href@noop {} {\bibfield  {journal} {\bibinfo  {journal} {Phys. Rev. A}\ }\textbf {\bibinfo {volume} {66}},\ \bibinfo {pages} {052322} (\bibinfo {year} {2002})}\BibitemShut {NoStop}%
\bibitem [{\citenamefont {Pirandola}\ \emph {et~al.}(2006)\citenamefont {Pirandola}, \citenamefont {Mancini}, \citenamefont {Vitali},\ and\ \citenamefont {Tombesi}}]{Pirandola2006}%
  \BibitemOpen
  \bibfield  {author} {\bibinfo {author} {\bibfnamefont {S.}~\bibnamefont {Pirandola}}, \bibinfo {author} {\bibfnamefont {S.}~\bibnamefont {Mancini}}, \bibinfo {author} {\bibfnamefont {D.}~\bibnamefont {Vitali}}, \ and\ \bibinfo {author} {\bibfnamefont {P.}~\bibnamefont {Tombesi}},\ }\bibfield  {title} {\enquote {\bibinfo {title} {{Continuous variable encoding by ponderomotive interaction}},}\ }\href {\doibase 10.1140/epjd/e2005-00306-3} {\bibfield  {journal} {\bibinfo  {journal} {Eur. Phys. J. D}\ }\textbf {\bibinfo {volume} {37}},\ \bibinfo {pages} {283--290} (\bibinfo {year} {2006})}\BibitemShut {NoStop}%
\bibitem [{\citenamefont {Svore}\ \emph {et~al.}(2013)\citenamefont {Svore}, \citenamefont {Hastings},\ and\ \citenamefont {Freedman}}]{svore2013faster}%
  \BibitemOpen
  \bibfield  {author} {\bibinfo {author} {\bibfnamefont {Krysta~M}\ \bibnamefont {Svore}}, \bibinfo {author} {\bibfnamefont {Matthew~B}\ \bibnamefont {Hastings}}, \ and\ \bibinfo {author} {\bibfnamefont {Michael}\ \bibnamefont {Freedman}},\ }\bibfield  {title} {\enquote {\bibinfo {title} {Faster phase estimation},}\ }\href@noop {} {\bibfield  {journal} {\bibinfo  {journal} {arXiv preprint arXiv:1304.0741}\ } (\bibinfo {year} {2013})}\BibitemShut {NoStop}%
\bibitem [{\citenamefont {Terhal}\ and\ \citenamefont {Weigand}(2016)}]{terhal2016encoding}%
  \BibitemOpen
  \bibfield  {author} {\bibinfo {author} {\bibfnamefont {BM}~\bibnamefont {Terhal}}\ and\ \bibinfo {author} {\bibfnamefont {Daniel}\ \bibnamefont {Weigand}},\ }\bibfield  {title} {\enquote {\bibinfo {title} {Encoding a qubit into a cavity mode in circuit {QED} using phase estimation},}\ }\href@noop {} {\bibfield  {journal} {\bibinfo  {journal} {Phys. Rev. A}\ }\textbf {\bibinfo {volume} {93}},\ \bibinfo {pages} {012315} (\bibinfo {year} {2016})}\BibitemShut {NoStop}%
\bibitem [{\citenamefont {Motes}\ \emph {et~al.}(2017)\citenamefont {Motes}, \citenamefont {Baragiola}, \citenamefont {Gilchrist},\ and\ \citenamefont {Menicucci}}]{motes2017encoding}%
  \BibitemOpen
  \bibfield  {author} {\bibinfo {author} {\bibfnamefont {Keith~R}\ \bibnamefont {Motes}}, \bibinfo {author} {\bibfnamefont {Ben~Q}\ \bibnamefont {Baragiola}}, \bibinfo {author} {\bibfnamefont {Alexei}\ \bibnamefont {Gilchrist}}, \ and\ \bibinfo {author} {\bibfnamefont {Nicolas~C}\ \bibnamefont {Menicucci}},\ }\bibfield  {title} {\enquote {\bibinfo {title} {Encoding qubits into oscillators with atomic ensembles and squeezed light},}\ }\href@noop {} {\bibfield  {journal} {\bibinfo  {journal} {Phys. Rev. A}\ }\textbf {\bibinfo {volume} {95}},\ \bibinfo {pages} {053819} (\bibinfo {year} {2017})}\BibitemShut {NoStop}%
\bibitem [{\citenamefont {Weigand}\ and\ \citenamefont {Terhal}(2020)}]{weigand2020realizing}%
  \BibitemOpen
  \bibfield  {author} {\bibinfo {author} {\bibfnamefont {Daniel~J}\ \bibnamefont {Weigand}}\ and\ \bibinfo {author} {\bibfnamefont {Barbara~M}\ \bibnamefont {Terhal}},\ }\bibfield  {title} {\enquote {\bibinfo {title} {Realizing modular quadrature measurements via a tunable photon-pressure coupling in circuit {QED}},}\ }\href@noop {} {\bibfield  {journal} {\bibinfo  {journal} {Physical Review A}\ }\textbf {\bibinfo {volume} {101}},\ \bibinfo {pages} {053840} (\bibinfo {year} {2020})}\BibitemShut {NoStop}%
\bibitem [{\citenamefont {Fl{\"u}hmann}\ \emph {et~al.}(2018)\citenamefont {Fl{\"u}hmann}, \citenamefont {Negnevitsky}, \citenamefont {Marinelli},\ and\ \citenamefont {Home}}]{fluhmann2018sequential}%
  \BibitemOpen
  \bibfield  {author} {\bibinfo {author} {\bibfnamefont {Christa}\ \bibnamefont {Fl{\"u}hmann}}, \bibinfo {author} {\bibfnamefont {Vlad}\ \bibnamefont {Negnevitsky}}, \bibinfo {author} {\bibfnamefont {Matteo}\ \bibnamefont {Marinelli}}, \ and\ \bibinfo {author} {\bibfnamefont {Jonathan~P}\ \bibnamefont {Home}},\ }\bibfield  {title} {\enquote {\bibinfo {title} {Sequential modular position and momentum measurements of a trapped ion mechanical oscillator},}\ }\href@noop {} {\bibfield  {journal} {\bibinfo  {journal} {Phys. Rev. X}\ }\textbf {\bibinfo {volume} {8}},\ \bibinfo {pages} {021001} (\bibinfo {year} {2018})}\BibitemShut {NoStop}%
\bibitem [{\citenamefont {Fl{\"u}hmann}\ \emph {et~al.}(2019)\citenamefont {Fl{\"u}hmann}, \citenamefont {Nguyen}, \citenamefont {Marinelli}, \citenamefont {Negnevitsky}, \citenamefont {Mehta},\ and\ \citenamefont {Home}}]{fluhmann2018encoding}%
  \BibitemOpen
  \bibfield  {author} {\bibinfo {author} {\bibfnamefont {Christa}\ \bibnamefont {Fl{\"u}hmann}}, \bibinfo {author} {\bibfnamefont {Thanh~Long}\ \bibnamefont {Nguyen}}, \bibinfo {author} {\bibfnamefont {Matteo}\ \bibnamefont {Marinelli}}, \bibinfo {author} {\bibfnamefont {Vlad}\ \bibnamefont {Negnevitsky}}, \bibinfo {author} {\bibfnamefont {Karan}\ \bibnamefont {Mehta}}, \ and\ \bibinfo {author} {\bibfnamefont {JP}~\bibnamefont {Home}},\ }\bibfield  {title} {\enquote {\bibinfo {title} {Encoding a qubit in a trapped-ion mechanical oscillator},}\ }\href@noop {} {\bibfield  {journal} {\bibinfo  {journal} {Nature}\ }\textbf {\bibinfo {volume} {566}},\ \bibinfo {pages} {513} (\bibinfo {year} {2019})}\BibitemShut {NoStop}%
\bibitem [{\citenamefont {Kapit}(2018)}]{kapit2018error}%
  \BibitemOpen
  \bibfield  {author} {\bibinfo {author} {\bibfnamefont {Eliot}\ \bibnamefont {Kapit}},\ }\bibfield  {title} {\enquote {\bibinfo {title} {Error-transparent quantum gates for small logical qubit architectures},}\ }\href@noop {} {\bibfield  {journal} {\bibinfo  {journal} {Physical review letters}\ }\textbf {\bibinfo {volume} {120}},\ \bibinfo {pages} {050503} (\bibinfo {year} {2018})}\BibitemShut {NoStop}%
\bibitem [{\citenamefont {Puri}\ \emph {et~al.}(2019)\citenamefont {Puri}, \citenamefont {Grimm}, \citenamefont {Campagne-Ibarcq}, \citenamefont {Eickbusch}, \citenamefont {Noh}, \citenamefont {Roberts}, \citenamefont {Jiang}, \citenamefont {Mirrahimi}, \citenamefont {Devoret},\ and\ \citenamefont {Girvin}}]{puri2019stabilized}%
  \BibitemOpen
  \bibfield  {author} {\bibinfo {author} {\bibfnamefont {Shruti}\ \bibnamefont {Puri}}, \bibinfo {author} {\bibfnamefont {Alexander}\ \bibnamefont {Grimm}}, \bibinfo {author} {\bibfnamefont {Philippe}\ \bibnamefont {Campagne-Ibarcq}}, \bibinfo {author} {\bibfnamefont {Alec}\ \bibnamefont {Eickbusch}}, \bibinfo {author} {\bibfnamefont {Kyungjoo}\ \bibnamefont {Noh}}, \bibinfo {author} {\bibfnamefont {Gabrielle}\ \bibnamefont {Roberts}}, \bibinfo {author} {\bibfnamefont {Liang}\ \bibnamefont {Jiang}}, \bibinfo {author} {\bibfnamefont {Mazyar}\ \bibnamefont {Mirrahimi}}, \bibinfo {author} {\bibfnamefont {Michel~H}\ \bibnamefont {Devoret}}, \ and\ \bibinfo {author} {\bibfnamefont {Steven~M}\ \bibnamefont {Girvin}},\ }\bibfield  {title} {\enquote {\bibinfo {title} {Stabilized cat in a driven nonlinear cavity: a fault-tolerant error syndrome detector},}\ }\href@noop {} {\bibfield  {journal} {\bibinfo  {journal} {Physical Review X}\ }\textbf {\bibinfo {volume} {9}},\ \bibinfo {pages} {041009} (\bibinfo {year}
  {2019})}\BibitemShut {NoStop}%
\bibitem [{\citenamefont {Shi}\ \emph {et~al.}(2019)\citenamefont {Shi}, \citenamefont {Chamberland},\ and\ \citenamefont {Cross}}]{shi2019fault}%
  \BibitemOpen
  \bibfield  {author} {\bibinfo {author} {\bibfnamefont {Yunong}\ \bibnamefont {Shi}}, \bibinfo {author} {\bibfnamefont {Christopher}\ \bibnamefont {Chamberland}}, \ and\ \bibinfo {author} {\bibfnamefont {Andrew}\ \bibnamefont {Cross}},\ }\bibfield  {title} {\enquote {\bibinfo {title} {Fault-tolerant preparation of approximate gkp states},}\ }\href@noop {} {\bibfield  {journal} {\bibinfo  {journal} {New Journal of Physics}\ }\textbf {\bibinfo {volume} {21}},\ \bibinfo {pages} {093007} (\bibinfo {year} {2019})}\BibitemShut {NoStop}%
\bibitem [{\citenamefont {Ma}\ \emph {et~al.}(2020)\citenamefont {Ma}, \citenamefont {Zhang}, \citenamefont {Wong}, \citenamefont {Noh}, \citenamefont {Rosenblum}, \citenamefont {Reinhold}, \citenamefont {Schoelkopf},\ and\ \citenamefont {Jiang}}]{ma2020path}%
  \BibitemOpen
  \bibfield  {author} {\bibinfo {author} {\bibfnamefont {Wen-Long}\ \bibnamefont {Ma}}, \bibinfo {author} {\bibfnamefont {Mengzhen}\ \bibnamefont {Zhang}}, \bibinfo {author} {\bibfnamefont {Yat}\ \bibnamefont {Wong}}, \bibinfo {author} {\bibfnamefont {Kyungjoo}\ \bibnamefont {Noh}}, \bibinfo {author} {\bibfnamefont {Serge}\ \bibnamefont {Rosenblum}}, \bibinfo {author} {\bibfnamefont {Philip}\ \bibnamefont {Reinhold}}, \bibinfo {author} {\bibfnamefont {Robert~J}\ \bibnamefont {Schoelkopf}}, \ and\ \bibinfo {author} {\bibfnamefont {Liang}\ \bibnamefont {Jiang}},\ }\bibfield  {title} {\enquote {\bibinfo {title} {Path-independent quantum gates with noisy ancilla},}\ }\href@noop {} {\bibfield  {journal} {\bibinfo  {journal} {Physical Review Letters}\ }\textbf {\bibinfo {volume} {125}},\ \bibinfo {pages} {110503} (\bibinfo {year} {2020})}\BibitemShut {NoStop}%
\bibitem [{\citenamefont {Vy}\ \emph {et~al.}(2013)\citenamefont {Vy}, \citenamefont {Wang},\ and\ \citenamefont {Jacobs}}]{vy2013error}%
  \BibitemOpen
  \bibfield  {author} {\bibinfo {author} {\bibfnamefont {Os}~\bibnamefont {Vy}}, \bibinfo {author} {\bibfnamefont {Xiaoting}\ \bibnamefont {Wang}}, \ and\ \bibinfo {author} {\bibfnamefont {Kurt}\ \bibnamefont {Jacobs}},\ }\bibfield  {title} {\enquote {\bibinfo {title} {Error-transparent evolution: the ability of multi-body interactions to bypass decoherence},}\ }\href@noop {} {\bibfield  {journal} {\bibinfo  {journal} {New Journal of Physics}\ }\textbf {\bibinfo {volume} {15}},\ \bibinfo {pages} {053002} (\bibinfo {year} {2013})}\BibitemShut {NoStop}%
\bibitem [{\citenamefont {Rosenblum}\ \emph {et~al.}(2018)\citenamefont {Rosenblum}, \citenamefont {Reinhold}, \citenamefont {Mirrahimi}, \citenamefont {Jiang}, \citenamefont {Frunzio},\ and\ \citenamefont {Schoelkopf}}]{rosenblum2018fault}%
  \BibitemOpen
  \bibfield  {author} {\bibinfo {author} {\bibfnamefont {Serge}\ \bibnamefont {Rosenblum}}, \bibinfo {author} {\bibfnamefont {Philip}\ \bibnamefont {Reinhold}}, \bibinfo {author} {\bibfnamefont {Mazyar}\ \bibnamefont {Mirrahimi}}, \bibinfo {author} {\bibfnamefont {Liang}\ \bibnamefont {Jiang}}, \bibinfo {author} {\bibfnamefont {Luigi}\ \bibnamefont {Frunzio}}, \ and\ \bibinfo {author} {\bibfnamefont {Robert~J}\ \bibnamefont {Schoelkopf}},\ }\bibfield  {title} {\enquote {\bibinfo {title} {Fault-tolerant detection of a quantum error},}\ }\href@noop {} {\bibfield  {journal} {\bibinfo  {journal} {Science}\ }\textbf {\bibinfo {volume} {361}},\ \bibinfo {pages} {266--270} (\bibinfo {year} {2018})}\BibitemShut {NoStop}%
\bibitem [{\citenamefont {Reinhold}\ \emph {et~al.}(2020)\citenamefont {Reinhold}, \citenamefont {Rosenblum}, \citenamefont {Ma}, \citenamefont {Frunzio}, \citenamefont {Jiang},\ and\ \citenamefont {Schoelkopf}}]{reinhold2020error}%
  \BibitemOpen
  \bibfield  {author} {\bibinfo {author} {\bibfnamefont {Philip}\ \bibnamefont {Reinhold}}, \bibinfo {author} {\bibfnamefont {Serge}\ \bibnamefont {Rosenblum}}, \bibinfo {author} {\bibfnamefont {Wen-Long}\ \bibnamefont {Ma}}, \bibinfo {author} {\bibfnamefont {Luigi}\ \bibnamefont {Frunzio}}, \bibinfo {author} {\bibfnamefont {Liang}\ \bibnamefont {Jiang}}, \ and\ \bibinfo {author} {\bibfnamefont {Robert~J}\ \bibnamefont {Schoelkopf}},\ }\bibfield  {title} {\enquote {\bibinfo {title} {Error-corrected gates on an encoded qubit},}\ }\href@noop {} {\bibfield  {journal} {\bibinfo  {journal} {Nature Physics}\ }\textbf {\bibinfo {volume} {16}},\ \bibinfo {pages} {822--826} (\bibinfo {year} {2020})}\BibitemShut {NoStop}%
\bibitem [{\citenamefont {Grimm}\ \emph {et~al.}(2020)\citenamefont {Grimm}, \citenamefont {Frattini}, \citenamefont {Puri}, \citenamefont {Mundhada}, \citenamefont {Touzard}, \citenamefont {Mirrahimi}, \citenamefont {Girvin}, \citenamefont {Shankar},\ and\ \citenamefont {Devoret}}]{grimm2020stabilization}%
  \BibitemOpen
  \bibfield  {author} {\bibinfo {author} {\bibfnamefont {Alexander}\ \bibnamefont {Grimm}}, \bibinfo {author} {\bibfnamefont {Nicholas~E}\ \bibnamefont {Frattini}}, \bibinfo {author} {\bibfnamefont {Shruti}\ \bibnamefont {Puri}}, \bibinfo {author} {\bibfnamefont {Shantanu~O}\ \bibnamefont {Mundhada}}, \bibinfo {author} {\bibfnamefont {Steven}\ \bibnamefont {Touzard}}, \bibinfo {author} {\bibfnamefont {Mazyar}\ \bibnamefont {Mirrahimi}}, \bibinfo {author} {\bibfnamefont {Steven~M}\ \bibnamefont {Girvin}}, \bibinfo {author} {\bibfnamefont {Shyam}\ \bibnamefont {Shankar}}, \ and\ \bibinfo {author} {\bibfnamefont {Michel~H}\ \bibnamefont {Devoret}},\ }\bibfield  {title} {\enquote {\bibinfo {title} {Stabilization and operation of a {Kerr}-cat qubit},}\ }\href@noop {} {\bibfield  {journal} {\bibinfo  {journal} {Nature}\ }\textbf {\bibinfo {volume} {584}},\ \bibinfo {pages} {205--209} (\bibinfo {year} {2020})}\BibitemShut {NoStop}%
\bibitem [{\citenamefont {Frattini}\ \emph {et~al.}(2022)\citenamefont {Frattini}, \citenamefont {Corti{\~n}as}, \citenamefont {Venkatraman}, \citenamefont {Xiao}, \citenamefont {Su}, \citenamefont {Lei}, \citenamefont {Chapman}, \citenamefont {Joshi}, \citenamefont {Girvin}, \citenamefont {Schoelkopf} \emph {et~al.}}]{frattini2022squeezed}%
  \BibitemOpen
  \bibfield  {author} {\bibinfo {author} {\bibfnamefont {Nicholas~E}\ \bibnamefont {Frattini}}, \bibinfo {author} {\bibfnamefont {Rodrigo~G}\ \bibnamefont {Corti{\~n}as}}, \bibinfo {author} {\bibfnamefont {Jayameenakshi}\ \bibnamefont {Venkatraman}}, \bibinfo {author} {\bibfnamefont {Xu}~\bibnamefont {Xiao}}, \bibinfo {author} {\bibfnamefont {Qile}\ \bibnamefont {Su}}, \bibinfo {author} {\bibfnamefont {Chan~U}\ \bibnamefont {Lei}}, \bibinfo {author} {\bibfnamefont {Benjamin~J}\ \bibnamefont {Chapman}}, \bibinfo {author} {\bibfnamefont {Vidul~R}\ \bibnamefont {Joshi}}, \bibinfo {author} {\bibfnamefont {SM}~\bibnamefont {Girvin}}, \bibinfo {author} {\bibfnamefont {Robert~J}\ \bibnamefont {Schoelkopf}},  \emph {et~al.},\ }\bibfield  {title} {\enquote {\bibinfo {title} {The squeezed {Kerr} oscillator: spectral kissing and phase-flip robustness},}\ }\href@noop {} {\bibfield  {journal} {\bibinfo  {journal} {arXiv preprint arXiv:2209.03934}\ } (\bibinfo {year} {2022})}\BibitemShut {NoStop}%
\bibitem [{\citenamefont {Glancy}\ and\ \citenamefont {Knill}(2006)}]{glancy2006error}%
  \BibitemOpen
  \bibfield  {author} {\bibinfo {author} {\bibfnamefont {Scott}\ \bibnamefont {Glancy}}\ and\ \bibinfo {author} {\bibfnamefont {Emanuel}\ \bibnamefont {Knill}},\ }\bibfield  {title} {\enquote {\bibinfo {title} {Error analysis for encoding a qubit in an oscillator},}\ }\href@noop {} {\bibfield  {journal} {\bibinfo  {journal} {Physical Review A}\ }\textbf {\bibinfo {volume} {73}},\ \bibinfo {pages} {012325} (\bibinfo {year} {2006})}\BibitemShut {NoStop}%
\bibitem [{\citenamefont {Duivenvoorden}\ \emph {et~al.}(2017)\citenamefont {Duivenvoorden}, \citenamefont {Terhal},\ and\ \citenamefont {Weigand}}]{duivenvoorden2017single}%
  \BibitemOpen
  \bibfield  {author} {\bibinfo {author} {\bibfnamefont {Kasper}\ \bibnamefont {Duivenvoorden}}, \bibinfo {author} {\bibfnamefont {Barbara~M}\ \bibnamefont {Terhal}}, \ and\ \bibinfo {author} {\bibfnamefont {Daniel}\ \bibnamefont {Weigand}},\ }\bibfield  {title} {\enquote {\bibinfo {title} {Single-mode displacement sensor},}\ }\href@noop {} {\bibfield  {journal} {\bibinfo  {journal} {Physical Review A}\ }\textbf {\bibinfo {volume} {95}},\ \bibinfo {pages} {012305} (\bibinfo {year} {2017})}\BibitemShut {NoStop}%
\bibitem [{Note2()}]{Note2}%
  \BibitemOpen
  \bibinfo {note} {With the definition $\protect \tilde {{\protect \bf p}}^{\perp L}_a=-\protect \tilde {{\protect \bf q}}^L_a$}\BibitemShut {NoStop}%
\bibitem [{\citenamefont {Zak}(1967)}]{zak1967finite}%
  \BibitemOpen
  \bibfield  {author} {\bibinfo {author} {\bibfnamefont {Joshua}\ \bibnamefont {Zak}},\ }\bibfield  {title} {\enquote {\bibinfo {title} {Finite translations in solid-state physics},}\ }\href@noop {} {\bibfield  {journal} {\bibinfo  {journal} {Physical Review Letters}\ }\textbf {\bibinfo {volume} {19}},\ \bibinfo {pages} {1385} (\bibinfo {year} {1967})}\BibitemShut {NoStop}%
\bibitem [{\citenamefont {Royer}\ \emph {et~al.}(2022)\citenamefont {Royer}, \citenamefont {Singh},\ and\ \citenamefont {Girvin}}]{royer2022encoding}%
  \BibitemOpen
  \bibfield  {author} {\bibinfo {author} {\bibfnamefont {Baptiste}\ \bibnamefont {Royer}}, \bibinfo {author} {\bibfnamefont {Shraddha}\ \bibnamefont {Singh}}, \ and\ \bibinfo {author} {\bibfnamefont {Steven~M}\ \bibnamefont {Girvin}},\ }\bibfield  {title} {\enquote {\bibinfo {title} {Encoding qubits in multimode grid states},}\ }\href@noop {} {\bibfield  {journal} {\bibinfo  {journal} {PRX Quantum}\ }\textbf {\bibinfo {volume} {3}},\ \bibinfo {pages} {010335} (\bibinfo {year} {2022})}\BibitemShut {NoStop}%
\bibitem [{\citenamefont {Milul}\ \emph {et~al.}(2023)\citenamefont {Milul}, \citenamefont {Guttel}, \citenamefont {Goldblatt}, \citenamefont {Hazanov}, \citenamefont {Joshi}, \citenamefont {Chausovsky}, \citenamefont {Kahn}, \citenamefont {{\c{C}}ifty{\"u}rek}, \citenamefont {Lafont},\ and\ \citenamefont {Rosenblum}}]{milul2023superconducting}%
  \BibitemOpen
  \bibfield  {author} {\bibinfo {author} {\bibfnamefont {Ofir}\ \bibnamefont {Milul}}, \bibinfo {author} {\bibfnamefont {Barkay}\ \bibnamefont {Guttel}}, \bibinfo {author} {\bibfnamefont {Uri}\ \bibnamefont {Goldblatt}}, \bibinfo {author} {\bibfnamefont {Sergey}\ \bibnamefont {Hazanov}}, \bibinfo {author} {\bibfnamefont {Lalit~M}\ \bibnamefont {Joshi}}, \bibinfo {author} {\bibfnamefont {Daniel}\ \bibnamefont {Chausovsky}}, \bibinfo {author} {\bibfnamefont {Nitzan}\ \bibnamefont {Kahn}}, \bibinfo {author} {\bibfnamefont {Engin}\ \bibnamefont {{\c{C}}ifty{\"u}rek}}, \bibinfo {author} {\bibfnamefont {Fabien}\ \bibnamefont {Lafont}}, \ and\ \bibinfo {author} {\bibfnamefont {Serge}\ \bibnamefont {Rosenblum}},\ }\bibfield  {title} {\enquote {\bibinfo {title} {A superconducting quantum memory with tens of milliseconds coherence time},}\ }\href@noop {} {\bibfield  {journal} {\bibinfo  {journal} {arXiv preprint arXiv:2302.06442}\ } (\bibinfo {year} {2023})}\BibitemShut {NoStop}%
\bibitem [{\citenamefont {Place}\ \emph {et~al.}(2021)\citenamefont {Place}, \citenamefont {Rodgers}, \citenamefont {Mundada}, \citenamefont {Smitham}, \citenamefont {Fitzpatrick}, \citenamefont {Leng}, \citenamefont {Premkumar}, \citenamefont {Bryon}, \citenamefont {Vrajitoarea}, \citenamefont {Sussman} \emph {et~al.}}]{place2021new}%
  \BibitemOpen
  \bibfield  {author} {\bibinfo {author} {\bibfnamefont {Alexander~PM}\ \bibnamefont {Place}}, \bibinfo {author} {\bibfnamefont {Lila~VH}\ \bibnamefont {Rodgers}}, \bibinfo {author} {\bibfnamefont {Pranav}\ \bibnamefont {Mundada}}, \bibinfo {author} {\bibfnamefont {Basil~M}\ \bibnamefont {Smitham}}, \bibinfo {author} {\bibfnamefont {Mattias}\ \bibnamefont {Fitzpatrick}}, \bibinfo {author} {\bibfnamefont {Zhaoqi}\ \bibnamefont {Leng}}, \bibinfo {author} {\bibfnamefont {Anjali}\ \bibnamefont {Premkumar}}, \bibinfo {author} {\bibfnamefont {Jacob}\ \bibnamefont {Bryon}}, \bibinfo {author} {\bibfnamefont {Andrei}\ \bibnamefont {Vrajitoarea}}, \bibinfo {author} {\bibfnamefont {Sara}\ \bibnamefont {Sussman}},  \emph {et~al.},\ }\bibfield  {title} {\enquote {\bibinfo {title} {New material platform for superconducting transmon qubits with coherence times exceeding 0.3 milliseconds},}\ }\href@noop {} {\bibfield  {journal} {\bibinfo  {journal} {Nature communications}\ }\textbf {\bibinfo {volume} {12}},\ \bibinfo {pages}
  {1779} (\bibinfo {year} {2021})}\BibitemShut {NoStop}%
\bibitem [{\citenamefont {Wang}\ \emph {et~al.}(2022)\citenamefont {Wang}, \citenamefont {Li}, \citenamefont {Xu}, \citenamefont {Li}, \citenamefont {Wang}, \citenamefont {Yang}, \citenamefont {Mi}, \citenamefont {Liang}, \citenamefont {Su}, \citenamefont {Yang} \emph {et~al.}}]{wang2022towards}%
  \BibitemOpen
  \bibfield  {author} {\bibinfo {author} {\bibfnamefont {Chenlu}\ \bibnamefont {Wang}}, \bibinfo {author} {\bibfnamefont {Xuegang}\ \bibnamefont {Li}}, \bibinfo {author} {\bibfnamefont {Huikai}\ \bibnamefont {Xu}}, \bibinfo {author} {\bibfnamefont {Zhiyuan}\ \bibnamefont {Li}}, \bibinfo {author} {\bibfnamefont {Junhua}\ \bibnamefont {Wang}}, \bibinfo {author} {\bibfnamefont {Zhen}\ \bibnamefont {Yang}}, \bibinfo {author} {\bibfnamefont {Zhenyu}\ \bibnamefont {Mi}}, \bibinfo {author} {\bibfnamefont {Xuehui}\ \bibnamefont {Liang}}, \bibinfo {author} {\bibfnamefont {Tang}\ \bibnamefont {Su}}, \bibinfo {author} {\bibfnamefont {Chuhong}\ \bibnamefont {Yang}},  \emph {et~al.},\ }\bibfield  {title} {\enquote {\bibinfo {title} {Towards practical quantum computers: Transmon qubit with a lifetime approaching 0.5 milliseconds},}\ }\href@noop {} {\bibfield  {journal} {\bibinfo  {journal} {npj Quantum Information}\ }\textbf {\bibinfo {volume} {8}},\ \bibinfo {pages} {3} (\bibinfo {year} {2022})}\BibitemShut {NoStop}%
\bibitem [{\citenamefont {Eickbusch}\ \emph {et~al.}(2022)\citenamefont {Eickbusch}, \citenamefont {Sivak}, \citenamefont {Ding}, \citenamefont {Elder}, \citenamefont {Jha}, \citenamefont {Venkatraman}, \citenamefont {Royer}, \citenamefont {Girvin}, \citenamefont {Schoelkopf},\ and\ \citenamefont {Devoret}}]{eickbusch2022fast}%
  \BibitemOpen
  \bibfield  {author} {\bibinfo {author} {\bibfnamefont {Alec}\ \bibnamefont {Eickbusch}}, \bibinfo {author} {\bibfnamefont {Volodymyr}\ \bibnamefont {Sivak}}, \bibinfo {author} {\bibfnamefont {Andy~Z}\ \bibnamefont {Ding}}, \bibinfo {author} {\bibfnamefont {Salvatore~S}\ \bibnamefont {Elder}}, \bibinfo {author} {\bibfnamefont {Shantanu~R}\ \bibnamefont {Jha}}, \bibinfo {author} {\bibfnamefont {Jayameenakshi}\ \bibnamefont {Venkatraman}}, \bibinfo {author} {\bibfnamefont {Baptiste}\ \bibnamefont {Royer}}, \bibinfo {author} {\bibfnamefont {SM}~\bibnamefont {Girvin}}, \bibinfo {author} {\bibfnamefont {Robert~J}\ \bibnamefont {Schoelkopf}}, \ and\ \bibinfo {author} {\bibfnamefont {Michel~H}\ \bibnamefont {Devoret}},\ }\bibfield  {title} {\enquote {\bibinfo {title} {Fast universal control of an oscillator with weak dispersive coupling to a qubit},}\ }\href@noop {} {\bibfield  {journal} {\bibinfo  {journal} {Nature Physics}\ }\textbf {\bibinfo {volume} {18}},\ \bibinfo {pages} {1464--1469} (\bibinfo {year}
  {2022})}\BibitemShut {NoStop}%
\bibitem [{\citenamefont {Tzitrin}\ \emph {et~al.}(2020)\citenamefont {Tzitrin}, \citenamefont {Bourassa}, \citenamefont {Menicucci},\ and\ \citenamefont {Sabapathy}}]{tzitrin2020progress}%
  \BibitemOpen
  \bibfield  {author} {\bibinfo {author} {\bibfnamefont {Ilan}\ \bibnamefont {Tzitrin}}, \bibinfo {author} {\bibfnamefont {J~Eli}\ \bibnamefont {Bourassa}}, \bibinfo {author} {\bibfnamefont {Nicolas~C}\ \bibnamefont {Menicucci}}, \ and\ \bibinfo {author} {\bibfnamefont {Krishna~Kumar}\ \bibnamefont {Sabapathy}},\ }\bibfield  {title} {\enquote {\bibinfo {title} {Progress towards practical qubit computation using approximate {Gottesman-Kitaev-Preskill} codes},}\ }\href@noop {} {\bibfield  {journal} {\bibinfo  {journal} {Physical Review A}\ }\textbf {\bibinfo {volume} {101}},\ \bibinfo {pages} {032315} (\bibinfo {year} {2020})}\BibitemShut {NoStop}%
\bibitem [{\citenamefont {Zhang}\ \emph {et~al.}(2019)\citenamefont {Zhang}, \citenamefont {Lester}, \citenamefont {Gao}, \citenamefont {Jiang}, \citenamefont {Schoelkopf},\ and\ \citenamefont {Girvin}}]{zhang2019engineering}%
  \BibitemOpen
  \bibfield  {author} {\bibinfo {author} {\bibfnamefont {Yaxing}\ \bibnamefont {Zhang}}, \bibinfo {author} {\bibfnamefont {Brian~J}\ \bibnamefont {Lester}}, \bibinfo {author} {\bibfnamefont {Yvonne~Y}\ \bibnamefont {Gao}}, \bibinfo {author} {\bibfnamefont {Liang}\ \bibnamefont {Jiang}}, \bibinfo {author} {\bibfnamefont {RJ}~\bibnamefont {Schoelkopf}}, \ and\ \bibinfo {author} {\bibfnamefont {SM}~\bibnamefont {Girvin}},\ }\bibfield  {title} {\enquote {\bibinfo {title} {Engineering bilinear mode coupling in circuit {QED}: Theory and experiment},}\ }\href@noop {} {\bibfield  {journal} {\bibinfo  {journal} {Physical Review A}\ }\textbf {\bibinfo {volume} {99}},\ \bibinfo {pages} {012314} (\bibinfo {year} {2019})}\BibitemShut {NoStop}%
\bibitem [{\citenamefont {Eriksson}\ \emph {et~al.}(2023)\citenamefont {Eriksson}, \citenamefont {S{\'e}pulcre}, \citenamefont {Kervinen}, \citenamefont {Hillmann}, \citenamefont {Kudra}, \citenamefont {Dupouy}, \citenamefont {Lu}, \citenamefont {Khanahmadi}, \citenamefont {Yang}, \citenamefont {Moreno} \emph {et~al.}}]{eriksson2023universal}%
  \BibitemOpen
  \bibfield  {author} {\bibinfo {author} {\bibfnamefont {Axel~M}\ \bibnamefont {Eriksson}}, \bibinfo {author} {\bibfnamefont {Th{\'e}o}\ \bibnamefont {S{\'e}pulcre}}, \bibinfo {author} {\bibfnamefont {Mikael}\ \bibnamefont {Kervinen}}, \bibinfo {author} {\bibfnamefont {Timo}\ \bibnamefont {Hillmann}}, \bibinfo {author} {\bibfnamefont {Marina}\ \bibnamefont {Kudra}}, \bibinfo {author} {\bibfnamefont {Simon}\ \bibnamefont {Dupouy}}, \bibinfo {author} {\bibfnamefont {Yong}\ \bibnamefont {Lu}}, \bibinfo {author} {\bibfnamefont {Maryam}\ \bibnamefont {Khanahmadi}}, \bibinfo {author} {\bibfnamefont {Jiaying}\ \bibnamefont {Yang}}, \bibinfo {author} {\bibfnamefont {Claudia~Castillo}\ \bibnamefont {Moreno}},  \emph {et~al.},\ }\bibfield  {title} {\enquote {\bibinfo {title} {Universal control of a bosonic mode via drive-activated native cubic interactions},}\ }\href@noop {} {\bibfield  {journal} {\bibinfo  {journal} {arXiv preprint arXiv:2308.15320}\ } (\bibinfo {year} {2023})}\BibitemShut {NoStop}%
\bibitem [{\citenamefont {Lu}\ \emph {et~al.}(2023)\citenamefont {Lu}, \citenamefont {Maiti}, \citenamefont {Garmon}, \citenamefont {Ganjam}, \citenamefont {Zhang}, \citenamefont {Claes}, \citenamefont {Frunzio}, \citenamefont {Girvin},\ and\ \citenamefont {Schoelkopf}}]{lu2023high}%
  \BibitemOpen
  \bibfield  {author} {\bibinfo {author} {\bibfnamefont {Yao}\ \bibnamefont {Lu}}, \bibinfo {author} {\bibfnamefont {Aniket}\ \bibnamefont {Maiti}}, \bibinfo {author} {\bibfnamefont {John~WO}\ \bibnamefont {Garmon}}, \bibinfo {author} {\bibfnamefont {Suhas}\ \bibnamefont {Ganjam}}, \bibinfo {author} {\bibfnamefont {Yaxing}\ \bibnamefont {Zhang}}, \bibinfo {author} {\bibfnamefont {Jahan}\ \bibnamefont {Claes}}, \bibinfo {author} {\bibfnamefont {Luigi}\ \bibnamefont {Frunzio}}, \bibinfo {author} {\bibfnamefont {SM}~\bibnamefont {Girvin}}, \ and\ \bibinfo {author} {\bibfnamefont {Robert~J}\ \bibnamefont {Schoelkopf}},\ }\bibfield  {title} {\enquote {\bibinfo {title} {A high-fidelity microwave beamsplitter with a parity-protected converter},}\ }\href@noop {} {\bibfield  {journal} {\bibinfo  {journal} {arXiv preprint arXiv:2303.00959}\ } (\bibinfo {year} {2023})}\BibitemShut {NoStop}%
\bibitem [{\citenamefont {Wiseman}\ and\ \citenamefont {Milburn}(2009)}]{wiseman2009quantum}%
  \BibitemOpen
  \bibfield  {author} {\bibinfo {author} {\bibfnamefont {Howard~M}\ \bibnamefont {Wiseman}}\ and\ \bibinfo {author} {\bibfnamefont {Gerard~J}\ \bibnamefont {Milburn}},\ }\href@noop {} {\emph {\bibinfo {title} {Quantum measurement and control}}}\ (\bibinfo  {publisher} {Cambridge university press},\ \bibinfo {year} {2009})\BibitemShut {NoStop}%
\bibitem [{\citenamefont {Rist{\`e}}\ \emph {et~al.}(2012)\citenamefont {Rist{\`e}}, \citenamefont {Bultink}, \citenamefont {Lehnert},\ and\ \citenamefont {DiCarlo}}]{riste2012feedback}%
  \BibitemOpen
  \bibfield  {author} {\bibinfo {author} {\bibfnamefont {D}~\bibnamefont {Rist{\`e}}}, \bibinfo {author} {\bibfnamefont {CC}~\bibnamefont {Bultink}}, \bibinfo {author} {\bibfnamefont {Konrad~W}\ \bibnamefont {Lehnert}}, \ and\ \bibinfo {author} {\bibfnamefont {L}~\bibnamefont {DiCarlo}},\ }\bibfield  {title} {\enquote {\bibinfo {title} {Feedback control of a solid-state qubit using high-fidelity projective measurement},}\ }\href@noop {} {\bibfield  {journal} {\bibinfo  {journal} {Physical review letters}\ }\textbf {\bibinfo {volume} {109}},\ \bibinfo {pages} {240502} (\bibinfo {year} {2012})}\BibitemShut {NoStop}%
\bibitem [{\citenamefont {Campagne-Ibarcq}\ \emph {et~al.}(2013)\citenamefont {Campagne-Ibarcq}, \citenamefont {Flurin}, \citenamefont {Roch}, \citenamefont {Darson}, \citenamefont {Morfin}, \citenamefont {Mirrahimi}, \citenamefont {Devoret}, \citenamefont {Mallet},\ and\ \citenamefont {Huard}}]{campagne2013persistent}%
  \BibitemOpen
  \bibfield  {author} {\bibinfo {author} {\bibfnamefont {Philippe}\ \bibnamefont {Campagne-Ibarcq}}, \bibinfo {author} {\bibfnamefont {Emmanuel}\ \bibnamefont {Flurin}}, \bibinfo {author} {\bibfnamefont {Nicolas}\ \bibnamefont {Roch}}, \bibinfo {author} {\bibfnamefont {David}\ \bibnamefont {Darson}}, \bibinfo {author} {\bibfnamefont {Pascal}\ \bibnamefont {Morfin}}, \bibinfo {author} {\bibfnamefont {Mazyar}\ \bibnamefont {Mirrahimi}}, \bibinfo {author} {\bibfnamefont {Michel~H}\ \bibnamefont {Devoret}}, \bibinfo {author} {\bibfnamefont {Fran{\c{c}}ois}\ \bibnamefont {Mallet}}, \ and\ \bibinfo {author} {\bibfnamefont {Benjamin}\ \bibnamefont {Huard}},\ }\bibfield  {title} {\enquote {\bibinfo {title} {Persistent control of a superconducting qubit by stroboscopic measurement feedback},}\ }\href@noop {} {\bibfield  {journal} {\bibinfo  {journal} {Physical Review X}\ }\textbf {\bibinfo {volume} {3}},\ \bibinfo {pages} {021008} (\bibinfo {year} {2013})}\BibitemShut {NoStop}%
\end{thebibliography}%

\end{document}